\documentclass{article}
\usepackage{cite}
\def\vep{\varepsilon}

\def\be{\begin{equation}}
\def\ee{\end{equation}}
\def\bea{\begin{eqnarray}}
\def\eea{\end{eqnarray}}

\begin{document}

\title{
Nonstationary Casimir Effect and analytical solutions 
for quantum fields in cavities with moving boundaries
}

\author{V. V. DODONOV\thanks{E-mail:vdodonov\@df.ufscar.br}\\
Departamento de F\'{\i}sica, Universidade Federal de S\~{a}o Carlos,\\
Via Washington Luiz km 235, 13565-905  S\~ao Carlos,  SP, Brazil\\
Moscow Institute of Physics and Technology\\
Lebedev Physics Institute of Russian Academy of Sciences
}
\date{}
\maketitle

\section{Introduction }

Electrodynamics in vacuum or in media with moving boundaries was the
subject of numerous studies in the XXth century. It is sufficient to
remember that the famous Einstein's paper which gave birth to the
relativity theory was entitled ``Electrodynamics of moving bodies''
\cite{Ein1905}. The total number of publications in this field is enormous,
since it includes, in particular, such important problem as radio-location.
In this review we confine ourselves to the problem of {\em cavities\/}
with (ideal) {\em reflecting\/} moving boundaries.
This means that we consider the fields confined
in some limited volume, thus leaving aside the problem of the field
propagation in the (semi)infinite space and reflection from single
boundaries (having in mind that numerous references to the publications
related to this ``full space'' problem can be found,
e.g., in \cite{Fran72,Ostr75,Kunz80,Bol89}).

The plan of the chapter is as follows. In the next section we
give a brief historical review of the relevant studies,
both in the fields of classical and quantum electrodynamics,
supplying it with an extensive list of the known
publications. Although we tried to give a more or less complete list,
it is clear that some publications have been (undeliberately) missed.
To excuse we could mention that for decades
different groups of physicists and mathematicians performed the studies
in their own fields of interest, not suspecting the existence of
analogous results found in other areas. We hope that our review will
serve to diminish this gap essentially.

It is clear that all results accumulated for several decades
cannot be collected in one
chapter. Therefore in the following sections we have decided to make
an emphasis mainly on the detailed exposition of our own results
concerning the recently obtained
{\em analytical solutions\/} for the cavities with {\em resonantly
oscillating\/} boundaries, since we hope that these solutions could be
important for the further studies on the problem known under the names
``Nonstationary Casimir Effect'' or ``Dynamical Casimir Effect.''

\section[Brief history]
{Brief history of studies on electrodynamics with moving boundaries}

\subsection{Classical fields in cavities with moving boundaries}

The first exact solution of the wave equation ($c=1$)
\begin{equation}
\frac {\partial^2A}{\partial t^2}\,-\frac {\partial^
2A}{\partial x^2}=0,
\label{we}
\end{equation}
in a time-dependent domain $0<x<L(t)$ (where
$L(t)$ is the given law of motion of the right boundary),
satisfying the boundary conditions
\begin{equation}
A(0,t)=A(L(t),t)=0
\label{boundcon}
\end{equation}
was obtained by Nicolai \cite{Nic1} for
\be
L(t)=L_0(1+\alpha t).
\label{linL}
\ee
The solution was interpreted in terms of the transverse vibrations of
a string with a variable length. A few years later, these results were
published in \cite{Nic2}, where the extension to the case of
electromagnetic field was also made. A similar treatment
was given by Havelock \cite{Hav} in connection with the problem
of radiation pressure.
Quarter a century later, the one-dimensional wave
equation in the time-dependent interval interval $0<x<a+bt$ was considered
in \cite{Carr49} under the name ``Spaghetti problem.''

A new wave of interest to the problem of electromagnetic field in a cavity
with moving boundaries has arisen only in the beginning
of 60s, being motivated, partially, by the experiments \cite{Zag61}
on the ``field compression,'' accompanied by the frequency multiplication
(for 2.3 times) due to multiple reflections of the initial $H_{011}$ wave
($\lambda_{in}=10\,$cm) from the opposite sides of a resonator,
one of which was a ``plasma piston'' moving uniformly with the velocity
$v\sim 2\cdot 10^7\,$cm/s.
Kurilko \cite{Kur60} studied
linearly polarized electromagnetic field between two ideal infinite
plates moving with equal constant velocities towards each other.
He considered
consecutive reflections of the waves from each boundary, writing
explicit expressions for the finite time-space intervals, corresponding
to zero, one, two, etc., reflections.
Balazs \cite{Bal} gave a detailed study of the string problem
with the aid of the method similar to that used by Nicolai and Havelock.
Besides considering the uniformly moving boundary, he has found an exact
solution for $L(t)=\left(t^2+1\right)^{1/2}$, and gave
some graphical method of finding the solution for an arbitrary law
of motion $L(t)$.
Greenspan \cite{Greens63} studied the one-dimensional string
with the uniformly moving right boundary, assuming the boundary condition
at the left point in the form $A(0,t)=\sin(\omega t)$.
The problem of the ``field compression'' between two ideal infinite
moving walls (the same geometry as in \cite{Kur60}) was studied by
Stetsenko \cite{Stet63a,Stet63b}.
An approximate solution for
the electromagnetic field in a rectangular waveguide
cavity with a uniformly moving boundary was obtained in \cite{Stet64}.
In this case one has to solve the equation ($c=1$)
\be
A_{xx} -A_{tt} = \kappa^2 A
\label{eq-kap}
\ee
with the same boundary condition as in (\ref{boundcon}).
The detailed paper by Baranov and Shirokov \cite{Bar67} can be considered,
in a sense, as a concluding study of the one-dimensional problem
with a {\em uniformly moving boundary\/} (although many publications
on this subject continued to appear up to last years).
The experiments on {\em laser cavities\/} with uniformly moving mirrors
were described, e.g., in \cite{Smith67,Peek67,Klin72,Mak75,Anokh78}.
In these experiments, the constant velocity of the mirror
varied from 7 cm/s \cite{Smith67} to 400 m/s \cite{Anokh78}.

In short note \cite{Ask62}, Askar'yan has pointed out two possible effects
of {\em oscillating\/} surfaces on the electromagnetic field inside
the (laser) resonator cavities. The first one is the influence of
oscillations on the generation and intensity of the laser radiation.
It was extensively studied in many experiments, devoted, in particular,
to such problems as the generation of optical pulses \cite{Henneb,Ober93},
phase locking of laser modes (where the frequencies of the mirror
oscillations varied from 50 Hz \cite{Bamb68} to 500 KHz
\cite{500KHza,500KHzb} and 1 MHz \cite{Ger64,Ger69}, see the review in
\cite{Bern92}), or
modulation of the laser radiation
\cite{Rus65,Bel71} (in \cite{Bel71} the frequencies varied from 17 to
70 KHz). The theory of these phenomena was considered, e.g., in
\cite{Bod78,Korn80,Long99}.

The second effect predicted by Askar'yan was the {\em field amplification\/}
inside the cavity under the parametric resonance condition, when the
mirror oscillates at twice the field eigenfrequency. It was not
observed yet, as far as we know, and the main part
of the present review is devoted just to the progress in the theoretical
treatment of this phenomenon achieved in recent years.

In 1967, Grinberg \cite{Grin} has proposed
a general method of solving the wave equation in the case of an arbitrary
law of motion of the boundary, based on expanding the solution
over the complete set of ``instantaneous modes.''
Krasil'nikov \cite{Kras3d} seems to be the first who gave rather detailed
study of the electromagnetic vibrations in
a {\em spherical\/} cavity with an {\em oscillating boundary\/}.
The one-dimensional cavity with a {\em resonantly oscillating\/} boundary
was considered with the aid of the method of characteristics in study
\cite{Kras}, where it was shown that the energy is `pumped' to the
high-frequency modes at the expense of the lower-frequency ones.

A significant contribution was made in a series
of papers by Vesnitskii and co-authors. In \cite{Ves69} he gave an
exact solution for the problem of a rectangular waveguide with a
uniformly moving lateral wall, i.e., for the equation and
the boundary conditions
\be
A_{xx} + A_{zz} -A_{tt}= 0, \quad
\left.A\right|_{x=0} = \left.A\right|_{x=L(t)}=0
\label{lat-guide}
\ee
with $L(t)=L_0(1+\alpha t)$ (in this case, $A$ is the $E_y$ component
of the field of the type $H_{n0}$).
A spherical resonator, whose radius linearly changed with time,
was considered in \cite{VesKos71}.
A general (although implicit, in some sense) solution of the one-dimensional
problem
with an arbitrary law of motion of the boundary was given in
\cite{Ves-app}, where it was shown that a complete set of solutions to
the problem (\ref{we})-(\ref{boundcon}) can be expressed through
the solution of some simple functional equation (see equation
(\ref{Moore}) in the next subsection).
The solutions of the {\em inhomogeneous\/} one-dimensional wave equation
for the law of motion $L(t)=L_0(1+\alpha t)^{\pm 1}$, with an arbitrary
inhomogeneity and arbitrary initial conditions, were given in
\cite{Ves71-31}.
A family of concrete laws of motion admitting simple explicit
expressions for the mode functions
was found in the framework of the {\em inverse problem\/}
in \cite{Ves-osc}.
The two-dimensional rectangular membrane with a single uniformly moving
boundary was considered in \cite{Bor-Ves}. A possibility of the frequency
modulation in the waveguide with a slowly oscillating boundary was
studied in \cite{Bol-Ves}. A review of the results obtained by
Vesnitskii and co-authors was made in \cite{Ves-Pot}.

Exact solutions for the waveguide with {\em nonuniformly\/} moving
boundary (problem (\ref{lat-guide})) have been found by Barsukov and
Grigoryan \cite{Bars76-2,Bars76-4}. They considered also the
electromagnetic resonator with moving boundary \cite{Bars76-res}.
Similar problems were studied in \cite{Bart79,Bart82}.
Periodic solutions of a one-dimensional wave equation with homogeneous
conditions on moving boundaries were considered in \cite{Balan80}.
Transition processes in one-dimensional systems with moving boundaries
were studied in \cite{VesPot82}. Oscillations of a round membrane
with a uniformly varying radius were considered in \cite{BorVes85}.
A nonlinear transformation was applied to solve the inhomogeneous
problem of the forced resonance oscillations in a one-dimensional
cavity with moving boundaries in \cite{VesVes85}.
The scaling transformation method, which reduces the problem with
moving boundaries to that with fixed ones by means of the transformations
such as $x \rightarrow x/L(t)$, was considered in \cite{Wil88,WilHas89}.

In the last decade of the century, the problem of vibrating string
with moving boundaries was studied in \cite{Ram96} (two supports
moving toward each other with constant velocities) and
\cite{Coop93,Dit1,Yama97,Yama98,Yama00} (oscillating supports).
The electromagnetic (better to say, massless scalar) field
in one dimensional ideal cavities with periodically moving boundaries
were considered in \cite{Coop,Ditt96,Ditt,Gonz98,Llave99}.
Analytical solutions for the field in circular waveguides with
(linearly) moving boundaries were obtained recently in
\cite{Gaff1,Gaff2}.
The ``dynamical'' field modes in the one-dimensional and spherical cavities
with uniformly moving boundaries were found once more time in \cite{Dol98}.
The same problem for the expanding/contracting ideal spherical cavity,
whose radius varies as $R(t)=R_0\sqrt{1+\alpha t}$, was solved recently
in \cite{Mkrt00}. (A family of the laws of motion of the boundary,
which includes, in particular, the dependences
such as $R(t)=R_0\sqrt{1+\alpha t +\beta t^2}$,
$R(t)=D t + E + F(At +B)^{-1}$, and their combinations, was considered
in the case of the {\em diffusion-type\/} equations in
\cite{Grin69,Grin71}, where it was shown that this family admits
{\em exact\/} solutions of the problem.)

\subsection{Quantum fields in the presence of moving boundaries}

Moore's paper \cite{Moore} seems to be the first one devoted to the
problem of {\em quantum fields\/} in cavities mith moving boundaries.
It was motivated by the studies of a more general problem of the
particle (in particular, photon) creation in the nonstationary
universe \cite{Par68} and in external intense fields
(see, e.g., the books \cite{BirDav,Grib-book} and reviews by
Ritus and Nikishov in \cite{168}),
which is closely related to the problem of field quantization in
the spaces with nontrivial (e.g., time-dependent) geometry.
Considering a model of the ``scalar electrodynamics'' (when the
field depends on a single space coordinate), Moore has found
a complete set of solutions to the problem (\ref{we})-(\ref{boundcon})
in the form
\begin{equation}
A_n(x,t)= C_n\left\{\exp\left[-i\pi nR(t-x)\right]
-\exp\left[-i\pi nR(t+x)\right]\right\},
\label{anzMoore}
\end{equation}
where function $R(\xi)$ must satisfy the functional equation
\begin{equation}
R(t+L(t))-R(t-L(t))=2.
\label{Moore}
\end{equation}
As a matter of fact, quite similar approach was
used in \cite{Nic1,Nic2,Hav} for a {\it linear\/} function $L(t)$.
Independently, equation (\ref{Moore}) was obtained by Vesnitskii
\cite{Ves-app}.
Moore's approach was developed in \cite{BirDav,FulDav,Hos79}.
However, the most efforts were applied to the case of a {\em single\/}
mirror (not necessarily plane) moving with a {\em relativistic\/} velocity
or with a great acceleration,
when the effect of particle production becomes significant
\cite{DeWit75,Cand77,DavFul,Horib79,Oku79,FrolSer,FrolSer2,Ford,Sin,%
WalDav,CastFer,Walk85,Carl87,Otte88,NikRi95,Rit96,Frol99,And99}.

The case of two or more boundaries, one of which moves with a
{\em constant relativistic\/} velocity, was analyzed in
\cite{Bordag,Petrov85,Bordag86,Petrov89}
(the case of constant relative acceleration was considered in \cite{Bey90}).
For the uniform law of motion of the boundary (\ref{linL}) the
solution to Eq. (\ref{Moore}), found by many authors cited above,
reads (remember that we assume $c=1$)
\be
R_{\alpha}(\xi)=\frac{2\ln|1+\alpha\xi|}{\ln|(1+v)/(1-v)|},
\quad v=\alpha L_0.
\label{R-unif}
\ee
Evidently, if $\alpha\to 0$, this function goes to $R_0(\xi)=\xi/L_0$.
For an arbitrary nonrelativistic law of motion one can find the
solution in the form of the expansion over subsequent time derivatives
of the wall displacement. We give it in the form obtained in
\cite{DKM92-208}:
\be
R(\xi)= \xi\lambda(\xi) -\frac12 \xi^2 \dot\lambda(\xi)
+\frac16 \xi \ddot\lambda(\xi)\left[\xi^2- L^2(\xi)\right] +\cdots,
\quad \lambda(\xi)\equiv L^{-1}(\xi).
\label{R-approx}
\ee
In the special case $L(t)=L_0/(1+\alpha t)$, when
$\ddot\lambda(\xi)\equiv 0$, Eq. (\ref{R-approx}) yields another
exact solution, $R(\xi)=L_0^{-1}\left(\xi +\frac12\alpha\xi^2\right)$.
Unfortunately, the expansions such as (\ref{R-approx}) cannot be used
in the long-time limit $\xi \to\infty$, since the terms proportional
to the derivatives of $\lambda(\xi)$
(which are supposed to be small corrections)
become bigger than the unperturbed term $\xi\lambda(\xi)$.

Castagnino and Ferraro \cite{CastFer} have found several solutions of the
Moore equation (\ref{Moore}) with the aid of the inverse method, which
was used earlier by Vesnitskii \cite{Ves-osc}. In this method,
one chooses some reasonable function $R(\xi)$ and determines the
corresponding law of motion of the boundary $L(t)$ using the
consequence of Eq. (\ref{Moore}),
\be
\dot{L}(t)= \frac{R'[t-L(t)] -R'[t+L(t)]}{R'[t-L(t)] +R'[t+L(t)]}.
\label{CastFer-eq}
\ee
To solve differential equation (\ref{CastFer-eq}) with some simple
functions $R(\xi)$ is more easy (at least this can be done numerically)
than to solve the functional equation (\ref{Moore})
for the given function
$L(t)$. However, not for any simple function $R(\xi)$ the dependence
$L(t)$ appears admissible from the point of view of physics (the velocity
may occur greater than the speed of light, or some discontinuities may
arise). Actually, the cases considered in \cite{CastFer} correspond to
some monotonous displacements of the mirror from the initial to final
positions. Typical functions $R(\xi)$ used in \cite{CastFer} were
some combinations of $\xi/L_0$ and some trigonometric functions
such as $\sin(m\pi\xi/L_0)$. A large list of simple functions $R(\xi)$
(rational, exponential, logarithmic, hyperbolic, trigonometrical
and inverse trigonometrical) and
corresponding to them functions $L(t)$ was given in \cite{Ves-Pot}.
However, no one of these functions can be used in the parametric
resonance case. The {\em asymptotical\/} solution of the Moore
equation in the parametric resonance case
$L(t)=L_0\left[ 1+ \epsilon\sin(\pi qt/L_0)\right]$, $q=1,2,\ldots$,
was found in \cite{DK92,DK92a,KD92,DKN93}. For $\epsilon t\gg1$
it has the form (here $L_0=1$)
\be
R(t)=t- \frac{2}{\pi q}\,{\rm Im}\left\{\ln\left[
1+ \zeta +\exp(i\pi qt)(1-\zeta)\right]\right\},
\label{sol-M-as}
\ee
\[
 \zeta=\exp\left[(-1)^{q+1}\pi q\epsilon t\right],
\]
which clearly demonstrates that the asymptotical mode structure in the
resonance case is quite different from the mode structure inside the
cavity with unmoving walls.
The solution (\ref{sol-M-as}) was improved and generalized to the
case of {\em two\/} vibrating boundaries in \cite{Dal98,Dal99}.
However, the form of this solution is not very convenient for the
calculation of various sums giving the mean numbers of photons, energy,
etc., so a lot of rather sophisticated calculations must be done
before the final physical results could be obtained.

One of the first estimations of the number of photons which could be
created from {\em vacuum\/} in a cavity whose boundary moves with
{\em nonrelativistic\/} velocity have been performed by Rivlin \cite{Riv79},
who considered the parametric amplification of the initial vacuum
field oscillations in the framework of the classical approach.
He gave an estimation of the number of created photons
${\cal N}\sim (\epsilon\omega_1 t)^2$, where
$\epsilon\sim \delta L/ L$ is the relative amplitude of the
variation of the distance between the walls and
$\omega_1$ is the fundamental unperturbed field eigenfrequency
(provided the frequency of the wall vibrations $\omega_w$ is
close to $2\omega_1$).
A similar estimation was given by Sarkar \cite{Sarkar}, who used
an approximate solution to the Moore equation (\ref{Moore})
in the form of the asymptotic series
with respect to a small parameter $\epsilon$, found in \cite{Sarkar88}
(such solutions were constructed earlier in \cite{Bar67,Ves-app}).
However, the tremendous numerical values obtained by Sarkar were
quite unrealistic,
since he used the value of $\epsilon$ which was many orders of magnitude
higher than those which can be achieved in the laboratory under
real conditions.
Moreover, the oversimplified approach of Rivlin and simple
perturbative solutions
of Sarkar are not valid, as a matter of fact, under the resonance
conditions, due to the presence of the secular terms (as it always
happens for parametric systems).
Actually, the structure of the `dynamical'
modes in the resonance case is completely different
(in the most interesting long-time limit) from the simple
standing waves existing in the cavity with unmoving walls.
The same is true for the estimation made by Askar'yan \cite{Ask62}
in the classical case. He evaluated the average work done by the moving wall
on the field as $A\sim \overline{\int pv\,dt}$, where
$v=v_0\sin(\omega_w t + \phi)$ is the wall velocity, and $p(t)$ is the
radiation pressure. Taking the monochromatic dependence
$p(t)=p_0\sin^2(2\omega_1 t)$, he obtained the {\em linear\/}
dependence on time in the resonance case $\omega_w=2\omega_1$:
$A\sim p_0 v_0 t \sin\phi + const$.
However, these evaluations can serve only
as some indication that in the resonance case the energy of the field
can grow at the expense of the mechanical work done by the vibrating wall.
The real time dependence can be quite different, because the relation
$p(t)=p_0\sin^2(2\omega_1 t)$ holds, as a matter of fact, only until
$\epsilon\omega_1 t \ll 1$, whereas for larger times the effect of
the mode reconstruction must be taken into account. One of the goals
of this review is to demonstrate what happens in reality in the
resonance case.

Approximate solutions to the Moore equation, such as (\ref{R-approx}), were
used in \cite{DKM89} to evaluate corrections to the famous
Casimir attractive force between infinite ideal walls \cite{Cas}
(for the reviews on the Casimir effect see, e.g.,
\cite{Plun1,Mil,Most,Lamor99})
due to the {\em nonrelativistic\/}  motion of the walls.
The leading term of these corrections turned out proportional to the
{\em square of velocity\/} of the boundary (i.e., of the order of
$(v/c)^2$). The Casimir force in the relativistic case was calculated
in \cite{Moore,FulDav,BirDav,Ford,Sin,CastFer,Bordag,Bordag86};
it depends in the generic case not only on the instantaneous velocity,
but on the whole
time dependence $L(t)$ (through the function $R(\xi)$, i.e., on
the acceleration and other time derivatives).
The first calculations of the forces acting on the single mirrors moving with
{\em nonrelativistic\/} velocities due to the vacuum or thermal
fluctuations of the field were performed in the framework of the
spectral approach (using the fluctuation-dissipation theorem) in
\cite{Brag} (three-dimensional case, the force proportional to the
fifth-order derivative of the coordinate) and in
\cite{Jaekel} (one-dimensional model, the force proportional to the
third-order derivative of the coordinate). It was mentioned
that the force could be significantly amplified under
the resonance conditions, either in the $LC$-contour \cite{Brag}
or in the Fabry--Perot cavity \cite{Jaekel} (earlier, such a possibility
was discussed in \cite{DKM90}).
These studies were continued in
\cite{JaRe92-1,JaRe92-2,JaRe93-1,JaRe93-2,Eber93,NetRey93,%
Neto,NetMa95,Mats96},
where it was assumed that the velocity of
the boundary is perpendicular to the surface. The Casimir force between two
parallel plates, when their relative velocity is also parallel to
the surfaces, was considered by Levitov \cite{Levit}.
Later, the theory of ``Casimir friction'' was developped in
\cite{Hoy92,Hoy93,Mkr95,Pend97,Vol99}.
The reviews of these approaches can be found in \cite{JaRe97,KarGol99}.

Various quantum effects arising due to the motion of
{\em dielectric boundaries\/},
including the modification of the Casimir force and creation of photons,
both in one and three dimensions, and for different orientations of
the velocity vector with respect to the surface,
have been studied in detail in the series of papers by Barton and
his collaborators
\cite{BE,BarCal95,CalBar95,SalBar95,Bar96,BarNor96,GuEb98}.
In the paper \cite{BE} the term {\em Mirror-Induced Radiation\/} (MIR)
has been introduced.

Another name, {\it Nonstationary Casimir Effect\/} (NSCE), was
introduced earlier in \cite{DKM89} for the class of phenomena
caused by the reconstruction of the quantum state
of field due to a time dependence of the geometrical configuration
\cite{Man90,Man91,DOM91,Man92}.
Its synonim is the term {\it Dynamical Casimir Effect\/},
which became popular after the series of articles
by Schwinger  \cite{Sch1,Sch2,Sch3,Sch4,Sch5}, in which he
tried to explain the phenomenon of {\em sonoluminescence\/}
by the creation of photons in bubbles with time-dependent radii,
oscillating under the action of acoustic pressure in the liquids
(see a brief discussion of this subject in the last section).

A possibility of generating the ``nonclassical''
(in particular, {\em squeezed\/}) states
of the electromagnetic field in the cavity with moving walls
was pointed out in
\cite{DKM92-208,Sarkar,DKM90,DKM-IGTMP,DKM91,DKN91,Sark92}.
The dynamical Casimir force
was interpreted as a mechanical signature of the squeezing effect
associated with the mirror's motion in \cite{Jaekel,JaRe92-1}
(see also \cite{Weig}).

It was suggested also in
\cite{DKM92-208,DKM90,DKM91,DKN91}
that a significant
amount of photons could be created from vacuum even for quite small
nonrelativistic velocities of the walls, provided the boundaries of
a high-Q cavity perform small oscillations at a frequency proportional 
to some cavity unperturbed eigenfrequency, due to an accumulation of
 small changes in the state of the field for a long time.
Indeed, using the asymptotical solutions of the Moore equation
(\ref{sol-M-as}), it was shown in
\cite{DK92,DK92a,KD92,DKN93} that the rate of photon generation
in each mode
becomes constant in the long-time limit, being linearly proportional to the
product $\epsilon\omega_1$, and that the photons are generated in a wide
frequency band whose width grows exponentially fast in time \cite{Klim}.
Other approximate or exact solutions of the Moore equation
for the specific periodical time dependences of the cavity dimensions
were found by different methods in studies
\cite{Dal98,Dal99,LawPRL,Cole,Mep,Fu97,Wu}, which
confirmed the effect of resonance generation of photons.

Moore's approach is based on the decomposition (\ref{anzMoore})
of the field over the mode functions satisfying automatically the
(one-dimensional) {\em wave equation} (\ref{we}).
There exists another approach (proposed in the framework of the
classical problem as far back as in \cite{Grin}),
when the mode functions are chosen in such
a way that they satisfy automatically the time-dependent
{\em boundary conditions\/} (\ref{boundcon}): see Eq. (\ref{psit}) below.
In this case it is possible to describe the behavior of the field
with the aid of some {\em effective Hamiltonian\/}, which is an
infinite-dimensional quadratic
form of the boson creation/annihilation operators with time-dependent
coefficients responsible for the coupling between different modes.
Such an approach was considered for the first time (in the quantum case)
by Razavy and Terning \cite{RazTer84,RazTer85,Raz85}
(the case of massive field was considered in \cite{Raz83},
see also \cite{Reut89}).
However, the resulting
infinite set of coupled evolution equations for the annihilation and
creation operators turned out rather complicated in the generic case.
and for this reason they were
treated only perturbatively \cite{RazTer85}. Later, a similar method
was used by Calucci \cite{Calucci}, who confined, however, only with the
case of adiabatically slow motion of the wall, when no photons could
be created (recently, this adiabatic case was studied in detail
in \cite{Jan}).
An essential progress in the development of the Hamiltonian
approach (we shall call it also the {\em instantaneous basis method\/} or
IBM-method) was achieved after the papers by Law \cite{Law,Law-new},
who has demonstrated that the effective Hamiltonian can be significantly
simplified under the resonance conditions.
Nonetheless, even
the reduced coupled equations of motion resulting from the
simplified Hamiltonians have been treated for some time
either perturbatively (i.e., in the form of
the Taylor expansion with respect to the time variable) or
numerically (actually, also in the
short-time limit corresponding to the initial stage of the process);
see the studies
\cite{Law,Law-new,CSZ,Ji97,Ji98,Ji98b} (for the 1D cavity model)
and \cite{Ji98a} (for a three-dimensional cavity and a waveguide).
The general structure of the
effective infinite-dimensional quadratic Lagrangians and Hamiltonians
arising in the canonical approach to the dynamical Casimir effect
was analyzed and classified in \cite{JoSar96,Plun,Plun99}.
The methods of diagonalization
of such Hamiltonians were considered in \cite{Wu2,Wu3}.

The first {\em analytical\/} solutions describing the field inside
the 1D cavity with resonantly oscillating boundaries have been found
in the simplest cases in \cite{D95,DK96,D96}. More general solutions
have been obtained in \cite{Djpa,DA,AD00}.
They hold for any moment of time (provided the amplitude of the wall
vibrations is small enough; this limitation, however, is quite unessential
under realistic conditions). Moreover, these solutions enable us not only
to calculate the number of photons created from an arbitrary initial state
(thus giving, for instance, the temperature corrections), but also to
account for the effects of detuning from a strict resonance.
Besides, they enable to calculate the degree of {\it squeezing\/} in the
field quadrature components, to find the photon distribution function
and the energy density distribution inside the cavity, etc..
Therefore, one of the main purposes of this chapter is to give
the detailed description of the analytical solutions found in
\cite{D95,DK96,D96,Djpa,DA,AD00} and to discuss their physical
consequences.

Another goal is to consider the simplest models of the
{\em three-dimensional\/} cavity, following the scheme given
in \cite{D95,DK96} (for ideal boundaries) and in \cite{D98,D98a}
(for lossy cavities).
A more detailed study of the quantum properties of the electromagnetic
field in rectangular 3D cavities, which takes into account the
polarization of the field, was performed
in \cite{JaViHa95,ViHaJa95}, but only for the uniform motion
of the walls. The periodic motion was considered in \cite{Mund},
also with account of the polarization and the influence of all
three dimensions, but in the framework of some approximations
equivalent to the short-time limit.
The case of a three-dimensional rectangular cavity
divided in two parts by an ideal mirror, which suddenly disappears,
was considered in \cite{Cir99}.

The name {\em Motion Induced Radiation\/} was given to the effect
of radiation emission outside the cavity with vibrating walls \cite{Lamb}.
Using the spectral approach, the authors of \cite{Lamb} showed that
the radiation can be essentially enhanced under the resonance conditions
(by the orders of magnitude, comparing with the case of a single mirror).
The problem of the photon generation by a single perfectly reflecting
mirror performing a bounded nonrelativistic motion was studied in
\cite{NetoMa96,Mend}, where the effects of polarization have been taken into
account, and the spectral and angular distributions of the emitted photons
have been found.
Arbitrary space-time deformations of a single moving mirror have been
treated in \cite{GolKar97,Miri}  with the aid of the path-integral
approach. The same approach was used in the case of a cavity with
deformable perfectly reflecting boundaries in \cite{Gol98}.

The creation of photons or specific (e.g., squeezed) states of the
electromagnetic field due to the motion of some effective mirrors
made of the free electrons moving with (ultra)relativistic velocities
was considered in \cite{Kul96,Kul97}.
Another kind of ``effective moving mirrors'' consisting of the
electron-hole plasma generated in semiconductors under the action of
powerful laser pulses was suggested in \cite{Loz1,Loz}.

The influence of temperature on the dynamical Casimir effect was
evaluated in \cite{DKN93,Lamb-puls}; a more detailed analysis
was given in \cite{Plun00} (in the framework of the Hamiltonian approach),
\cite{Hui00} (using the thermofield dynamics), and especially in
\cite{DA,AD00}.
The energy density distribution inside the cavity under the resonance
conditions is no more uniform, on the contrary,
the main part of the energy is concentrated
in several sharp peaks which move from one
boundary to another, becoming more and more narrow with
the course of time.
This effect was studied in the framework of numerical calculations in
\cite{Dal99,LawPRL,Cole,Mep,Wu} (earlier, it was discussed in
\cite{Ves-osc} for the classical field). The analytical form of the
pulses, including their ``fine structure'' in the case of initial
states different from the vacuum or thermal ones, was found in
\cite{AD00}.
A similar pulse structure
of radiation emitted from the high finesse vibrating cavity
with partially transparent mirrors was studied in
\cite{Lamb-puls,Lamb98}.

The evolution of {\em classical\/} fields in the cavities filled in
with media whose dielectric properties vary in time was considered,
e.g., as far back as in \cite{Yar66}.
Yablonovitch \cite{Yabl89} proposed to use a medium with a rapidly
 decreasing in time refractive index (``plasma window'')
 to simulate the so-called  {\em Unruh effect\/} \cite{Unruh},
i.e., creation of quanta in an accelerated frame of reference.
More rigorous and detailed studies of quantum phenomena in
nonstationary (deformed) media have been performed in
\cite{DKM91,DKN91,BB87,Lob91a,Lob91b,DGMUY92,Hizh92,DKNPR,Kiel,Oku95,%
Art96,Cir97,Art98,Pen,Eber99}.
The case when the dielectric constant changes simultaneously with
the distance between mirrors (in one dimension) was considered in
\cite{SaHy96a,SaHy96b}.
A comparison of the spectra of photons created due to the motion of
mirrors and
due to the time variations of the dielectric permeability was made
in \cite{JoSar95}.
An analog of the nonstationary Casimir effect in the superfluid $^3$He,
namely, the friction force on the moving interface between two different
phases, was discussed in \cite{Vol96}.

The problem of the interaction between the electromagnetic field created
due to the NSCE and various detectors (harmonic oscillators,
two-level systems, Rydberg atoms, etc.,) placed inside the cavity with
moving walls was studied by means of different methods in
\cite{D95,DK96,Law95,JaVi96,TaKo98,Janow98,Fedot00}. It is also discussed
in this chapter.

The first experiments on the interaction between the powerful laser
radiation and freely suspended light mirrors have been reported
in \cite{Dor,Coz85}, where the effect of the optical bistability,
similar to that observed usually in the so-called Kerr media, was
observed. The first theoretical study of this phenomenon in the cavities
whose walls can move under the action of the radiation pressure
force was given in \cite{Mey}.
This subject received much attention for the last decade in connection
with different problems, such as, e.g., the attenuation or elimination
of noise in interferometers
\cite{Sol91,Luis92,Fabre94,Jac,Manc94,HeiRey94,PinFab95,%
Heid97,JacTit99,Pinar99,Coha99,Brif00}.
It is important, in particular, for the gravitational wave detectors
and for the general problem of measuring weak
forces acting on a quantum system: see, e.g.,
\cite{Pace,BoOn96} for more details.
Another possible application could be the
generation of the so-called ``nonclassical states''
of the field and the mirror itself (when it is considered as a quantum
object, too) \cite{Brif00,Manci,Bose,Zheng98}.
The stability of such states, in turn, is closely related to the general
problem of {\em decoherence\/} in quantum mechanics, and one of the
mechanisms which can destroy the coherence of pure quantum superpositions
is just the dynamical Casimir effect
\cite{DaNe00,NeDa00}.
The back reaction of the dynamical Casimir effect was recently studied in
 \cite{Naga00}.
The influence of fluctuations of the positions of the walls
on the field inside the cavity was considered in
\cite{Sarkar88,Gol98,Ford98},
whereas the Brownian motion of the walls due
to the field fluctuations was studied in \cite{JaRe97,Gour99}.
The role of the dynamical Casimir effect in the cosmological problems
was studied recently in \cite{Anast00,Brev00}.

\section{1D cavity with oscillating boundaries}

Let us start with the case of a single space dimension.
Consider a cavity formed by two infinite ideal plates moving in
accordance with the prescribed laws
\[
x_{left}(t)=u(t), \quad
x_{right}(t)=u(t)+L(t)
\]
where $L(t)>0$ is the time dependent length of the cavity.
Taking into account only the electromagnetic modes
whose vector potential is directed along $z$-axis
(``scalar electrodynamics'' \cite{Moore}),
one can write down
the  field operator {\em in the Heisenberg representation\/}
$\hat {A}(x,t)$ at $t\le 0$ (when both the plates were at rest at the
positions $x_{left}=0$ and $x_{right}=L_0$) as (we assume $c=\hbar=1$)
\begin{equation}
\hat {A}_{in}=2\sum_{n=1}^{\infty}\frac 1{\sqrt {n}}\sin\frac {
n\pi x}{L_0}\hat b_n\exp\left(-i\omega_nt\right)+\mbox{h.c.}
 \label{Ast}
 \end{equation}
where $\hat {b}_n$ means the usual annihilation photon
operator and $\omega_n=\pi n/L_0$.
The choice of coefficients in equation (\ref{Ast})
corresponds to the standard form of the field Hamiltonian
\begin{equation}
\hat {H}\equiv\frac 1{8\pi}\int_0^{L_0}\mbox{d}x\,\left
[\left(_{}\frac {\partial A}{\partial t}\right)^2+\left(_{}\frac {
\partial A}{\partial x}\right)^2\right]
=\sum_{n=1}^{\infty}\omega_n\left(\hat
b^{\dag}_n\hat b_n+\frac 12\right). \label{Ham}
\end{equation}
For $t>0$ the field operator can be written as
\be
\hat {A}(x,t)=2\sum_{n=1}^{\infty}\frac 1{\sqrt {n}}\left[
\hat b_n\psi^{(n)}(x,t)\,+\,\mbox{h.c.}\,\right].
\label{At>0}
\ee
To find the explicit form of functions $\psi^{(n)}(x,t)$, $n=1,2,\ldots$,
one should take into account that the field operator must satisfy

i) the wave equation (\ref{we}),

ii) the boundary conditions (\ref{boundcon}) or their generalization
\begin{equation}
A(u(t),t)=A(u(t)+L(t),t)=0,
\label{boundcon1}
\end{equation}

iii) the initial condition (\ref{Ast}), which is equivalent to
\begin{equation}
\psi^{(n)}\left(x,t<0\right)=\sin\frac {n\pi x}{L_0}
\exp\left(-i\omega_nt\right).
\label{init}
\end{equation}
Following the approach of Refs.
\cite{Grin,RazTer85,Calucci,Law,Djpa}
we expand the function $\psi^{(n)}(x,t)$
in a series with respect to the {\em instantaneous basis\/}:
\begin{equation}\psi^{(n)}(x,t>0)=\sum_{k=1}^{\infty}
Q_k^{(n)}(t)\sqrt {\frac {
L_0}{L(t)}}\sin\left(\frac {\pi k[x-u(t)]}{L(t)}\right),
\quad n=1,2,\ldots
\label{psit}
\end{equation}
with the initial conditions
\[
Q_k^{(n)}(0)=\delta_{kn},\quad\dot {Q}_k^{(n)}(0)=-i\omega_n\delta_{kn},
\quad k,n=1,2,\ldots
\]
This way we satisfy automatically both the boundary conditions
(\ref{boundcon1}) and the initial condition (\ref{init}).
Putting expression  (\ref{psit}) into the wave equation (\ref{we}),
one can arrive after some algebra at an infinite set of coupled
differential equations \cite{Ji98,Plun,Djpa}
\begin{equation}
\ddot {Q}_k^{(n)}+\omega_k^2(t)Q_k^{(n)}
=2\sum_{j=1}^{\infty}  g_{kj}(t)\dot {Q}_j^{(n)}+
\sum_{j=1}^{\infty}  \dot{g}_{kj}(t) Q_j^{(n)}
+{\cal O}\left(g_{kj}^2\right),
\label{Qeq}
\end{equation}
where
\[
\omega_k(t)= {k\pi}/{L(t)}
\]
and the time dependent antisymmetric coefficients $g_{kj}(t)$ read
(for $j\neq k$)
\begin{equation}
g_{kj}=-g_{jk}=(-1)^{k-j}\frac {2kj
\left(\dot {L} +\dot {u}\epsilon_{kj}\right)}{\left(j^2-k^2\right)L(t)},
\quad
\epsilon_{kj}= 1-(-1)^{k-j}.
\label{gkj}
\end{equation}
For $u=0$ (the left wall at rest) the equations like
(\ref{Qeq})-(\ref{gkj}) were derived in \cite{Calucci,Law-new}.

If the wall comes back to its initial position $L_0$ after some interval of
time $T$, then the right-hand side of equation (\ref{Qeq}) disappears, so
at $t>T$ one can write
\begin{equation}
Q_k^{(n)}(t)=\xi_k^{(n)}e^{-i\omega_k(t+\delta T)}
+\eta_k^{(n)}e^{i\omega_k(t+\delta T)},
\quad k,n=1,2,\ldots
\label{ksi}
\end{equation}
where $\xi_k^{(n)}$ and $\eta_k^{(n)}$ are some constant complex
coefficients.
Consequently, at $t>T$ the initial annihilation operators $\hat {b}_n$
cease to be ``physical'', due to the contribution of the terms with
``incorrect signs'' in the exponentials $\exp(i\omega_kt)$. Introducing
a new set of ``physical'' operators $\hat {a}_m$
and $\hat {a}_m^{\dag}$, which give the decomposition of the vector
potential operator at $t>T$ in the form analogous to (\ref{Ast}),
\be
 \hat {A}(x,t)=\sum_{n=1}^{\infty}\frac 2{\sqrt {n}}
\sin\left(\pi nx/L_0\right)\left[
\hat a_n
e^{-i\omega_n(t+\delta T)}
\,+\,\mbox{h.c.}\,\right]
\label{vecpotfin}
\ee
one can easily check that the two sets
of operators are related by means of the Bogoliubov transformation
\begin{equation}
\hat {a}_m=\sum_{n=1}^{\infty} \left(\hat b_n\alpha_{nm}+\hat b_
n^{\dag}\beta_{nm}^{*}\right), \quad m=1,2,\ldots
\label{Bogol}
\end{equation}
with the coefficients
\begin{equation}\alpha_{nm}=\sqrt {\frac mn}\xi_m^{(n)},\qquad\beta_{
nm}=\sqrt {\frac mn}\eta_m^{(n)}.\label{al-ksi}
\end{equation}
The unitarity of the transformation (\ref{Bogol}) implies the following
constraints:
\begin{eqnarray}
&&\sum_{m=1}^{\infty}\left(\alpha_{nm}^*\alpha_{km} -
\beta_{nm}^*\beta_{km}\right)
= \sum_{m=1}^{\infty} \frac{m}{n}
\left(\xi_{m}^{(n)*}\xi_{m}^{(k)} -
\eta_{m}^{(n)*}\eta_{m}^{(k)} \right) =\delta_{nk}
\label{cond1} \\[2mm]
&&\sum_{n=1}^{\infty}\left(\alpha_{nm}^*\alpha_{nj} -
\beta_{nm}^*\beta_{nj}\right)
= \sum_{n=1}^{\infty}\frac{m}{n}
\left(\xi_{m}^{(n)*}\xi_{j}^{(n)} -
\eta_{m}^{(n)*}\eta_{j}^{(n)} \right) =\delta_{mj}
\label{cond2}\\[2mm]
&&\sum_{n=1}^{\infty}\left(\beta_{nm}^*\alpha_{nk} -
\beta_{nk}^*\alpha_{nm}\right)
= \sum_{n=1}^{\infty}\frac{1}{n}
\left(\eta_{m}^{(n)*}\xi_{k}^{(n)} -
\eta_{k}^{(n)*}\xi_{m}^{(n)} \right) =0
\label{cond3}
\end{eqnarray}

The mean number of photons in the $m$th mode equals the
average value of the operator $\hat {a}_m^{\dag}\hat {a}_m$ in the initial
state $|{\rm in}\rangle$ (remember that we use the Heisenberg picture),
since just this operator has a physical meaning at $t>T$:
\begin{eqnarray}
&&{\cal N}_m \equiv \langle {\rm in}|\hat {a}_m^{\dag}
\hat {a}_m|{\rm in}\rangle \nonumber\\[2mm]
&&=\sum_n|\beta_{nm}|^2 +\sum_{n,k}\left[\left(\alpha_{nm}^*\alpha_{km}
+\beta_{nm}^*\beta_{km}\right)\langle\hat {b}_n^{\dag}\hat {b}_k\rangle
+ 2 {\rm Re}\left(\beta_{nm}\alpha_{km}\langle\hat {b}_n\hat {b}_k
\rangle\right)\right]
\nonumber\\[2mm]
&&= \sum_{n=1}^{\infty}\frac{m}{n}|\eta_m^{(n)}|^2
+ \sum_{n,k=1}^{\infty} \frac{m}{\sqrt{nk}}
\left(\xi_{m}^{(n)*}\xi_{m}^{(k)} +
\eta_{m}^{(n)*}\eta_{m}^{(k)} \right)
\langle\hat {b}_n^{\dag}\hat {b}_k\rangle
\nonumber\\[2mm]
&&+ 2{\rm Re}\sum_{n,k=1}^{\infty} \frac{m}{\sqrt{nk}}
\eta_{m}^{(n)}\xi_{m}^{(k)}\langle\hat {b}_n\hat {b}_k \rangle .
\label{number}
\end{eqnarray}
The first sum in the right-hand sides of the relations above describes the
effect of the photon creation from vacuum due to the NSCE,
while the other sums are different from zero only in the case of a
nonvacuum initial state of the field.

To find the coefficients $\xi_k^{(n)}$ and $\eta_k^{(n)}$ one has to solve
an infinite set of coupled equations (\ref{Qeq}) ($k=1,2,\ldots$)
with time-dependent coefficients, moreover, each
equation also contains an infinite number of terms.
However, the problem can be essentially simplified, if the walls
perform small oscillations at the frequency $\omega_w$ close to some
unperturbed field eigenfrequency:
\[
L(t)=L_0\left(1+\varepsilon_L \sin\left[p\omega_1(1+\delta)t\right]
\right), \quad
u(t)=\varepsilon_u L_0\sin\left[p\omega_1(1+\delta)t +\varphi\right].
\]
Assuming $|\varepsilon_L|,|\varepsilon_u|\sim \varepsilon\ll 1$,
it is natural to look for
the solutions of equation (\ref{Qeq}) in the form similar to (\ref{ksi}),
\begin{equation}
Q_k^{(n)}(t)=\xi_k^{(n)}e^{-i\omega_k(1+\delta)t}
+\eta_k^{(n)}e^{i\omega_k(1+\delta)t},
\label{ksitime}
\end{equation}
but now we allow the
coefficients $\xi_k^{(n)}$ and $\eta_k^{(n)}$ to be
{\em slowly varying functions of time}.
The further procedure is well known in the theory of
parametrically excited systems \cite{Louis,Land,Bogol}.
First we put
expression (\ref{ksitime}) into equation  (\ref{Qeq}) and neglect the terms
$\ddot{\xi },\ddot{\eta}$ (having in mind that $\dot{\xi },\dot{\eta}
\sim\varepsilon$, while $\ddot{\xi },\ddot{\eta}\sim\varepsilon^2$),
as well as the terms proportional to
$\dot{L}^2\sim\dot{u}^2\sim\varepsilon^2$.
Multiplying the resulting equation for $Q_k$ by the factors
$\exp\left[i\omega_k(1+\delta)t\right]$ and
$\exp\left[-i\omega_k(1+\delta)t\right]$ and performing
averaging over fast oscillations with the frequencies proportional to
$\omega_k$ (since the functions $\xi ,\eta$ practically do not change
their values at the time scale of $2\pi /\omega_k$)
one can verify that only the terms with the difference $j-k=\pm p$
survive in the right-hand side. Consequently, for {\it even\/} values of
$p$ the term $\dot{u}$ in $g_{kj}(t)$ does not make any contribution to
the simplified equations of motion, thus only the rate of change of the
cavity length $\dot{L}/L_0$ is important in this case.
On the contrary, if $p$ is an {\it odd\/} number, then the
field evolution depends on the velocity of the {\it centre of the cavity\/}
$v_c=\dot{u}+\dot{L}/2$ and does not depend on $\dot{L}$ alone.
These {\it interference effects\/} were discussed
in the short time limit $\varepsilon\omega_1 t\ll 1$ in \cite{Ji98}
(see also \cite{Lamb}).
We assume hereafter that $u=0$ (i.e. that the left wall is at rest),
since this assumption does not change anything if $p$ is an even number,
whereas one should simply replace $\dot{L}/L_0$ by $2v_c/L_0$
if $p$ is an odd number.

The final equations for the coefficients
$\xi_k^{(n)}$ and $\eta_k^{(n)}$ contain only three terms with simple
{\it time independent\/} coefficients in the right-hand sides:
\begin{eqnarray}
\frac {\mbox{d}}{\mbox{d}\tau}\xi_k^{(n)}&=&
(-1)^p\left[(k+p)\xi_{k+p}^{(n)}- (k-p)\xi_{k-p}^{(n)}\right]
+2i\gamma k \xi_{k}^{(n)} ,
\label{pksik}\\
 \frac {\mbox{d}}{\mbox{d}\tau}\eta_k^{(n)}&=&
(-1)^p\left[(k+p)\eta_{k+p}^{(n)} -(k-p)\eta_{k-p}^{(n)}\right]
- 2i\gamma k \eta_{k}^{(n)}.
\label{petak}
\end{eqnarray}
The dimensionless parameters $\tau$ (a ``slow'' time) and $\gamma$ read
($\varepsilon\equiv\varepsilon_L$)
\begin{equation}
\tau =\frac 12\varepsilon\omega_1t, \qquad
\gamma=\delta/\varepsilon.
\label{tau}
\end{equation}
The initial conditions are
\begin{equation}\xi_k^{(n)}(0)=\delta_{kn},\qquad\eta_k^{(n)}(0)=0.
\label{ini}
\end{equation}
Note, however, that uncoupled equations (\ref{pksik})-(\ref{petak}) hold
only for $k\ge p$. This means that they describe the evolution of
{\it all\/} the Bogoliubov coefficients only if $p=1$. Then
{\it all\/} the functions $\eta_k^{(n)}(t)$ are {\it identically
equal to zero} due to the initial conditions (\ref{ini}),
consequently, no photon can be created from vacuum.
Moreover, in the next section we show that
the total number of photons (but not the total energy)
is an integral of motion in this specific case.

\section{``Semi-resonance'' case ($p=1$)}\label{semi}

If $p=1$,
one has to solve the set of equations ($k,n=1,2,\ldots$)
\begin{equation}
\frac {\mbox{d}}{\mbox{d}\tau}\xi_k^{(n)}=
(k-1)\xi_{k-1}^{(n)} - (k+1)\xi_{k+1}^{(n)} +2i\gamma k \xi_{k}^{(n)}\,.
\label{1ksik}
\end{equation}
An immediate consequence of these equations
 and the condition $\xi_k^{(n)}(0)=\delta_{kn}$  is the identity
\begin{equation}
\sum_{m}m \xi_{m}^{(n)}(\tau)\xi_{m}^{(k)*}(\tau) \equiv
n\delta_{nk},
\label{invar}
\end{equation}
which is nothing but the unitarity condition of the Bogoliubov
transformation in this special case.
Taking into account this identity, one can easily verify that
the total average number of photons in all modes is conserved in time:
\begin{equation}
{\cal N}=\sum_{n k m}\frac{m}{\sqrt{nk}}\xi_{m}^{(n)*}
\xi_{m}^{(k)}\langle\mbox{in}|\hat {b}_n^{\dag}\hat {b}_k|\mbox{in}\rangle
=\sum_n \langle\mbox{in}|\hat {b}_n^{\dag}\hat {b}_n|\mbox{in}\rangle.
\label{num}
\end{equation}
A similar phenomenon in the classical case was discussed in \cite{Ves-osc},
whereas the quantum case was considered in \cite{Law} and
especially in \cite{D96}.
Also, using Eqs. (\ref{1ksik}) and (\ref{invar}) one can verify
that the total energy (normalized by $\omega_1$)
\[
{\cal E}=\sum_{n k m}\frac{m^2}{\sqrt{nk}}\xi_{m}^{(n)*}
\xi_{m}^{(k)}\langle\mbox{in}|\hat {b}_n^{\dag}\hat {b}_k|\mbox{in}\rangle
\]
satisfies the simple equation
(hereafter overdots mean the differentiation with
respect to the dimensionless time $\tau$)
\begin{equation}
\ddot{\cal E}=4a^2{\cal E} +4\gamma^2{\cal E}(0)
-2\gamma {\rm Im}({\cal G}_1),
\label{eqEtot}
\end{equation}
where
\begin{equation}
 a= \sqrt{1-\gamma^2}\;,
\label{def-a}
\end{equation}
\begin{equation}
{\cal G}_1 = 2\sum_{n=1}^{\infty}
\sqrt{n(n+1)}\langle\hat {b}_n^{\dag}\hat {b}_{n+1}\rangle.
\label{defcalG1}
\end{equation}
The quantum averaging is performed over the initial state of the field
(no matter pure or mixed).
The initial values of the total energy and its first derivative
(with respect to $\tau$) are given by
\be
{\cal E}(0)= \sum_{n=1}^{\infty} n
\langle\hat {b}_n^{\dag}\hat {b}_{n}\rangle , \quad
\dot{\cal E}(0)= {\rm Re}({\cal G}_1).
\label{inconE}
\end{equation}
Consequently, the solution to equation (\ref{eqEtot}) can be expressed as
\be
{\cal E}(\tau)= {\cal E}(0) +\frac{2\sinh^2(a\tau)}{a^2}
 \left[{\cal E}(0) -\frac{\gamma }{2}
 {\rm Im}({\cal G}_1)\right]
+{\rm Re}({\cal G}_1) \frac{\sinh(2a\tau)}{2a}.
\label{ansEtot1}
\ee
One can easily prove that $|\dot{\cal E}(0)|\le{\cal E}(0)$. Thus the total
energy grows exponentially when $\tau\gg 1$ (provided $\gamma< 1$),
although it can decrease at $\tau\ll 1$, if $\dot{\cal E}(0)<0$.
Since the total number of
photons is constant, such a behaviour is explained by the effect of pumping
the highest modes at the expense of the lowest ones (in the classical case
this effect was noticed in \cite{Kras}).

We see that the total energy can be found without any knowledge of
the Bogoliubov coefficients. However, these coefficients are necessary,
if one wants to know the distribution of the energy or the mean photon
numbers over the modes.
To solve the infinite set of equations (\ref{1ksik}) we
introduce the {\em generating function}
\begin{equation}
X^{(n)}(z,\tau)=\sum_{k=1}^{\infty}\xi_k^{(n)}(\tau )z^k
\label{1defX}
\end{equation}
where $z$ is an auxiliary variable.
Using the relation $kz^k=z(\mbox{d}z^k/\mbox{d}z)$ one obtains
the first-order partial differential equation
\begin{equation}
\frac{\partial X^{(n)}}{\partial\tau}=\left(z^2-1 +2i\gamma z\right)
\frac{\partial X^{(n)}}{\partial z}+\xi_1^{(n)}(\tau)
\label{eqG}
\end{equation}
whose solution satisfying the initial condition $X^{(n)}(0,z)=z^n$ reads
\begin{equation}
X^{(n)}(z,\tau)=\left[\frac{z g(\tau) -S(\tau)}
{g^*(\tau)- zS(\tau)}\right]^n
+\int_0^{\tau} \xi_1^{(n)}(x)\,\mbox{d}x
\label{solG}
\end{equation}
where
\begin{equation}
S(\tau)=\sinh(a\tau)/a\,, \quad g(\tau)= \cosh(a\tau) + i\gamma S(\tau)\,.
\label{def-gS}
\end{equation}
Differentiating (\ref{solG}) over $z$ we find
\begin{equation}
\xi_1^{(n)}(\tau)=\frac{n[-S(\tau)]^{n-1}}
{[g^*(\tau)]^{n+1}}.
\label{sol-1n}
\end{equation}
Putting this expression into the integral in the right-hand side of
equation (\ref{solG}) we arrive at the final form of the generating function
\begin{equation}
X^{(n)}(z,\tau)=\left[\frac{z g(\tau) -S(\tau)}
{g^*(\tau)- zS(\tau)}\right]^n
-\left[\frac{ -S(\tau)}{g^*(\tau)}\right]^n
\label{solGfin}
\end{equation}
which satisfies automatically the necessary boundary condition
$X^{(n)}(\tau,0)=0$.
The right-hand side of (\ref{solGfin}) can be expanded into the power
series of $z$
with the aid of the formula (\cite{Bateman}, vol. 3, section 19.6,
equation (16))
\[
(1-t)^{b-c}(1-t+xt)^{-b}=\sum_{m=0}^\infty \frac{t^m}{m!} (c)_m
F(-m,b;c;x),
\]
where $F(a,b;c;x)$ is the Gauss hypergeometric function, and
$(c)_k\equiv \Gamma(c+k)/\Gamma(c)$.
In turn, the function $(c)_m F(-m,b;c;x)$  with an integer $m$
is reduced to the Jacobi polynomial in accordance with the formula
(\cite{Bateman}, vol. 2, section 10.8, equation (16))
\[
(c)_m F(-m,b;c;x)=m!(-1)^m P_m^{(b-m-c,\,c-1)}(2x-1).
\]
Consequently,
\begin{equation}
(1-t)^{b-c}(1-t+xt)^{-b}=\sum_{m=0}^\infty
(-t)^m P_m^{(b-m-c,\,c-1)}(2x-1)
\label{genBatJac}
\end{equation}
and the coefficient $\xi_m^{(n)}(\tau)$ reads
\begin{equation}
\xi_m^{(n)}(\tau)=(-\kappa)^{n-m}\lambda^{n+m}
P_m^{(n-m,\,-1)}\left(1-2\kappa^2\right)
\label{sol-mnJac}
\end{equation}
where
\begin{eqnarray}
\kappa(\tau)&=&\frac{S}{\sqrt{gg^*}}
\equiv \frac{S(\tau)}
{\sqrt{1+S^2(\tau)}}
\label{def-kap}\\
\lambda(\tau)&=&\sqrt{g(\tau)/g^*(\tau)}\equiv
\sqrt{1-\gamma^2\kappa^2}+i\gamma\kappa, \quad |\lambda|=1.
\label{def-lam}
\end{eqnarray}
The form (\ref{sol-mnJac}) is useful for $n\ge m$. To find a convenient
formula in the case of $n\le m$ we introduce the {\it two-dimensional\/}
generating function
\begin{eqnarray}
&&X(\tau,z,y)=\sum_{m=1}^\infty\sum_{n=1}^\infty z^m y^n
\xi_m^{(n)}(\tau)=\sum_{n=1}^\infty X^{(n)}(z,\tau)y^n \nonumber\\
&&=
\frac{ yz}{[g^*(\tau)+yS(\tau)]
[g^*(\tau) -g(\tau)yz+ S(\tau)(y-z)]}.
\label{G}
\end{eqnarray}
The coefficient at $z^m$ in (\ref{G}) yields another
one-dimensional generating function
\begin{equation} X_{m}(\tau,y)=\sum_{n=1}^\infty y^n \xi_m^{(n)}(\tau)
= y\frac{[g(\tau)y+S(\tau)]^{m-1}}
{[g^*(\tau) +yS(\tau)]^{m+1}}.
\label{Gm}\end{equation}
Then equation (\ref{genBatJac}) results in the expression
\begin{equation}
\xi_m^{(n)}=
(1-\kappa^2)\kappa^{m-n}\lambda^{n+m}
P_{n-1}^{(m-n,\,1)} \left(1-2\kappa^2\right).
\label{sol-nmJac}
\end{equation}
Note that the functions $S(\tau)$, $\cosh(a\tau)$ and $\kappa(\tau)$ are real
for any value of $\gamma$. For $\gamma>1$ it is convenient to use instead
of (\ref{def-gS}) the equivalent expressions in terms of the
trigonometrical functions:
\begin{equation}
\tilde{S}(\tau)=\sin(\tilde{a}\tau)/\tilde{a}\,, \quad
\tilde{g}(\tau)= \cos(\tilde{a}\tau) + i\gamma \tilde{S}(\tau), \quad
\tilde{a}= \sqrt{\gamma^2-1}
\label{def-tilgS}
\end{equation}
In the special case $\gamma=1$ one has
$S(\tau)=\tau$ and $g(\tau)=1 + i\tau$. In particular,
\begin{equation}
\xi_m^{(n)}(\tau;\gamma=1)
= \frac{\tau^{m-n}(1+i\tau)^{n-1}}
{(1-i\tau)^{m+1}}
P_{n-1}^{(m-n,\,1)}
\left(\frac{1-\tau^2}{1+\tau^2}\right).
\label{1a0}
\end{equation}

The knowledge of the two-dimensional generating function enables to
verify the unitarity condition (\ref{cond2}). Consider
the product $X^*(\tau,z_1,y_1)X(\tau,z_2,y_2)$, which is a four--variable
generating function for the products $\xi_m^{(n)*}\xi_l^{(k)}$.
Taking $y_1=\sqrt{u}\exp(i\varphi)$, $y_2^*=\sqrt{u}\exp(-i\varphi)$ and
integrating over $\varphi$ from $0$ to $2\pi$ one obtains
a three--variable generating function
$\sum z_1^{*m} z_2^l u^n \xi_m^{(n)*}\xi_l^{(n)}$.
Dividing it by $u$ and
integrating the ratio over $u$ from $0$ to $1$ one arrives finally at
the relation
\begin{equation}
\sum_{n,m,l=1}^{\infty} z_1^{*m} z_2^l \frac1n \xi_m^{(n)*}
\xi_l^{(n)}=-\ln\left(1-z_1^* z_2\right)=\sum_{k=1}^{\infty} \frac1k
\left(z_1^* z_2\right)^k,
\label{ident-2}
\end{equation}
which is equivalent to the special case of (\ref{cond2}) for
$\eta_{m}^{(k)}\equiv 0$:
\begin{equation}
\sum_{n}\; \frac1n \xi_{m}^{(n)*}(\tau)\xi_{j}^{(n)}(\tau)
\equiv \frac1m \delta_{mj}.
\label{cond2-1}
\end{equation}

\subsection{Examples}

Suppose that initially there was a single excited mode labeled with an
index $n$. Due to the linearity of the process one may assume that the mean
number of photons in this mode was $\nu_n=1$.
 Then the mean occupation number of the $m$-th mode at $\tau>0$ equals
\begin{equation}
{\cal N}_m^{(n)}=\frac{m}{n}\left[\xi_m^{(n)}\right]^2
= \frac{m}{n}\left[(1-\kappa^2) \kappa^{m-n}
P_{n-1}^{(m-n,\,1)}\left( 1-2\kappa^2 \right)\right]^2,
\label{num-nm}
\end{equation}
where $\kappa$ is given by (\ref{def-kap}).
For example, in the special case $\gamma=0$ we have
\[{\cal N}_m^{(1)}=\frac{m(\tanh\tau)^{2m-2}}{(\cosh\tau)^4},
\]
\[
{\cal N}_m^{(2)}=\frac{m(\tanh\tau)^{2m-4}}{2(\cosh\tau)^4}
\left[(m-1)-(m+1)\tanh^2\tau\right]^2.
\]
The maximum of function ${\cal N}_m^{(1)}(\tau)$ is
achieved at $\sinh\tau_{max}=\sqrt{(m-1)/2}$. For $m\gg 1$ it equals
${\cal N}_m^{(1)}(\tau_{max})\approx 4/\left(me^4\right)$. For a fixed value
of $\tau\gg 1$, the occupation number distribution ${\cal N}_m^{(1)}$
reaches its maximum at $m_{max{\cal N}}^{(1)}=\cosh^2\tau$, and
${\cal N}_{max}^{(1)}=\left(e\cosh^2\tau\right)^{-1}\ll 1$. The maximum of
the energy distribution is shifted to the right, $m_{max{\cal E}}^{(1)}=
2\cosh^2\tau$, and its value is not decreased with time: ${\cal E}_{max}
^{(1)}=4/e^2$. This explains the exponential growth of the total energy.

Although formula  (\ref{num-nm}) seems asymmetrical
with respect to the indices $m$ and $n$, actually the relation
\begin{equation}
{\cal N}_m^{(n)}={\cal N}_n^{(m)}
\label{nm}
\end{equation}
holds. To prove it we calculate the generating function
\begin{equation}
Q(u,v)\equiv \sum_{m,n=1}^{\infty}v^m u^n {\cal N}_m^{(n)}.
\label{defgen-N}
\end{equation}
It is related to the function $X(z,y)$ (\ref{G}) as follows
\[
Q(u,v)= v\frac{d}{dv}\int_0^{u}dr\int_0^{2\pi}\int_0^{2\pi}
\frac{d\varphi d\psi}{(2\pi)^2}
X\left(\sqrt{r} e^{i\varphi},\sqrt{v} e^{i\psi}\right)
X^*\left(\sqrt{r} e^{i\varphi},\sqrt{v} e^{i\psi}\right).
\]
Having performed all the calculations we arrive at the expression
\begin{equation} 2Q(u,v)=
\frac{1+uv -\kappa^2(u+v)}
{\left\{ \left[1+uv -\kappa^2(u+v)\right]^2 -4uv(1-\kappa^2)^2\right\}^{1/2}}  -1.
\label{gen-N}
\end{equation}
Then (\ref{nm}) is a consequence of the relation $Q(u,v)=Q(v,u)$.

The initial stage of the evolution of
${\cal N}_m^{(n)}(\tau)$ does not depend on the detuning parameter $\gamma$,
since the principal term of the expansion of (\ref{num-nm}) with respect to
$\tau$ yields
\[
{\cal N}_{n\pm q}^{(n)}(\tau\to 0)= \frac{n\pm q}{n}
\left[\frac{n(n\pm 1)\ldots(n\pm q \mp 1)}{q!}\right]^2\tau^{2q}.
\]
However, the further evolution is sensitive to the value of $\gamma$.
If $\gamma\le 1$, then the function ${\cal N}_m^{(n)}(\tau)$ has
many maxima and minima (especially for large values of
$m$ and $n$), but finally it decreases asymptotically as
$mna^4/\cosh^4(a\tau)$. On the contrary, if $\gamma>1$, then
the function ${\cal N}_m^{(n)}(\tau)$ is periodic with the period
$\pi/\tilde{a}$, and it turns into zero for $\tau=k\pi/\tilde{a}$,
$k=1,2,\ldots$ (excepting the case $m=n$). The magnitude of
the coefficient ${\cal N}_m^{(n)}(\tau)$  decreases approximately
as $\gamma^{-2|m-n|}$ for $\gamma\gg 1$.

Now let us assume for simplicity that $\gamma=0$.
Then Eq. (\ref{sol-nmJac}) can be represented in the equivalent form
\be
 \xi_m^{(n)}=n\frac{(\tanh\tau)^{m-n}}{\cosh^2\tau}(-1)
^{n-1}F\left(1-n,m+1;2;\frac{1}{\cosh^2\tau}\right).
\label{sol-nmF-Ga}
\ee
If $m\gg n\sim{\cal O}(1)$, then $(m)_k\approx m^k$ ($k\le n$), so the Gauss
hypergeometric function in Eq. (\ref{sol-nmF-Ga}) can be replaced by the
confluent hypergeometric function with a negative integral first index,
which is reduced to the associated Laguerre polynomial \cite{Bateman}
$L_{n-1}^{(1)}(\mu)$ of the scaled variable $\mu=m/\cosh^2\tau$.
Using the approximation $(\tanh\tau)^{2m}\approx\exp\left(-m/\cosh^2\tau
\right)$ valid for $\tau\gg 1$, we arrive at a simplified expression
\begin{equation} {\cal E}_m^{(n)}=m{\cal N}_m^{(n)}=\frac1n \mu^2 e^{-\mu}
\left[L_{n-1}^{(1)}(\mu)\right]^2, \label{Lag}\end{equation}
describing the energy distribution over the modes with large number $m$.

The fluctuations of the occupation numbers can be calculated
with the aid of the formula (again for $\gamma=0$)
\[\langle\hat{\cal N}^2\rangle_m(\tau) -\langle\hat{\cal N}\rangle_m(\tau)
=\frac{m^2}{n^2}\left[\xi_m^{(n)}(\tau)\right]^4\left[\langle\hat{\cal N}^2
\rangle_n(0) -\langle\hat{\cal N}\rangle_n(0)\right],\]
which is an immediate consequence of Eq. (\ref{cond2-1}).
Consequently, the type of the photon statistics (sub-- or super--Poissonian)
is conserved. In particular, if the initial mode was in a coherent state,
then all other modes will be excited in the coherent states, too.

The assumption on the equidistant eigenmode spectrum can be justified to
certain extent for the longitudinal modes of a Fabry-Perot resonator
with perfect mirrors, if the order of interference (the mode number)
is high enough.
So let us suppose that initially some single mode with $n\gg 1$ was excited.
Although the lowest modes are not equidistant in this case, for a limited
period of time in the beginning of the process (until $\tau\ll \log(2n)$),
the
evolution of the chosen mode and its neighbours can be described by the
solutions obtained above, since the influence of the ``remote'' modes becomes
essential at sufficiently large times. Then the populations of the modes
exhibit strong oscillations, since they are proportional to the squares of
the Jacobi polynomials of large degrees. The asymptotics of these polynomials
(\cite{Bateman}, Eq. (10.14.10)) yields
\[{\cal N}_m^{(n)}\approx \frac{2}{n\pi\sin(2\varphi)}\cos^2\left\{(m+n)
\varphi-[2|m-n|+1]\frac{\pi}{4}\right\},\qquad \sin\varphi=\tanh\tau.\]
This formula holds provided that $n\gg|m-n|\sim{\cal O}(1)$ and
$n\sin(2\varphi)\gg 1$. For $\tau\ll 1$ we observe fast oscillations, which
amplitude is modulated with the period $\pi/\tau$, whereas for $\tau\gg 1$
we have more slow periodic variations of ${\cal N}_m^{(n)}$ as the function
of $m$ with the period $\pi e^{\tau}/2$.

An interesting problem is the field evolution in a cavity which was initially
in the equilibrium state at a finite temperature, when the initial occupation
numbers were given by the Planck distribution $\nu_n=[\exp(\beta n)-1]^{-1}$.
Let us consider two limit cases. The first one corresponds to the
low--temperature approximation $\nu_n=\exp(-\beta n)$. Then the occupation
number of the $m$-th mode is nothing but the coefficient at $v^m$ in the
expansion (\ref{gen-N}) with $u=\exp(-\beta)$. Using the well known generating
function of the Legendre polynomials $P_m(z)$ (\cite{Bateman}, Eq. (10.10.39)),
one can obtain the following expression (for $\gamma=0$):
\[ {\cal N}_m^{\{\beta\}}=\frac12\left(\frac{e^{-\beta}-\rho^2}
{1-e^{-\beta}\rho^2}\right)^m \left[P_m(x) + P_{m-1}(x)\right],
\quad \rho\equiv \tanh(\tau),
\]
\[
x=\frac{e^{-\beta}\left(1-\rho^2\right)^2 +\rho^2(1-e^{-\beta})^2}
{\left(e^{-\beta}-\rho^2\right)\left(1-e^{-\beta}\rho^2\right)}.
\]
In particular, ${\cal N}_1^{\{\beta\}}=e^{-\beta}(\cosh\tau)^{-4}
\left(1-e^{-\beta}\tanh^2\tau\right)^{-2}$.

In the special case of a cavity filled in with a {\it high-temperature
thermal radiation\/},
the initial distribution over modes reads $\nu_n(\Theta)=\Theta/n$,
constant $\Theta$ being proportional to the temperature. Then
${\cal N}_m^{\{\Theta\}}=\sum_{n}\nu_n(\Theta) {\cal N}_m^{(n)}$.
This sum is nothing but $\Theta$ multiplied by the
coefficient at $v^m$ in the Taylor expansion of the function
\[ \tilde{Q}(v)=\int_0^1\frac{\mbox{d}u}{u}Q(u,v)=
\ln\frac{1-v\kappa^2(\tau)}{1-v}.\]
Thus we have
\[
{\cal E}_m^{\{\Theta\}}=m{\cal N}_m^{\{\Theta\}}=
\Theta\left(1-[\kappa(\tau)]^{2m}\right).
\]
We see that the resonance vibrations of
the wall cause an effective cooling of the lowest electromagnetic modes
(provided $|\gamma|<1$).
The total number of quanta and the total energy in this example are
formally infinite,
due to the equipartition law of the classical statistical mechanics. In
reality both these quantities are finite, since
$\nu_n(\Theta) < \Theta/n$ at $n\to\infty$
due to the quantum corrections.

\section{Generic resonance case $p\ge 2$}
\label{generic}

If $p\ge 2$, we have $p-1$ pair of {\it coupled\/} equations for the
coefficients with lower indices $1\le k\le p-1$
\begin{eqnarray}
\frac {\mbox{d}}{\mbox{d}\tau}\xi_k^{(n)}&=&
(-1)^p\left[(k+p)\xi_{k+p}^{(n)}- (p-k)\eta_{p-k}^{(n)}\right]
+2i\gamma k \xi_{k}^{(n)} ,
\label{pksikin}\\
 \frac {\mbox{d}}{\mbox{d}\tau}\eta_k^{(n)}&=&
(-1)^p\left[(k+p)\eta_{k+p}^{(n)} -(p-k)\xi_{p-k}^{(n)}\right]
- 2i\gamma k \eta_{k}^{(n)}.
\label{petakin}
\end{eqnarray}
In this case some functions $\eta_{k}^{(n)}(t)$ are not equal to zero at
$t>0$, thus we have the effect of photon creation from the vacuum.

It is convenient to introduce a new set of coefficients $\rho_k^{(n)}$,
whose lower indices run over all integers from $-\infty$ to $\infty$:
\begin{equation}
\rho_k^{(n)}=\left\{
\begin{array}{ll}
\xi_k^{(n)}\,, & k>0\\
0\,, & k=0\\
-\eta_{-k}^{(n)}\,, & k<0
\end{array}\right.
\label{defrho}
\end{equation}
Then one can verify that equations (\ref{pksik})-(\ref{petak}) and
(\ref{pksikin})-(\ref{petakin}) can be combined in a {\it single\/} set of
equation ($k=\pm 1, \pm 2, \ldots$) \cite{Djpa}
\begin{equation}
\frac {\mbox{d}}{\mbox{d}\tau}\rho_k^{(n)}=
(-1)^p\left[(k+p)\rho_{k+p}^{(n)}- (k-p)\rho_{k-p}^{(n)}\right]
+2i\gamma k \rho_{k}^{(n)}
\label{prhok}
\end{equation}
with the initial conditions ($n=1,2,\ldots$)
\begin{equation}
\rho_k^{(n)}(0)=\delta_{kn}.
\label{inirho}
\end{equation}
A remarkable feature of the set of equations (\ref{prhok}) is that its
solutions satisfy {\it exactly\/}
the unitarity conditions (\ref{cond1})-(\ref{cond3})
(although the coefficients $\xi_k^{(n)}$ and $\eta_k^{(n)}$ introduced via
equation (\ref{ksitime}) have additional phase
factors in comparison with the coefficients defined in equation (\ref{ksi}),
these phases do not affect the identities concerned), which
can be rewritten as
\begin{eqnarray}
&& \sum_{m=-\infty}^{\infty}
m\rho_{m}^{(n)*}\rho_{m}^{(k)}
=n\delta_{nk}\,, \quad n,k=1,2,\ldots
\label{rhocond1} \\[2mm]
&&
 \sum_{n=1}^{\infty}\frac{m}{n}
\left[\rho_{m}^{(n)*}\rho_{j}^{(n)} -
\rho_{-m}^{(n)*}\rho_{-j}^{(n)} \right] =\delta_{mj}\,,
\quad m,j=1,2,\ldots
\label{rhocond2}\\[2mm]
&&
 \sum_{n=1}^{\infty}\frac{1}{n}
\left[\rho_{m}^{(n)*}\rho_{-j}^{(n)} -
\rho_{j}^{(n)*}\rho_{-m}^{(n)} \right] =0\,,
\quad m,j=1,2,\ldots
\label{rhocond3}
\end{eqnarray}
For example, calculating the derivative
$I=(d/d\tau)\sum_{m=-\infty}^{\infty}\, m\rho_{m}^{(n)*}\rho_{m}^{(k)}$
with the aid of equation (\ref{prhok}) and its complex conjugated
counterpart one can easily verify that $I=0$. Then the value of the
right-hand side of (\ref{rhocond1}) is a consequence of the initial
conditions (\ref{inirho}).
The identities (\ref{rhocond2}) and (\ref{rhocond3}) can be verified
in a similar way, if one uses instead of (\ref{prhok}) the recurrence
relations between the coefficients $\rho_{m}^{(n)}$ with the same lower
index $m$ but with different
{\it upper\/} indices: see
equations (\ref{recrho}) and (\ref{recrho1}).

Due to the initial conditions (\ref{inirho}) the solutions to (\ref{prhok})
satisfy the relation
\begin{equation}
\rho_{j+mp}^{(k+np)}\equiv 0 \quad {\rm if} j\neq k
\label{identrho}
\end{equation}
\[
j,k=0,1,\ldots,p-1, \quad m=0,\pm 1,\pm 2, \ldots\,,
\quad n=0, 1, 2, \ldots
\]
Consequently, the nonzero coefficients $\rho_m^{(n)}$ form
$p$ independent subsets
\begin{equation}
y_k^{(q,j)}\equiv\rho_{j+kp}^{(j+qp)}
\label{subsets}
\end{equation}
\[
j=0,1,\ldots,p-1, \quad q=0,1,2,\ldots\,,
\quad k=0,\pm 1,\pm 2, \ldots
\]
The subset $y_k^{(q,0)}$ is distinguished, because
$y_k^{(q,0)}\equiv 0$ for $k\le 0$ and the upper index $q$ begins at $q=1$.
One can verify that the functions $y_m^{(n,0)}(\tau)$
 with $m\ge 1$ are given by the formulae for
$\xi_m^{(n)}(\tau)$ found in the preceding section, provided
one replaces $\tau$ by $(-1)^p p\tau$ and $\gamma$ by $(-1)^p\gamma$,
whereas $y_m^{(n,0)}(\tau)\equiv 0$ for $m\le 0$.

In the generic case $j\neq 0$
it is reasonable to introduce a
generating function in the form of the {\it Laurent series\/} of an
auxiliary variable $z$
\begin{equation}
R^{(n,j)}(z,\tau)=\sum_{m=-\infty}^{\infty}y_m^{(n,j)}(\tau)z^m
\label{defR}
\end{equation}
since the lower index of the coefficient $y_m^{(n,j)}$ runs over all
integers from $-\infty$ to $\infty$.
One can verify that the function (\ref{defR}) satisfies the {\it homogeneous\/}
equation
\begin{equation}
\frac{\partial R^{(n,j)}}{\partial\tau}=\left[\sigma\left(
\frac1z -z\right) +2i\gamma \right]
\left(j+pz\frac{\partial }{\partial z}\right)R^{(n,j)},
 \quad \sigma=(-1)^p.
\label{eqR}
\end{equation}
The solution to (\ref{eqR}) satisfying
the initial condition $R^{(n,j)}(z,0)=z^n$ reads
\begin{equation}
R^{(n,j)}(z,\tau)=z^{-j/p}\left[\frac{z g(p\tau)
+\sigma S(p\tau)}{g^*(p\tau)+ z\sigma S(p\tau)}\right]^{n+j/p}
\label{solR}
\end{equation}
where the functions
$S(\tau)$ and $g(\tau)$ were defined in (\ref{def-gS}).
The coefficients of the Laurent series (\ref{defR}) can be calculated with
the aid of the Cauchy formula
\begin{equation}
y_m^{(n,j)}(\tau)=\frac{1}{2\pi i}\oint_{\cal C}\frac{dz}{z^{m+1}}
R^{(n,j)}(z,\tau)
\label{Cauchy}
\end{equation}
where the closed curve ${\cal C}$ rounds the point $z=0$ in the complex
plane in the counterclockwise direction. Making a scale transformation one
can reduce the integral (\ref{Cauchy}) with the integrand (\ref{solR})
to the integral representation of the Gauss hypergeometric function
(\cite{Bateman}, vol~1, section~2.1.3)
\begin{equation}
F(a,b;c;x)=\frac{-i\Gamma(c)\exp(-i\pi b)}{2\sin(\pi b)\Gamma(c-b)\Gamma(b)}
\oint_{1}^{(0+)}\frac{t^{b-1}(1-t)^{c-b-1}}{(1-tx)^a}dt,
\label{intpred}
\end{equation}
where ${\rm Re}(c-b)>0$, $b\neq 1,2,3,\ldots$, and the integration contour
begins at the point $t=1$ and passes around the point $t=0$ in the positive
direction. After some algebra one can obtain the expression
\begin{eqnarray}
y_m^{(n,j)}&=&
-\,\frac{\Gamma\left(-m-j/p\right)\Gamma\left(1+n+j/p\right)
\sin\left[\pi\left(m+j/p\right)\right]}
{\pi\Gamma\left(1+n-m\right) }
\nonumber\\
&\times&(\sigma\kappa)^{n-m}\lambda^{m+n+2j/p}
F\left(n+j/p\,,\,-m -j/p\,;\, 1+n-m\,;\, \kappa^2\right).
\label{solrhogen}
\end{eqnarray}
We assume hereafter $\kappa\equiv\kappa(p\tau)$ and
$\lambda\equiv\lambda(p\tau)$, the functions $\kappa(x)$ and
$\lambda(x)$ being defined as in (\ref{def-kap}) and (\ref{def-lam}).
Using the known formula
\begin{equation}
\Gamma(-z)\sin(\pi z)=-\pi/\Gamma(z+1)
\label{gammapm}
\end{equation}
one can eliminate the gamma-function of a negative argument:
\begin{eqnarray}
y_m^{(n,j)}&=&
\frac{\Gamma\left(1+n+j/p\right)
(\sigma\kappa)^{n-m}\lambda^{m+n+2j/p}}
{\Gamma\left(1+m+j/p\right)\Gamma\left(1+n-m\right) }
\nonumber\\
&\times&
F\left(n+j/p\,,\,-m -j/p\,;\, 1+n-m\,;\, \kappa^2\right).
\label{solrhogen1}
\end{eqnarray}
The form (\ref{solrhogen1}) gives an explicit expression for the
coefficient $\xi_{j+pm}^{(j+pn)}$ with $0\le m\le n$.
Moreover, it clearly shows the fulfilment of the initial
condition $y_m^{(n,j)}(\tau=0)=\delta_{mn}$.
Transforming the hypergeometric
function with the aid of the formula \cite{Bateman,Abram}
\[
\lim_{c\to -n}\frac{F(a,b;c;x)}{\Gamma(c)}=
\frac{(a)_{n+1}(b)_{n+1}x^{n+1}}{(n+1)!}
 F(a+n+1,b+n+1;n+2;x)
\]
($n=0,1,2,\ldots$) and the identity (\ref{gammapm})
one obtains an equivalent expression
\begin{eqnarray}
y_m^{(n,j)}&=&
\frac{ \Gamma\left(m+j/p\right)
(-\sigma\kappa)^{m-n}\lambda^{m+n+2j/p}}
{\Gamma\left(n+j/p\right) \Gamma\left(1+m-n\right) }
\nonumber\\
&\times&
F\left(m+j/p\,,\,-n -j/p\,;\, 1+m-n\,;\, \kappa^2\right)
\label{solrhogen2}
\end{eqnarray}
which gives a convenient form of the
coefficient $\xi_{j+pm}^{(j+pn)}$ for $m\ge n$.
Formula (\ref{solrhogen}) with negative values of the lower index gives an
explicit expression for the nonzero coefficients $\eta_{pk-j}^{(pn+j)}$
($k\ge 1,n\ge 0$):
\begin{eqnarray}
\eta_{pk-j}^{(pn+j)}&=&
-\,\frac{\Gamma\left(k-j/p\right)\Gamma\left(1+n+j/p\right)
\sin\left[\pi\left(k-j/p\right)\right]}
{\pi\Gamma\left(1+n+k\right) }
\nonumber\\
&\times&(\sigma\kappa)^{n+k}\lambda^{n-k+2j/p}
F\left(n+j/p\,,\,k -j/p\,;\, 1+n+k\,;\, \kappa^2\right).
\label{solrhogen3}
\end{eqnarray}
 Note that the expressions (\ref{solrhogen1})-(\ref{solrhogen3}) are valid
for $j=0$, too. In this case they coincide with the formulae obtained in
the preceding section. Formulas (\ref{solrhogen1})-(\ref{solrhogen3})
immediately give the short-time behaviour of the Bogoliubov coefficients
at $\tau\to 0$: it is sufficient to put $\kappa\approx  p\tau$,
$\lambda\approx 1$ and to replace the hypergeometric functions by $1$.
In this limit the detuning parameter $\gamma$ drops out of the expressions
(in the leading terms of the Taylor expansions).

At $\tau\to\infty$ we have the following asymptotics of the functions
$\kappa(p\tau)$ and $\lambda(p\tau)$ (if $\gamma\le 1$)
\[
\kappa\approx 1-\frac12 S^{-2}(p\tau) \to 1, \quad
\lambda\to a+i\gamma, \quad \tau\to\infty .
\]
Then
equation (\ref{solrhogen}) together with the known asymptotics of the
hypergeometric function $F(a,b;a+b+1;1-x)$ at $x\ll 1$
\cite{Bateman,Abram}
\begin{equation}
F(a,b;a+b+1;1-x)=\frac{\Gamma(a+b+1)}{\Gamma(a+1)\Gamma(b+1)}
\left[1+abx\ln(x) +{\cal O}(x)\right]
\label{F1}
\end{equation}
lead to the asymptotical expression for the Bogoliubov coefficients
\begin{eqnarray}
y_m^{(n,j)}(\tau\gg 1)&=& \frac{\sin[\pi(m+j/p)]}{\pi(m+j/p)}
(a+i\gamma)^{m+n+2j/p}\sigma^{n-m} \nonumber\\
&\times&\left[ 1+{\cal O}\left( \frac{mn}{S^2}\ln S\right)\right]
\label{asxieta}
\end{eqnarray}
For $\gamma<1$ the correction has an order $mn\tau\exp(-2ap\tau)$, while for
$\gamma=1$ it has an order $mn\ln(\tau)/\tau^2$.

One can verify that the generating function (\ref{solR})
satisfies the recurrence relation
\begin{equation}
\frac{\partial R^{(q,j)}}{\partial\tau} =(j+qp)\left\{\sigma\left[
R^{(q-1,j)} -R^{(q+1,j)}\right] +2i\gamma R^{(q,j)}\right\}
\label{recR}
\end{equation}
Its immediate consequence is an analogous relation for the
Bogoliubov coefficients with the same lower indices:
\begin{equation}
\frac{d }{d\tau}\rho_m^{(n)} = n\left\{\sigma\left[
\rho_m^{(n-p)} -\rho_m^{(n+p)}\right] +2i\gamma \rho_m^{(n)}\right\}.
\label{recrho}
\end{equation}
Equation (\ref{recrho}) is valid for $n>p$ (when
$q\ge 1$ and $j\ge 1$ in (\ref{recR})), since the coefficients
$\rho_m^{(n)}$ are not defined when $n< 0$.
However, using the chain of identities
\begin{eqnarray*}
&&R^{(-1,j)}(z)=
z^{-j/p}\left[\frac{S+gz}{g^*+ Sz}\right]^{j/p-1}
=\frac1z\left(\frac1z\right)^{j/p-1}\left[\frac{S +g^*/z}{g+ S/z}\right]
^{1-j/p} \\
&&= \frac1z\left[R^{(0,p-j)}(1/z^*)\right]^*
=\frac1z \sum_{k=-\infty}^{\infty} y_k^{(0,p-j)*}\left(\frac1z\right)^k
=\sum_{k=-\infty}^{\infty} y_{-k-1}^{(0,p-j)*}z^k
\end{eqnarray*}
one can obtain the first $p-1$ recurrence relations
\begin{equation}
\frac{d }{d\tau}\rho_m^{(n)} = n\left\{\sigma\left[
\rho_{-m}^{(p-n)*} -\rho_m^{(p+n)}\right] +2i\gamma \rho_m^{(n)}\right\},
\quad n=1,2,\ldots,p-1.
\label{recrho1}
\end{equation}
To treat the special case $n=p$ (it corresponds to the distinguished
subset with $j=0$) one should take
into account that $R^{(0,0)}(z)\equiv 1$, which means formally that
$\rho_m^{(0)}=\delta_{m0}$.
So the last recurrence relation reads
\[
\frac{d }{d\tau}\rho_m^{(p)} = p\left\{-\sigma\rho_m^{(2p)}
+2i\gamma \rho_m^{(p)}\right\}, \quad m\ge 1
\]
(remember that $\rho_m^{(p)}\equiv 0$ for $m\le 0$).
Now one can verify that the unitarity conditions
(\ref{rhocond2})-(\ref{rhocond3})
are the consequencies of the equations (\ref{recrho}) and (\ref{recrho1}).

\section{Photon statistics}

To evaluate the mean number of photons and the statistical properties
of the quantum field created in the cavity at $t>T$
 we introduce the Hermitian quadrature component operators
\[
\hat{q}_m=\left(\hat {a}_m +\hat {a}_m^{\dagger}\right)/\sqrt2, \quad
\hat{p}_m=\left(\hat {a}_m -\hat {a}_m^{\dagger}\right)/(i\sqrt2).
\]
Their variances are defined as
\[
U_m=\langle \hat{q}_m^2\rangle -\langle \hat{q}_m\rangle^2 , \quad
V_m=\langle \hat{p}_m^2\rangle -\langle \hat{p}_m\rangle^2,
\]
whereas the covariance is given by
\[
Y_m=\frac12\langle \hat{p}_m\hat{q}_m +\hat{q}_m\hat{p}_m\rangle -
\langle \hat{p}_m\rangle\langle \hat{q}_m\rangle.
\]
The average values must be calculated in the state defined
{\it with respect to the initial operators\/} $\hat{b}_n$
(remember that we use here the Heisenberg picture).

\subsection{Initial vacuum state}
The vacuum state is defined by means of the relations
 $\hat{b}_n |0\rangle=0$.
In this case, $U_m+V_m=2{\cal N}_m^{(vac)} +1$,
where ${\cal N}_m^{(vac)}$ is the mean number of photons created
from vacuum in the $m$th mode. It is given by the single sums over
index $n$ in Eq. (\ref{number}). Initially,
$U_m(0)=V_m(0)=1/2$, $Y_m(0)= 0$.
Using (\ref{Bogol}) and assuming for simplicity $\omega_1=1$,
we obtain the following expressions for $\tau>0$,
\begin{equation}
U_m= \frac{m}{2}\sum_{n=1}^{\infty} \frac1n
\left|\rho_m^{(n)} -\rho_{-m}^{(n)}\right|^2, \quad
V_m= \frac{m}{2}\sum_{n=1}^{\infty} \frac1n
\left|\rho_m^{(n)} +\rho_{-m}^{(n)}\right|^2
\label{var}
\end{equation}
\begin{equation}
Y_m = \sum_{n=1}^{\infty} \frac{m}{n}
\mathrm{Im} \left[\rho_m^{(n)*}
\rho_{-m}^{(n)}\right]
\label{covar}
\end{equation}
where the coefficients $\rho_{\pm m}^{(n)}$ should be taken at the moment
$T$, thus their argument is
$\tau_T\equiv \frac12 \varepsilon \omega_1 T$.
Strictly speaking, the expressions (\ref{var})-(\ref{covar})
have physical meanings at those
moments of time $T$ when the wall returns to its initial position,
i.e. for $T=N\pi/[p(1+\delta)]$ with an integer $N$. Consequently,
the argument $\tau_T$ of the coefficients $\rho_{\pm m}^{(n)}$
in (\ref{var})-(\ref{covar}) assumes discrete values
$\tau^{(N)}=N\varepsilon\pi/[2p(1+\delta)]$. One should remember,
however, that something interesting in our problem happens for the values
$\tau\sim 1$ (or larger). Then $N\sim\varepsilon^{-1}\gg 1$, and the
minimal increment $\Delta\tau\sim\varepsilon$ is so small that
$\tau_T$ can be considered as a continuous variable (under the realistic
conditions, $\varepsilon \le 10^{-8}$ \cite{DK96}). For this reason,
we omit hereafter the subscript $T$, writing simply $\tau$ instead of
$\tau_T$ or $\tau^{(N)}$.

Differentiating the right-hand sides of equations (\ref{var}) and
(\ref{covar}) with respect to the `slow time' $\tau$,
one can remove the fraction $1/n$ with the aid of the
recurrence relations (\ref{recrho}) and (\ref{recrho1}).
After that, changing if necessary the summation index $n$ to $n \pm p$,
one can verify that almost all terms in the right-hand sides are cancelled,
and the infinite series are reduced to the finite sums:
\[
\begin{array}{l}
\mbox{d}U_m/\mbox{d}\tau\\
\mbox{d}V_m/\mbox{d}\tau
\end{array}\Bigg\}=
\sigma m\sum_{n=1}^{p-1}{\rm Re}\left(
\left[\rho_{m}^{(p-n)} \mp \rho_{-m}^{(p-n)}\right]
\left[\rho_{-m}^{(n)} \mp \rho_{m}^{(n)}\right]\right)
\]
\[
\mbox{d}Y_m/\mbox{d}\tau=
\sigma m\sum_{n=1}^{p-1} {\rm Im}
\left(\rho_{m}^{(n)*}\rho_{m}^{(p-n)*} +
\rho_{-m}^{(n)}\rho_{-m}^{(p-n)}\right)
\]
Now one should take into account the structure
of the coefficients $\rho_m^{(n)}$: they are different from
zero provided the difference between the upper index $n$ and the lower one
$m$ is some multiple of the number $p$. If $m=j+pk$
with $j=1,\ldots, p-1$ and $k=0,1,2,\ldots$, then only the terms with
$n=j$ or $n=p-j$ survive in the sums above.
Depending on whether $j=p/2$ or $j \neq p/2$,
we obtain two different sets of explicit expressions for the derivatives
of the (co)variances.

1) If $m=j+pk$
but $j\neq p/2$ (in particular, for all {\it odd\/} values of $p$), then
\begin{equation}
\frac{\mbox{d}U_m}{\mbox{d}\tau}=
\frac{\mbox{d}V_m}{\mbox{d}\tau}=
2\sigma m{\rm Re}\left(\rho_{m}^{(j)}\rho_{-m}^{(p-j)}\right), \quad
\frac{\mbox{d}Y_m}{\mbox{d}\tau}= 0.
\label{pneqj}
\end{equation}
In this case $Y_m\equiv 0$ and $U_m=V_m={\cal N}_m^{(vac)} +1/2$.

2) A different situation happens in the distinguished
modes with the numbers $\mu=p(k+1/2)$, $k=0,1,2,\ldots$:
\begin{equation}
\frac{\mbox{d}U_{\mu}}{\mbox{d}\tau}=
- {\mu}{\rm Re}\left(\left[\rho_{\mu}^{(p/2)}-\rho_{-\mu}^{(p/2)}
\right]^2\right), \quad
\frac{\mbox{d}V_{\mu}}{\mbox{d}\tau}=
 {\mu}{\rm Re}\left(\left[\rho_{\mu}^{(p/2)}+\rho_{-\mu}^{(p/2)}
\right]^2\right)
\label{p2j}
\end{equation}
\begin{equation}
\mbox{d}Y_{\mu}/\mbox{d}\tau = {\mu}\mathrm{Im}\left(
\left[\rho_{\mu}^{(p/2)*}\right]^2
+ \left[\rho_{-\mu}^{(p/2)}\right]^2 \right).
\label{derY}
\end{equation}
We shall call such modes as ``principal'' ones; they exist only if
$p$ is an {\it even\/} number.
In the strict resonance case ($\gamma=0$)
all the coefficients $\rho_{\mu}^{(p/2)}$ are real, so $Y_{\mu}=0$
and $dU_{\mu}/d\tau\le 0$ in the whole interval $0\le \tau<\infty$,
resulting in the inequality $U_{\mu}(\tau)<1/2$, which tells us that
the field occurs in the {\em squeezed\/} quantum state.

\subsubsection{Squeezing in the ``principal'' modes}

Note that the coefficients $\rho_{pm+p/2}^{(pn+p/2)}$ depend on the
parameter $p$ only through the dependence of the
the variable $\kappa$ on the product $p\tau$: see
equations (\ref{solrhogen}) or (\ref{solrhogen1}).
Thus to study the squeezing properties of the field created due to the NCE,
it is sufficient to consider the most important special case
of the parametric resonance at the {\it double\/} fundamental frequency
$2\omega_1$ (i.e. $p=2$), since the formulae for $p>2$ can be
obtained by a simple rescaling of the `slow time'
(for the `principal' modes).
In this case, only the odd modes can be excited from
the vacuum, and they do exhibit some squeezing.

Using equations (\ref{p2j}) and (\ref{derY}) one can immediately find
the Taylor expansions of
the (co)variances at $\tau\to 0$
(assuming $(-1)!!\equiv 1$):
\be
\begin{array}{l}
U_{2m+1}\\
V_{2m+1}
\end{array}\Bigg\}
= \frac12 \mp\tau^{2m+1} \left[\frac{(2m-1)!!}{m!}\right]^2
\left[1 \mp\frac{2m+1}{(m+1)^2}\tau +{\cal O}(\tau^2)\right]
\label{UVsmall}
\ee
\be
Y_{2m+1}= - 2\gamma(2m+1)\tau^{2(m+1)} \left[\frac{(2m-1)!!}{m!}\right]^2
+\cdots
\label{Y0}
\ee
We see that the $U$-variances are always less than $1/2$ at the
initial stage, but the degree of their squeezing rapidly decreases
with increase of the number $m$.
Note that the dependence on the detuning parameter $\gamma$ in the
short-time limit appears only in terms of the order of $\tau^{2m+3}$ 
(and higher).

In the opposite limit $\tau\to \infty$ (or $\kappa\to 1$),
using equations (\ref{p2j}), (\ref{derY}) and the asymptotics of the
Bogoliubov coefficients (\ref{asxieta}) we obtain
{\it constant\/} time derivatives
\bea
\left.\mbox{d}U_{2m+1}/\mbox{d}\tau\right|_{\tau\to\infty}&=&
\frac{16a}{\pi^2(2m+1)}\sin^2\left[\left(m+\frac12\right)\phi\right]
\label{Ubig}\\
\left.\mbox{d}V_{2m+1}/\mbox{d}\tau\right|_{\tau\to\infty}&=&
\frac{16a}{\pi^2(2m+1)}\cos^2\left[\left(m+\frac12\right)\phi\right]
\label{Vbig}\\
\left.\mbox{d}Y_{2m+1}/\mbox{d}\tau\right|_{\tau\to\infty}&=&
-\,\frac{8a}{\pi^2(2m+1)}\sin\left[\left(2m+1\right)\phi\right]
\label{Ybig}
\eea
where $\phi \equiv\arcsin\gamma$. Consequently, all the (co)variances
increase with time linearly, giving the constant photon generation rate
in the `principal' (odd) modes
\be
\left.\mbox{d}{\cal N}_{2m+1}/\mbox{d}\tau
\right|_{\tau\to\infty}
= \frac{8a}{\pi^2(2m+1)}.
\label{Ntot}
\ee
Equation (\ref{Ntot}) results in a simple estimation of the mean photon
number in the $\mu$th mode at $\tau>1$:
${\cal N}_{\mu}(\tau)\approx a\tau/\mu$.

Since the covariance $Y_{\mu}$ is different from zero if $\gamma\neq 0$,
the initial vacuum state of the field is transformed to the
{\em correlated\/} quantum state \cite{Kiel,Kur,205}.
One should remember, however, that
the values of $U_{\mu}$, $V_{\mu}$ and $Y_{\mu}$ yield the
(co)variances of the field quadratures only at the moment $t=T$ (when the
wall stopped to oscillate). At the
subsequent moments of time the quadrature variances exhibit fast
oscillations with twice the frequency of the mode. For example (omitting
the mode index),
\[
\sigma_q(t')=U\cos^2(\omega t') +V\sin^2(\omega t')
+Y\sin(2\omega t'), \quad t'=t-T.
\]
Therefore the physical meanings have not the values $U_{\mu}$, $V_{\mu}$
and $Y_{\mu}$ themselves, but rather the {\em minimal\/}
$\sigma_{min}\equiv u_{\mu}$ and
{\em maximal\/} $\sigma_{max}\equiv v_{\mu}$
values of the quadrature variances during the period of fast
oscillations \cite{LuPeHr,Pol}
\be
\left.
\begin{array}{l}
u_{\mu}\\
v_{\mu}
\end{array}
\right\}=
 \frac12\left(U_{\mu} +V_{\mu} \mp
\sqrt{\left(U_{\mu} -V_{\mu}\right)^2 +4Y_{\mu}^2}\right).
\label{def-uv}
\ee
Only in the special case of the strict resonane ($\gamma=0$) we have
$u_{\mu}=U_{\mu}$ and $v_{\mu}=V_{\mu}$.
In the generic case $\gamma\neq 0$,
all three (co)variances, $U_{\mu}$,
$V_{\mu}$, and $Y_{\mu}$, linearly increase with the interaction time $T$
if $\tau_T\equiv \tau\gg 1$, due to Eqs. (\ref{Ubig})-(\ref{Ybig}).
Nonetheless, the {\em minimal variance\/} $u_{\mu}$
tends to a {\em constant\/} value at $\tau\to\infty$.
This is shown in the subsection \ref{arbitr}.
The examples of explicit time dependences of the coefficients
$u_{\mu}$ and $v_{\mu}$ are given in subsection \ref{p=2}.

\subsubsection{Mean photon number}

Differentiating the ``vacuum'' part of sum (\ref{number}) with respect to
$\tau$ and performing
the summation over the upper index $n$ with the aid of
(\ref{recrho})-(\ref{recrho1}) (remembering that the
coefficients $\rho_m^{(n)}$
are different from zero provided the difference $n-m$ is a multiple of $p$)
one can obtain the formula for the photon generation rate from vacuum in
each mode ($0\le j\le p-1$, $q=0,1,2,\ldots$)
\begin{eqnarray}
&&\frac{d}{d\tau}{\cal N}_{j+pq}^{(vac)}=
-2\sigma(j+pq){\rm Re}\left[\xi_{j+pq}^{(j)}\eta_{j+pq}^{(p-j)}\right]
\nonumber\\[2mm]
&&=2 p\sqrt{1-\gamma^2\kappa^2}\,\frac{\sin(\pi j/p)\Gamma(q+j/p)
\Gamma(1+q+j/p)\Gamma(2-j/p)}{\pi \Gamma(j/p)\Gamma(q+1)\Gamma(q+2)}
\kappa^{2q+1}
\nonumber\\[2mm]
&&\times  F\left(q+j/p\,,\,-j/p\,;\,1+q\,;\,\kappa^2\right)
F\left(q+j/p\,,\,1-j/p\,;\,2+q\,;\,\kappa^2\right)
\label{ratejps}
\end{eqnarray}
We see that there is no photon creation in the modes with numbers
$p,2p,\ldots$.
In the short-time limit,
\[
\dot{\cal N}_{j+pq}^{(vac)}\sim \tau^{2q+1}, \quad
\tau\ll 1.
\]
In the
long-time limit the photon generation rate tends to the constant value
(if $\gamma<1$)
\begin{equation}
\frac{d}{d\tau}{\cal N}_{j+pq}^{(vac)}=
\frac{2ap^2 \sin^2(\pi j/p)}{\pi^2 (j+pq)}
\left[1+{\cal O}\left(\frac{pq}{S^2}\ln S\right)\right], \quad ap\tau\gg 1
\label{asrate}
\end{equation}
For $q\gg 1$ and for a fixed value of $\kappa$ one can simplify the
right-hand side of (\ref{ratejps}) using Stirling's formula for the
Gamma-functions and the easily verified asymptotical formula
\[
F(a,b;c;z)\approx (1-az/c)^{-b}, \quad a,c\gg 1\, .
\]
In this case
\begin{equation}
\frac{d}{d\tau}{\cal N}_{j+pq}^{(vac)}\approx
2 p \sqrt{1-\gamma^2\kappa^2}\,
\frac{\sin(\pi j/p)\Gamma(2-j/p)\kappa^{2q+1}}
{\pi\Gamma(j/p)q^{2(1-j/p)}\left(1-\kappa^2\right)^{1-2j/p}},
\quad q\gg 1.
\label{asratebigq}
\end{equation}
In particular, if $q\gg S^2(p\tau)|\gg 1$, then
\begin{equation}
\frac{d}{d\tau}{\cal N}_{j+pq}^{(vac)}\approx
2 pa \,
\frac{\sin(\pi j/p)\Gamma(2-j/p)\left(S^2/q\right)^{2(1-j/p)} }
{\pi\Gamma(j/p) S^{2}}
 \exp\left(-q/S^2\right).
\label{asratebigqS}
\end{equation}
Comparing (\ref{asrate}) and (\ref{asratebigqS}) one can conclude
 that the number of the effectively excited modes
(i.e. the modes with a time independent photon generation rate) increases
in time exponentially, approximately as $S^2(\tau)/\ln S(\tau)$.

The total number of photons generated from vacuum in all the modes equals
\begin{equation}
{\cal N}^{(vac)}=\sum_{m,n=1}^{\infty}\frac{m}{n}|\eta_m^{(n)}|^2 .
\label{Nvac}
\end{equation}
Differentiating (\ref{Nvac})  with respect to $\tau$
and performing the summation over $m$ with the help of equations
(\ref{pksik})-(\ref{petakin}) or (\ref{prhok}) one can obtain the formula
\begin{equation}
\frac {\mbox{d}{\cal N}^{(vac)}}{\mbox{d}\tau}=
2\sigma{\rm Re}\sum_{n=1}^{\infty}\frac 1n \sum_{m=1}^p m(p-m)
\rho_{-m}^{(n)*}(\tau )\rho_{p-m}^{(n)}(\tau ).
\label{ratetot}
\end{equation}
Evidently, the right-hand side of this equation equals zero in the
``semi-resonance'' case $p=1$.

Differentiating equation (\ref{ratetot} ) once again over $\tau$
one can perform the summation over the upper index $n$ with the aid of
equations (\ref{recrho})-(\ref{recrho1}) to obtain a closed expression
for the {\it second derivative\/} of the total number of ``vacuum'' photons
\begin{eqnarray}
&&\frac{d^2}{d\tau^2}{\cal N}^{(vac)}=
2{\rm Re}\sum_{m=1}^{p-1} m(p-m)\left[\xi_{m}^{(m)}\xi_{p-m}^{(p-m)}
+\eta_{m}^{(p-m)*}\eta_{p-m}^{(m)*}\right]
\nonumber\\[2mm]
&&=2 \sum_{m=1}^{p-1} m(p-m)\left\{m(p-m)\left[\frac{\kappa}{p}
F\left(\frac{m}{p}\,,\,1-\frac{m}{p}\,;\,2\,;\,\kappa^2\right)\right]^2
\right.
\nonumber\\[2mm]
&&\left.+ \left(1-2\gamma^2\kappa^2\right)
F\left(\frac{m}{p}\,,\,-\frac{m}{p}\,;\,1\,;\,\kappa^2\right)
F\left(\frac{m}{p}-1\,,\,1-\frac{m}{p}\,;\,1\,;\,\kappa^2\right)\right\}
\label{ratetot2}
\end{eqnarray}
In the short-time limit one obtains
\begin{equation}
\ddot{\cal N}^{(vac)}=\frac13 p(p^2-1), \quad |ap\tau|\ll 1
\label{2dersmall}
\end{equation}
In the long-time limit the formulae (\ref{gammapm}), (\ref{F1}) and
$\sum_{m=1}^{p-1} \sin^2(\pi m/p)=p/2$ lead to another simple expression
(provided $p\ge 2$)
\begin{equation}
\ddot{\cal N}^{(vac)}=2a^2p^3/\pi^2, \quad ap\tau\gg 1, \quad a>0
\label{2derbig}
\end{equation}
Consequently, the total number of photons created from vacuum due to NSCE
increases in time quadratically both in the short-time and in the long-time
limits (although with different coefficients).

\subsection{Arbitrary initial conditions}\label{arbitr}

For an arbitrary initial state of the field one can write
 $U_m=U_m^{(vac)} +\Delta U_m$, where $U_m^{(vac)}$ is given by
equation (\ref{var}); similar expressions can be written for $V_m$ and $Y_m$.
The corrections due to the nonvacuum initial states are given by
\bea
\begin{array}{l}
\Delta U_m \\
\Delta V_m
\end{array}
\Bigg\} &=&
\mathrm{Re} \sum_{n,j} \frac{m}{\sqrt{nj}}\Bigg(
\left[\rho_m^{(n)} \mp \rho_{-m}^{(n)}\right]^*
\left[\rho_m^{(j)} \mp \rho_{-m}^{(j)}\right]
\left[\langle \hat{b}_n^{\dag} \hat{b}_j \rangle
-\langle \hat{b}_n^{\dag} \rangle\langle\hat{b}_j \rangle \right]
 \nonumber\\
&& \pm
\left[\rho_m^{(n)} \mp \rho_{-m}^{(n)}\right]
\left[\rho_m^{(j)} \mp \rho_{-m}^{(j)}\right]
\left[\langle \hat{b}_n \hat{b}_j \rangle
-\langle \hat{b}_n \rangle\langle\hat{b}_j \rangle \right]\Bigg)
\label{deltavar}
\eea
\bea
\Delta Y_m  &=&
\mathrm{Im} \sum_{n,j} \frac{m}{\sqrt{nj}}\Bigg(
\left[\rho_m^{(n)*} \rho_{-m}^{(j)} - \rho_m^{(j)}\rho_{-m}^{(n)*}\right]
\left[\langle \hat{b}_n^{\dag} \hat{b}_j \rangle
-\langle \hat{b}_n^{\dag} \rangle\langle\hat{b}_j \rangle \right]
 \nonumber\\
&& +
\left[\rho_m^{(n)} \rho_m^{(j)} - \rho_{-m}^{(n)} \rho_{-m}^{(j)}\right]
\left[\langle \hat{b}_n \hat{b}_j \rangle
-\langle \hat{b}_n \rangle\langle\hat{b}_j \rangle \right]\Bigg)
\label{deltacovar}
\eea
where the average values like $\langle\hat{b}_n^{\dag} \hat{b}_j \rangle$
are calculated in the initial state.
All the corrections disappear in the case of the initial
coherent state,
$\hat{b}_n \vert \alpha\rangle =\alpha_n \vert \alpha\rangle$.
If the initial density matrix is {\em diagonal\/} in the Fock basis
(as happens e.g. for the Fock or thermal states)
then $\langle \hat{b}_n^{\dag} \hat{b}_j \rangle =\nu_n \delta_{nj}$
($\nu_n \ge 0$),
all other average values in (\ref{deltavar}) and (\ref{deltacovar})
being equal to zero. In this case the double sums are reduced to the
single ones:
\be
\Delta U_m =
m \sum_{n} \frac{\nu_n}{n}
\left|\rho_m^{(n)} - \rho_{-m}^{(n)}\right|^2 , \quad
\Delta V_m =
m \sum_{n} \frac{\nu_n}{n}
\left|\rho_m^{(n)} + \rho_{-m}^{(n)}\right|^2
\label{vartherm}
\ee
\be
\Delta Y_m =
2m \sum_{n} \frac{\nu_n}{n}
\mathrm{Im}\left[\rho_m^{(n)*} \rho_{-m}^{(n)}\right].
\label{covartherm}
\ee
We see that the initial fluctuations always increase both the
variances $U_m$ and $V_m$ (for the diagonal density matrix).
However, asymptotically at $\tau\to\infty$
the corrections are bounded for the {\em physical\/} initial
states having finite total numbers of photons,
because the coefficients
$\left|\rho_m^{(n)} \pm \rho_{-m}^{(n)}\right|^2$ and
$\mathrm{Im}\left[\rho_m^{(n)*} \rho_{-m}^{(n)}\right]$ do not depend on the
summation index $n$ in this limit.
For example, if $p=2$, then
\be
\rho_{2m+1}^{(2n+1)}(\tau\gg 1) \approx \frac{2(-1)^m}{\pi(2m+1)}
(a+i\gamma)^{m+n+1},
\label{asrho+}
\ee
\be
\rho_{-2m-1}^{(2n+1)}(\tau\gg 1) \approx \frac{2(-1)^m}{\pi(2m+1)}
(a+i\gamma)^{n-m},
\label{asrho-}
\ee
and the exponent $n$ disappears in the sums, because $|a+i\gamma|=1$.
Thus we have in the `principal' $\mu$-modes
(taking $p=2$ for the sake of simplicity)
\be
\left.
\begin{array}{l}
\Delta U_{\mu}^{(\infty)}\\
\Delta V_{\mu}^{(\infty)}\\
\Delta Y_{\mu}^{(\infty)}
\end{array}\right\}=
\frac{8{\cal Z}}{\pi^2\mu}\times \left\{
\begin{array}{l}
2\sin^2\left(\mu\phi/2\right)\\
2\cos^2\left(\mu\phi/2\right)\\
-\sin\left(\mu\phi\right)
\end{array},\right.
\label{delUinf}
\ee
where
\[
\phi \equiv\arcsin\gamma, \quad
{\cal Z}=\sum_{k=0}^{\infty} \frac{\nu_{2k+1}}{2k+1} .
\]
The expressions in (\ref{delUinf})
are very similar to those in equations (\ref{Ubig})-(\ref{Ybig}).
The consequence of Eq. (\ref{delUinf}) is the important result that
in the limit $\tau\to\infty$
the minimal variance $u_{\mu}$ {\em does not depend\/} on the initial state
of the field inside the cavity,
provided the initial density matrix was diagonal in the Fock basis.
Indeed, in this case,
combining the equations (\ref{Ubig})-(\ref{Ybig}) and
(\ref{delUinf}), we can write the variances at $\tau\gg 1$ as
(we omit the subscript $\mu$)
\be
\left(
\begin{array}{l}
U(\tau)\\
V(\tau)\\
Y(\tau)
\end{array}\right)=
\left(
\begin{array}{l}
2F\sin^2(\chi/2) + f\\
2F\cos^2(\chi/2) + g\\
-F\sin\chi+ h
\end{array}\right), \quad
F=\frac{8(a\tau +{\cal Z})}{\pi^2\mu}, \quad \chi=\mu\phi.
\label{vartherminf}
\ee
The corrections $f$, $g$ and $h$ can be found
by integrating equations (\ref{p2j})-(\ref{derY}), therefore
they {\em do not depend on the initial state}.
At $\tau\to\infty$ these corrections tend to finite limits, so they
are much smaller than $F$.
Evidently, $U +V= 2F +f +g$, whereas
\[
(U -V)^2 +4Y^2=
4F^2 +4F[(g-f)\cos\chi -2h\sin\chi] +(f-g)^2 +4h^2 .
\]
For $F\gg f,g,h$ we have
\[
\sqrt{(U -V)^2 +4Y^2} = 2F + (g-f)\cos\chi -2h\sin\chi +{\cal O}(1/F)
\]
so the minimal variance $u(\tau)$ (\ref{def-uv}) tends to the finite limit
\be
u(\infty)= f\cos^2(\chi/2) + g\sin^2(\chi/2) +h\sin\chi,
\label{u-inf}
\ee
which does not depend on ${\cal Z}$, i.e. on the initial state.

The correction to the mean number of photons in the $m$-th mode
for the ``diagonal'' initial distributions is given by the sum
\[
\Delta {\cal N}_m=
m \sum_{n} \frac{\nu_n}{n} \left(
\left|\rho_m^{(n)}\right||^2 + \left|\rho_{-m}^{(n)}\right|^2 \right).
\]
It tends to the limit
$\Delta {\cal N}_{\mu}^{(\infty)}= 8{\cal Z}/\left(\pi^2\mu\right)$.

The total number of photons in all the modes equals
${\cal N}={\cal N}^{(vac)}+{\cal N}^{(cav)}$, where
\begin{equation}
{\cal N}^{(cav)}= {\cal N}(0)
+ 2\sum_{m,n,k=1}^{\infty} \frac{m}{\sqrt{nk}}
\left[ \eta_{m}^{(n)*}\eta_{m}^{(k)}
\langle\hat {b}_n^{\dag}\hat {b}_k\rangle
+ {\rm Re}\left( \eta_{m}^{(n)}\xi_{m}^{(k)}
\langle\hat {b}_n\hat {b}_k \rangle\right)\right]
\label{Ncav}
\end{equation}
(to obtain this formula  one should use
the identity (\ref{cond1})).
Differentiating (\ref{Ncav}) with respect to $\tau$
and performing the summation over $m$ with the help of equations
(\ref{pksik}), (\ref{petak}), (\ref{pksikin}), (\ref{petakin}),
or (\ref{prhok}), one can obtain the formula
\begin{eqnarray}
\frac{d{\cal N}^{(cav)}}{d\tau}&=& 2\sigma\sum_{n,k=1}^{\infty}
\frac{\langle\hat {b}_n^{\dag}\hat {b}_k\rangle }
{\sqrt{nk}} \sum_{m=1}^p m(p-m)
\left[ \rho_{-m}^{(n)*}\rho_{p-m}^{(k)}+
\rho_{-m}^{(k)}\rho_{p-m}^{(n)*}\right]\nonumber\\
&-& 2\sigma{\rm Re}\sum_{n,k=1}^{\infty}
\frac{\langle\hat{b}_n\hat{b}_k\rangle}
{\sqrt{nk}} \sum_{m=1}^p m(p-m)
\left[ \rho_{-m}^{(n)}\rho_{m-p}^{(k)}+
\rho_{m}^{(n)}\rho_{p-m}^{(k)}\right].
\label{extradot}
\end{eqnarray}
Using equation (\ref{extradot}) and
replacing the coefficients $\rho_m^{(n)}$ by their asymptotical values
(\ref{asxieta}) one can obtain the expression
\begin{eqnarray}
&&\frac{d{\cal N}^{(cav)}}{d\tau}= \frac{4ap^2}{\pi^2}
\sum_{m=1}^{p-1}\sin^2(\pi m/p) \sum_{n,k=0}^{\infty}
\frac{\sigma^{n+k}}{\sqrt{(m+pn)(m+pk)}}\nonumber\\[2mm]
&&\times \left\{
\langle\hat {b}_{m+pn}^{\dag}\hat {b}_{m+pk}\rangle ( a+i\gamma)^{k-n}
-\sigma{\rm Re}\left[\langle\hat{b}_{m+pn}\hat{b}_{m+pk}\rangle
( a+i\gamma)^{k+n+1} \right]\right\}
\label{extradotas}
\end{eqnarray}
which holds provided $ap\tau\gg 1$ and  $a>0$.
For the physical initial states the sum in the right-hand side
of (\ref{extradotas}) is finite. This is obvious if a finite
number of modes was excited initially. But even if the cavity was initially
in a high-temperature thermal state, so that
$\langle\hat {b}_{n}^{\dag}\hat {b}_{k}\rangle=\delta_{nk}\Theta/n$,
$\langle\hat {b}_{n}\hat {b}_{k}\rangle=0$, the sum over $n,k$ yields a
finite value
$
\Theta \sum_{n=0}^{\infty}\,(m+pn)^{-2}
$.
Consequently, the total number of ``nonvacuum''
photons increases in time {\it linearly\/} at $ap\tau\gg 1$, whereas the
total number of quanta generated from vacuum increases {\it quadratically\/}
in the long time limit.

\subsection{The ``principal resonance'' ($p=2$)}\label{p=2}

Many formulas obtained above can be
simplified in the special case $p=2$. In this case there are two subsets
of nonzero Bogoliubov coefficients.
The first one consists of the coefficients with even upper
and lower indices $\xi_{2k}^{(2q)}$ which are reduced to the coefficients
$\xi_{k}^{(q)}$ of the ``semi-resonance'' case.
However, since $\eta_{2k}^{(2q)}\equiv 0$,
this subset does not contribute to the generation of new photons.
The second subset is formed by the ``odd'' coefficients which can be
written as [$\kappa\equiv \kappa(2\tau)$]
\begin{eqnarray}
\xi_{2m+1}^{(2n+1)}&=&
\frac{\Gamma\left(n+3/2\right)
\kappa^{n-m}\lambda^{m+n+1}}
{\Gamma\left(m+3/2\right)\Gamma\left(1+n-m\right) }
\nonumber\\
&\times&
F\left(n+1/2\,,\,-m -1/2\,;\, 1+n-m\,;\, \kappa^2\right),
\quad n\ge m
\label{xinm}
\end{eqnarray}
\begin{eqnarray}
\xi_{2m+1}^{(2n+1)}&=&
\frac{(-1)^{m-n} \Gamma\left(m+1/2\right)
\kappa^{m-n}\lambda^{m+n+1}}
{\Gamma\left(n+1/2\right) \Gamma\left(1+m-n\right) }
\nonumber\\
&\times&
F\left(m+1/2\,,\,-n -1/2\,;\, 1+m-n\,;\, \kappa^2\right),
\quad m\ge n
\label{ximn}
\end{eqnarray}
\begin{eqnarray}
\eta_{2k+1}^{(2n+1)}&=&
\frac{(-1)^{k-1}\Gamma\left(k+1/2\right)\Gamma\left(n+3/2\right)
\kappa^{n+k+1}\lambda^{n-k} }
{\pi\Gamma\left(2+n+k\right) }
\nonumber\\
&\times&
F\left(n+1/2\,,\,k +1/2\,;\, 2+n+k\,;\, \kappa^2\right).
\label{etank}
\end{eqnarray}

It is known \cite{BrMar} that the hypergeometric function $F(a,b;c;z)$
with `half-integral' parameters $a,b$ and an integral parameter $c$ can be
expressed in terms of the complete elliptic integrals
\be
{\bf K}(\kappa )=\int_0^{\pi /2}\frac {\mbox{d}\alpha}{\sqrt {1
-\kappa^2\sin^2\alpha}}
=\frac{\pi}{2} F\left(\frac12\,,\,\frac12\,;\,1\,;\,\kappa^2\right)
\label{defK}
\ee
\be
{\bf E}(\kappa )=\int_0^{\pi /2}\mbox{d}\alpha
\sqrt {1-\kappa^2\sin^2\alpha}
=\frac{\pi}{2} F\left(-\frac12\,,\,\frac12\,;\,1\,;\,\kappa^2\right).
\label{defE}
\ee
In particular,
\begin{equation}
\xi_1^{(1)}=\frac{2}{\pi}\lambda(\kappa){\bf E}(\kappa), \quad
\eta_1^{(1)}=\frac{2}{\pi\kappa}\left[\tilde{\kappa}^2{\bf K}(\kappa )
-{\bf E}(\kappa )\right],
\label{xietell}
\end{equation}
\begin{equation}
\rho_3^{(1)}=
\frac{2\lambda^2(\kappa)}{3\pi\kappa}\left[\left(1-2\kappa^2\right)
{\bf E}(\kappa) -\tilde{\kappa}^2{\bf K}(\kappa ) \right]
\label{xi3}
\ee
\be
\rho_{-3}^{(1)}=
-\,\frac{2}{3\pi\kappa^2\lambda(\kappa)}\left[\left(2-\kappa^2\right)
{\bf E}(\kappa) -2\tilde{\kappa}^2{\bf K}(\kappa ) \right]
\label{et3}
\end{equation}
where
\begin{equation}
\tilde{\kappa}\equiv\sqrt{1-\kappa^2}=
\left[1 +S^2(2\tau)\right]^{-1/2},
\label{deftilkappa}
\end{equation}
and $\lambda(\kappa)$ was defined in (\ref{def-lam}).

The general structure of the coefficients $\rho_{\mu}^{(1)}$
(we confine ourselves to the case $p=2$) is as follows
\begin{equation}
\rho_{2m+1}^{(1)}=
\frac{2\lambda^{m+1}(\kappa)}{\pi\kappa^m}\left[
f_m\left(\kappa^2\right) {\bf E}(\kappa)
+\tilde{\kappa}^2 g_m\left(\kappa^2\right) {\bf K}(\kappa ) \right]
\label{xim}
\ee
\be
\rho_{-2m-1}^{(1)}=
\frac{2}{\pi\kappa^{m+1}\lambda^m(\kappa)}\left[
r_m\left(\kappa^2\right) {\bf E}(\kappa)
+\tilde{\kappa}^2 s_m\left(\kappa^2\right){\bf K}(\kappa ) \right]
\label{etm}
\end{equation}
where $f_m(x), g_m(x), r_m(x), s_m(x)$ are the polynomials of the degree $m$
which can be found from the recurrence relations (\ref{prhok}).

The photon generation rate from vacuum in the
principal cavity mode ($m=1$) reads
\begin{equation}
\frac {\mbox{d}{\cal N}_1^{(vac)}}{\mbox{d}\tau}=
-2{\rm Re}\left[\eta_1^{(1)} \xi_1^{(1)}\right]
=\frac {8\sqrt{1-\gamma^2\kappa^2}}{\pi^2\kappa}
{\bf E}(\kappa )\left[{\bf E}(\kappa )-
\tilde{\kappa}^2{\bf K}(\kappa )\right].
\label{rate1}
\end{equation}
The average number of photons in the first mode can be obtained by
integrating this equation.
It is convenient to integrate with respect to the variable $\kappa$,
taking into account the relation (for $p=2$, for instance)
\[
\mbox{d}\kappa=2\beta\tilde{\kappa}^2\mbox{d}\tau, \quad
\beta=
\mbox{Re}\lambda =
\sqrt{1-\gamma^2\kappa^2}.
\]
For example, in the case of the quadrature variance $U_1$
we arrive at the equation
\bea
\frac{\mbox{d}U_1}{\mbox{d}\kappa}&=& -\,
\frac2{\pi^2 \tilde{\kappa}^2 \kappa^2 \beta} \left\{
\left[\kappa^2\left(1-2\gamma^2\kappa^2\right) +1 -2\beta\kappa\right]
{\bf E}^2(\kappa )
\right.\nonumber\\  && \left.
 - 2\tilde{\kappa}^2(1 -\beta\kappa)
{\bf E}(\kappa ){\bf K}(\kappa ) + \tilde{\kappa}^4 {\bf K}^2(\kappa )
\right\}.
\label{eqU1}
\eea
Let us consider first the case $\gamma=0$, when $\beta=1$.
Taking into account the differentiation rules \cite{Grad}
\begin{equation}
\frac {\mbox{d}{\bf K}(\kappa )}{\mbox{d}\kappa}=\frac {
{\bf E}(\kappa )}{\kappa\tilde{\kappa}^2}-\frac {{\bf K}(\kappa )}{
\kappa},\quad
\frac {\mbox{d}{\bf E}(\kappa )}{\mbox{d}\kappa}=\frac {
{\bf E}(\kappa )-{\bf K}(\kappa )}{\kappa}
\label{difrul}
\end{equation}
we may suppose that the factor $\tilde{\kappa}^2$ in the denominator
of the right-hand side of equation (\ref{eqU1}) comes from the derivative
$\mbox{d}{\bf K}/\mbox{d}\kappa $. Thus it is natural to look for the
solution in the form
\be
U_1= \frac 2{\pi^2\kappa}
\left[ A(\kappa){\bf K}^2(\kappa )
+B(\kappa){\bf K}(\kappa ) {\bf E}(\kappa) +
C(\kappa) {\bf E}^2(\kappa)\right],
\label{try}
\ee
where $A(\kappa)$, $B(\kappa)$ and $C(\kappa)$ are some polynomials of
$\kappa$. Putting the expression (\ref{try}) into equation (\ref{eqU1})
we obtain a set of coupled equations for the unknown functions
$A,B,C$. Writing $A(\kappa)=a_0 +A_1(\kappa)$,
$B(\kappa)=b_0 +B_1(\kappa)$, $C(\kappa)=c_0 +C_1(\kappa)$
we determine the constant coefficients
$a_0$, $b_0$ and $c_0$ by putting $\kappa=0$ in that equations.
Then we obtain new
equations for the functions $A_1(\kappa)$, $B_1(\kappa)$ and $C_1(\kappa)$
and repeat the procedure. After a few steps we arrive at the equations
which have obvious trivial solutions $A_n=B_n=C_n=0$. This confirms our
hypothesis on the polynomial structure of the functions
$A(\kappa)$, $B(\kappa)$ and $C(\kappa)$ and gives the final answer.
The equations for the variances $U_{\mu}$, $V_{\mu}$, etc. with $\mu\ge 3$
can be integrated in the same manner,
the only difference is that one should write $\kappa^{\mu}$ instead of
$\kappa$ in the denominator of the expression like (\ref{try}).
In the generic case $\gamma\neq 0$ we notice that the factor $\beta$
can appear in the denominator of the expression (\ref{eqU1}) as a result
of differentiating the function $\beta(\kappa)$, since
$\mbox{d}\beta/\mbox{d}\kappa= -\gamma^2\kappa/\beta$. Therefore we split
each function, $A,B,C$ in the `$\beta$-even' and `$\beta$-odd' parts like
$A=A_e(\kappa) +\beta(\kappa) A_o(\kappa)$.
The equations for the `even' and `odd'
coefficients turn out independent, and we solve them using the procedure
described above.

The results of the integrations are as follows,
\begin{equation}{\cal N}_1^{(vac)}(\kappa )=
\frac 2{\pi^2}{\bf K}(\kappa )
\left[2{\bf E}(\kappa) -\tilde{\kappa}^2{\bf K}(\kappa )\right]
-\frac 12.
\label{num1EK}
\end{equation}
\[
U_1= \frac 2{\pi^2\kappa}
\left[ \tilde{\kappa}^2(\beta-\kappa){\bf K}^2(\kappa )
-2(\beta-\kappa){\bf K}(\kappa ) {\bf E}(\kappa) +
\beta {\bf E}^2(\kappa)\right],
\]
\[
V_1= \frac 2{\pi^2\kappa}
\left[ 2(\beta+\kappa){\bf K}(\kappa ) {\bf E}(\kappa)
-\tilde{\kappa}^2(\beta+\kappa){\bf K}^2(\kappa ) -
\beta {\bf E}^2(\kappa)\right],
\]
\[
Y_1= \frac {2\gamma}{\pi^2}
\left[
\tilde{\kappa}^2{\bf K}^2(\kappa ) -
2{\bf K}(\kappa ) {\bf E}(\kappa)
+ {\bf E}^2(\kappa)\right].
\]
Making the transformation \cite{Bateman,Abram}
\[
{\bf K}\left(\frac{1-\tilde{\kappa}}{1+\tilde{\kappa}}\right)
=\frac{1+\tilde{\kappa}}{2}{\bf K}(\kappa), \quad
{\bf E}\left(\frac{1-\tilde{\kappa}}{1+\tilde{\kappa}}\right)
=\frac{{\bf E}(\kappa)+\tilde{\kappa}{\bf K}(\kappa)}
{1+\tilde{\kappa}}
\]
one can rewrite formulae (\ref{xietell}) and (\ref{num1EK}) in the form
found for the first time in \cite{DK96} (in the special case of $\gamma=0$).
Using the asymptotical expansions of the elliptic integrals at $\kappa\to 1$
\cite{Grad}
\begin{eqnarray*}
{\bf K}(\kappa )&\approx&\ln\frac 4{\tilde{\kappa}}
+\frac 14\left(\ln\frac 4{\tilde{\kappa}}-1\right)\tilde{\kappa}^
2+\cdots \\
{\bf E}(\kappa )&\approx& 1+\frac 12\left(\ln\frac
4{\tilde{\kappa}}-\frac 12\right)\tilde{\kappa}^2 +\cdots
\end{eqnarray*}
one can obtain the formula
\begin{equation}
{\cal N}_1^{(vac)}(\tau\gg 1)=\frac {8a}{\pi^2}\tau +
\frac4{\pi^2}\ln\left(\frac{2}{a}\right)-\frac 12 +
{\cal O}\left(\tau e^{-4a\tau}\right), \quad a>0.
\label{num1as}
\end{equation}
In the special case of $\gamma=1$ one can obtain the expansion
\[
{\cal N}_1^{(vac)}(\tau\gg 1)=\frac {4}{\pi^2}\ln\tau +
\frac{12}{\pi^2}\ln2-\frac 12 + {\cal O}\left(\tau^{-2}\right)
\]
If $\gamma>1$, the number of photons in the principal mode oscillates with
the period $\pi/(2\tilde{a})$. For $\gamma\gg 1$ one can write
$\kappa\approx\sin(2\tilde{a}\tau)/\tilde{a}$, i.e. $|\kappa|\ll 1$.
In this case
\[
{\cal N}_1^{(vac)}\approx \frac{\kappa^2}{4}\approx
\frac{\sin^2(2\tilde{a}\tau)}
{4\tilde{a}^2} \ll 1.
\]

Equation (\ref{def-uv}) yields
the minimal and maximal {\em invariant variances\/}
\be
u_1= \frac 2{\pi^2\kappa}
\left[ \tilde{\kappa}^2(1-\kappa){\bf K}^2(\kappa )
-2(1-\kappa){\bf K}(\kappa ) {\bf E}(\kappa) +{\bf E}^2(\kappa)\right],
\label{u1min}
\ee
\be
v_1= \frac 2{\pi^2\kappa}
\left[ 2(1+\kappa){\bf K}(\kappa ) {\bf E}(\kappa)
-\tilde{\kappa}^2(1+\kappa){\bf K}^2(\kappa ) - {\bf E}^2(\kappa)\right],
\label{v1min}
\ee
which depend on the detuning
parameter $\gamma$ only implicitly, through the dependence on $\gamma$
of the function $\kappa(\tau)$.
In the short time limit $\tau\ll 1$ (then $\kappa\approx 2\tau$) we obtain,
using the Taylor expansions of the complete elliptic integrals,
$u_1=\frac12 -\tau +\tau^2 +\cdots$ and
$v_1=\frac12 +\tau +\tau^2 +\cdots$
in accordance with \cite{DKM90}.
More precisely,
\[
\begin{array}{l}
u_1\\
v_1
\end{array}\Bigg\}
=\frac12\left( 1 \mp \kappa +\frac12\kappa^2 \mp \frac14\kappa^3
+\frac{7}{32}\kappa^4 +\cdots\right)
\]
The minimal variance $u_1$ monotonously decreases from the value $1/2$
at $t=0$ to the constant asymptotical value $2/\pi^2$ at $\tau\gg 1$,
confirming qualitatively the evaluations of \cite{DK92,DKN93}
and giving almost 50\% squeezing in the initial vacuum state.
The variance of the conjugate quadrature monotonously increases,
and for $\tau\gg 1$ it becomes practically linear function of time:
$v_1(\tau\gg 1)\approx 16\tau/\pi^2$.  The asymptotical minimal value
$u_1(\tau=\infty)$ does not depend on $\gamma$ provided $\gamma\le 1$
(only the rate of reaching this asymptotical value decreases with $\gamma$
as $\sqrt{1-\gamma^2}$). In the strongly detuned case, $\gamma >1$,
the minimal variance oscillates as a function of $\tau$
(being always greater than $2/\pi^2$),
since in this case the function $\kappa(\tau)$ oscillates between
$-\gamma^{-1}$ and $\gamma^{-1}$.

The minimal variance does not go to zero when $\tau\to\infty$ due to the
{\em strong intermode interaction\/}, which results in a high degree
of {\em quantum mixing\/} for each mode.
Since the state originating from the initial
vacuum state belongs to the class of {\em Gaussian\/} states
(see the next subsection), the quantum `purity'
$\chi_m\equiv\mbox{Tr}\hat\rho_m^2$
of the $m$th field mode (described by means of the density matrix
$\hat\rho_m$) can be expressed in
terms of the (co)variances as \cite{167}
$\chi_m=\left[4\left(U_m V_m -Y_m^2\right)\right]^{-1/2}$.
Using equations (\ref{vartherminf})-(\ref{u-inf}) one can check that
for $\tau\gg 1$,
$UV-Y^2 =2Fu(\infty) +{\cal O}(1) \sim \tau$.
Consequently, the purity factor $\chi$ asymptotically goes to zero as
$\tau^{-1/2}$.
For instance, for $m=1$ we have (writing simply ${\bf K}$ and ${\bf E}$
instead of ${\bf K}(\kappa)$ and ${\bf E}(\kappa)$)
\be
\chi_1=\frac{\pi^2}{4}\kappa\left[4{\bf K}{\bf E}^3
+4\tilde{\kappa}^4{\bf K}^3{\bf E}
-6\tilde{\kappa}^2{\bf K}^2{\bf E}^2
-{\bf E}^4
-\tilde{\kappa}^6{\bf K}^4 \right]^{-1/2}
\label{chi}
\ee
The initial dependence on $\kappa$ is rather weak:
$\chi(\kappa\ll 1) = 1-\frac{3}{32}\kappa^4 +\cdots$.
But when $\kappa\to 1$, $\chi$ rapidly goes to zero:
$\chi(\tilde{\kappa}\ll 1)\approx (8/\pi^2)
\left[\ln\left(4/\tilde{\kappa}\right)\right]^{-1/2}$, with
$\mbox{d}\chi/\mbox{d}\kappa \to -\infty$.

The expressions for the variances in the modes with numbers $\mu\ge 3$
are rather involved.
Here we give only one explicit example -- the variance $U_3$ for $\gamma=0$:
\bea
U_3 &=& \frac 2{9\pi^2\kappa^3}
\left[ \tilde{\kappa}^2(1-\kappa)
\left( 4 +10\kappa +9\kappa^2\right){\bf K}^2(\kappa )\right. \nonumber\\
&&\left.+(1-\kappa)\left( 4\kappa^3 -14\kappa^2 -20\kappa -8\right)
{\bf K}(\kappa ) {\bf E}(\kappa) \right. \nonumber\\
&&\left.+\left( 4\kappa^4 +6\kappa^3 -\kappa^2 +6\kappa +4\right)
{\bf E}^2(\kappa)\right]
\label{U3}
\eea
The Taylor expansion of the right-hand side of
(\ref{U3}) coincides with the expansion (\ref{UVsmall}). The asymptotical
value at $\tau\to\infty$ equals
$U_3(\kappa=1) =38/(9\pi^2)\approx 0.43$. We see that the squeezing
rapidly disappears with increase of the mode number $\mu$.
The variance $V_3$ can be obtained from (\ref{U3}) by means of a simple
substitution $\kappa\rightarrow -\kappa$. Therefore the mean number of
photons in the third mode is given by
\begin{equation}
{\cal N}_3=
\frac 2{3\pi^2\kappa^2}\left[\left(3\kappa^2 -2\right){\bf K}
\left(2{\bf E} -\tilde{\kappa}^2{\bf K}\right)
+2\left(1+\kappa^2\right){\bf E}^2\right]
-\frac 12\,.
\label{num3}
\end{equation}

The second derivative
of the total mean number of photons created from vacuum in all the modes
(\ref{ratetot2}) takes the form
\be
\frac {\mbox{d}^2{\cal N}^{(vac)}}{\mbox{d}\tau^2}=
\frac 8{\pi^2\kappa^2}
\left[\tilde{\kappa}^4{\bf K}^2
-2\tilde{\kappa}^2{\bf K}{\bf E}
+\left(1+\kappa^2 -2\gamma^2\kappa^4\right){\bf E}^2\right].
\label{secder}
\ee
Integrating this equation with the account of the condition
$\mbox{d}{\cal N}^{(vac)}/\mbox{d}\tau=0$ at $\tau=0$
[which is a trivial consequence of Eq. (\ref{ratetot})],
we obtain very simple expression
\be
{\cal N}^{(vac)}=
\frac 2{\pi^2} {\bf K}(\kappa)
\left[{\bf K}(\kappa) -{\bf E}(\kappa) \right].
\label{intsecder}
\ee
In the limiting cases this formula yields
\[
{\cal N}^{(vac)}(\tau\ll 1)\approx\tau^2
\]
\[
{\cal N}^{(vac)}(\tau\gg 1)= 8a^2\tau^2/{\pi^2} +{\cal O}(\tau),
 \quad a>0.
 \]
If $\gamma\gg 1$, then $|\kappa|\ll 1$, but
$\gamma^2\kappa^2\approx\sin^2(2\tilde{a}\tau)\sim {\cal O}(1)$. In this
case the Taylor expansion of the second derivative yields
$\ddot{{\cal N}}^{(vac)}=2\cos(4\tilde{a}\tau) + {\cal O}(\gamma^{-2})$.
Integrating this equation with account of the initial conditions
$\dot{{\cal N}}^{(vac)}(0)={\cal N}^{(vac)}(0)=0$ one obtains
$ {\cal N}^{(vac)}\approx {\cal N}_1^{(vac)}\approx
\sin^2(2\tilde{a}\tau)/(4\tilde{a}^2)$.

\subsection{Photon distribution}\label{PDF}

Now let us turn to the {\it photon distribution function\/} (PDF)
$
f({\bf n})\equiv \langle {\bf n}|\hat\rho_m(t)|{\bf n}\rangle
$,
where $|{\bf n}\rangle$ is the multimode Fock state,  
${\bf n}\equiv\left(n_1,n_2,\ldots\right)$, and $\hat\rho_m(t)$ is
the time-dependent density matrix of the $m$th field mode in the
{\it Schr\"odinger\/} picture.
Note that all the calculations in the
preceding sections were performed in the framework of the
{\it Heisenberg\/} picture.
Nonetheless, the available information  is sufficient  to calculate
the PDF for the special (but very important)
class of {\it Gaussian\/} initial states (defined as the states whose
density matrices, or wave functions, or Wigner functions, are
described by some Gaussian exponentials). This class includes
coherent, squeezed and thermal states; in particular, it includes
the vacuum state.

The solution is based on two key points. The first one is
the statement \cite{RazTer85,Law,JoSar96,Plun,Plun99}
that the field evolution in a cavity with moving boundaries
can be described not only in the Heisenberg picture, but, 
equivalently, in the
framework of the Schr\"odinger picture, with a {\it quadratic\/}
multidimensional time-dependent Hamiltonian. 
The second key point is the fact \cite{Int-75,183}
that the evolution governed by quadratic Hamiltonians
transforms any Gaussian state to another Gaussian state. 
Knowing these facts,
it remains to take into account that the photon distribution function of
any Gaussian state is determined completely by the
average values of quadratures and by their variances
\cite{1mod,DM94},
which obviously do not depend on the quantum mechanical representation.

In the most compact form the information on the photon distribution
$f(n)$ in some mode (we suppress here the mode index) is contained in
the {\it generating function\/}
\[
G(z)=\sum_{n=0}^{\infty}\,f(n)z^n
\]
For a generic one-mode Gaussian state it
can be expressed as \cite{DA,1mod,DM94}
\be
G(z)=[{\cal G}(z)]^{-1/2}\exp\left(\frac1{D}\left[
\frac{zg_1 -z^2 g_2}{{\cal G}(z)} - g_0\right] \right)
\label{GF}
\ee
where
\be
{\cal G}(z)=\frac14\left[(1+z)^2 +4\left(UV-Y^2\right) (1-z)^2 +
2(U+V)\left(1-z^2\right)\right]
\label{calG}
\ee
\[
D=1+2(U+V) +4 \left(UV-Y^2\right) = 4{\cal G}(0)
\]
\[
g_0= \langle \hat{p}\rangle^2(2U+1) +\langle \hat{q}\rangle^2(2V+1)
-4\langle \hat{p}\rangle\langle \hat{q}\rangle Y
\]
\begin{eqnarray*}
g_1&=& 2\langle \hat{p}\rangle^2\left(U^2 +Y^2 +U +\frac14\right)
+2\langle \hat{q}\rangle^2\left(V^2 +Y^2 +V +\frac14\right)\\
&&-4\langle \hat{p}\rangle\langle \hat{q}\rangle Y(U+V+1)
\end{eqnarray*}
\[
g_2= 2\langle \hat{p}\rangle^2\left(U^2 +Y^2 -\frac14\right)
+2\langle \hat{q}\rangle^2\left(V^2 +Y^2 -\frac14\right)
-4\langle \hat{p}\rangle\langle \hat{q}\rangle Y(U+V).
\]
In the generic case $f(n)$ is related to the two-dimensional `diagonal'
Hermite polynomials \cite{1mod}:
\be
f(n)=\frac{{\cal F}_0}{n!}H_{nn}^{\{{\cal R}\}}\left( x, x^*\right)
\label{f-Hnn}
\ee
where
\[
{\cal F}_0=f(0)= 2D^{-1/2}\exp\left(-g_0/D\right)
\]
\[
x=\frac{\sqrt{2}\left\{(2V-1)\langle \hat{q}\rangle -2Y\langle \hat{p}\rangle
+i\left[(1-2U)\langle \hat{p}\rangle +2Y\langle \hat{q}\rangle\right]\right\}}
{2(U+V) -4\left(UV-Y^2\right) -1}
\]
and $2\times2$ symmetric matrix ${\cal R}$ has the elements
\[
{\cal R}_{11}={\cal R}_{22}^* = \frac2{D}(V-U-2iY), \quad
{\cal R}_{12}={\cal R}_{21} =\frac1{D}\left[1-4\left(UV-Y^2\right)\right]
\]
The two-dimensional Hermite polynomials are defined via  the
expansion \cite{Bateman}
\be
\exp\left(-\frac12 {\bf a}{\cal R}{\bf a} +{\bf a}{\cal R}{\bf x}\right)
=\sum_{m,n=0}^{\infty} \frac{a_1^m a_2^n}{m!n!}
H_{mn}^{\{{\cal R}\}}\left( x_1, x_2\right)
\label{def-herm}
\ee
where $ {\bf x}=\left( x_1, x_2\right)$, ${\bf a}=\left( a_1, a_2\right)$.
The properties of these polynomials were studied in
\cite{DM94,D94}. In particular, they can be expressed as finite sums of the
products of the usual (one-dimensional) Hermite polynomials. The
corresponding formula for the probabilities reads \cite{1mod}
\be
f(n)={\cal F}_0\left(\frac{\Delta}{D}\right)^n
\sum_{k=0}^n \left(\frac{S}{\Delta}\right)^k \frac{n!}{[(n-k)!]^2 k!}
\left|H_{n-k}(\xi)\right|^2
\label{f-Herm}
\ee
where
\[
\Delta=\sqrt{(U-V)^2 +4Y^2}, \quad S=4\left(UV-Y^2\right)-1
\]
\[
\xi=\frac{(2V+1)\langle \hat{q}\rangle -2Y\langle \hat{p}\rangle
+i\left[(1+2U)\langle \hat{p}\rangle -2Y\langle \hat{q}\rangle\right]}
{\left[2D(V-U-2iY)\right]^{1/2}}
\]

If $\langle \hat{p}\rangle=\langle \hat{q}\rangle=0$, then
the generating function (\ref{GF}) is reduced to
$[{\cal G}(z)]^{-1/2}$, i.e. it has the same structure
as the known generating function of the Legendre polynomials $P_n(x)$.
In this case, we have the following expression for the
photon distribution in the $m$th field mode:
\begin{equation}
f_m(n)= \frac{2\left[(2u_m-1)(2v_m-1)\right]^{n/2}}
{\left[(2u_m+1)(2v_m+1)\right]^{(n+1)/2}}
P_n\left(\frac {4u_m v_m- 1}{
\sqrt{\left(4u_m^2-1\right)\left(4v_m^2-1\right)}}\right)
\label{fotdis}
\end{equation}
It depends only on the invariant minimal and maximal variances
$u_m$ and $v_m$.
Note that the argument of the polynomial in (\ref{fotdis}) is always
{\it outside\/} the `traditional' interval $(-1,1)$ 
(in particular, this argument is pure imaginary if $2u_m <1$), being exactly
equal to $1$ for the `non-principal' modes with
$u_m=v_m={\cal N}_m +\frac12$,
when formula (\ref{fotdis}) transforms to the
 time-dependent Planck's distribution
\[
f_m(n;\tau)=\frac{{\cal N}_m^n(\tau)}
{\left[{\cal N}_m(\tau)+1\right]^{n+1}}.
\]
Only for the `principal' $\mu$-modes the spectrum of photons is
different from Planck's one due to the squeezing effect. 

The function (\ref{fotdis}) can be simplified
in the long-time limit $\tau\gg 1$,
when the average number of created photons
${\cal N}\equiv \bar{n}\approx (V+U)/2$ exceeds $1$.
Then the mean-square fluctuation of the photon number 
has the same
order of magnitude as the mean photon number itself, 
$\sqrt{\sigma_n}\approx \sqrt2 {\cal N}$, and
 the most significant part
of the spectrum corresponds to the values $n\gg 1$.
Using the Laplace--Heine asymptotical formula for the
Legendre polynomial \cite{Szego}
\[
P_n(z)\approx \frac{\left(z+\sqrt{z^2-1}\right)^{n+1/2}}
{\sqrt{2\pi n}\left(z^2-1\right)^{1/4}}\,, \quad n\gg 1
\]
one can simplify (\ref{fotdis}) for the fixed values of the invariant
variances $u$ and $v$ as
\begin{equation}
f(n) \approx \frac1{\sqrt{\pi n(v-u)}}
\left(\frac{2v-1}{2v+1}\right)^{n+1/2}
\label{asf}
\end{equation}
provided the positive difference $v-u$ is not too small.
Another approximate formula can be used if $v\gg 1$ but
$u\sim 1$:
\be
f(n)\approx \frac{\sqrt2(2u-1)^{n/2}}
{\sqrt{v}(2u+1)^{(n+1)/2}} e^{-n/(2v)}
P_n\left(\frac {2u}{\sqrt{4u^2-1}}\right),
\quad n\ll 8v^2.
\label{aprdis}
\end{equation}

The first and second derivatives of the generating function (\ref{GF})
at $z=1$ yield the first two moments of the photon distribution
(hereafter we suppress subscript $m$)
\be
\bar{n} = \frac12(u+v-1), \quad
\sigma_n\equiv \overline{n^2} -(\bar{n})^2=
\frac14\left(2u^2 +2v^2  -1\right)
\label{var-n}
\ee
which result in the Mandel parameter \cite{Man-Q}
\be
Q\equiv \sigma_{n}/\bar{n} -1=
\frac{u^2 +v^2  - u -v + 1/2}{u+v-1}.
\label{Mand}
\ee
This parameter appears positive for all values of $\tau$, so the photon
statistics is super-Poissonian, with strong bunching of photons
(the pair creation of photons in the NSCE was discussed in
\cite{BE,Mund,Lamb,Lamb98}).
In particular,
\[
Q_{2m+1}(\tau\to 0)\approx \left[(m+1)(2m-1)!!/m!\right]^2 \tau^{2m}/(2m+1),
\quad Q_1(0)=1,
\]
whereas $Q_m\approx V_m(\tau)\gg 1$ for $\tau\gg 1$  (if $\gamma\ll 1$).

\section{Energy and formation of packets}

\subsection{Energy density}

The mean value of the energy density operator in one space dimension
\be
\hat{W}(x,t)=\frac1{8\pi}\left[\left(
\partial \hat{A}/\partial t\right)^2+
\left(\partial \hat{A}/\partial x\right)^2\right]
\label{hatW}
\ee
at $t\ge T$ equals (hereafter we assume $L_0=1$, i.e. $\omega_1=\pi$)
\bea
&&\tilde{W}(x,t)=\pi\sum_{m,j=1}^{\infty}\sqrt{mj}\Bigg\{
\cos[\pi(m+j)x]\mbox{Re}\left[\langle \hat{a}_m\hat{a}_j
\rangle e^{-i\pi(m+j)t'}\right]
\nonumber\\ && +
\frac12
\cos[\pi(m-j)x]\left[
\langle \hat{a}_m^{\dag}\hat{a}_j\rangle e^{i\pi(m-j)t'}
+\langle \hat{a}_m\hat{a}_j^{\dag}\rangle e^{-i\pi(m-j)t'}
\right]\Bigg\}
\label{tildeW}
\eea
where the quantum mechanical averaging $\langle\cdots\rangle$
is performed over the initial state of the field
(the Heisenberg picture) and $t'\equiv t+\delta T$.

\subsubsection{Regularization and Casimir's energy}
Due to the commutation relations
$ \left[\hat a_m\,,\,\hat a_j^{\dagger}\right]=\delta_{mj}$
the series (\ref{tildeW}) contains the vacuum divergent `diagonal'
 part with $m=j$:
\be
\tilde{W}^{(vac)} =(\pi/2)\sum_{m=1}^{\infty}
m\exp(-i\pi mt' +i\pi mt').
\label{ap1}
\ee
The recipe how to regularize this divergence was given in \cite{FulDav}.
One should write the first term in the argument of the
exponential in (\ref{ap1}) as it stands, but to replace $t'$ in the
second term by $t' + i\eta$, $\eta >0$
(this is the so-called `point-splitting method').
Then the sum becomes convergent, giving
\[
\tilde{W}^{(vac)}(\eta) =
(\pi/2)\sum_{m=1}^{\infty}m e^{-m\pi\eta}=
(\pi/8)\left[\sinh(\pi\eta/2)\right]^{-2}.
\]
The Taylor expansion of this function reads
$\tilde{W}^{(vac)}(\eta) = (2\pi\eta^2)^{-1} -\pi/24 +{\cal O}(\eta^2)$.
According to \cite{FulDav}, one should remove the divergent term
$(2\pi\eta^2)^{-1}$ and after that proceed to the limit $\eta\to 0$.
This limit value gives us the known expression for
the one-dimensional negative vacuum Casimir energy
\cite{FulDav,Cas,Plun1,Mil,Most}
(which does not depend on the coordinate $x$ in the case involved)
\be
\tilde{W}^{(Cas)}= -\pi/24 \quad (\mbox{or} \;\; -\frac{\pi\hbar c}{24L_0^2}
\; \mbox{in the dimensional units})
\label{WCas}
\ee

Extracting this vacuum energy from $\tilde{W}$ we arrive at the expression
\bea
W&\equiv& \tilde{W}- \tilde{W}^{(Cas)}
=\pi\sum_{m,j=1}^{\infty} \sqrt{mj}
\mbox{Re}\Big[\langle \hat{a}_m\hat{a}_j
\rangle \cos[\pi(m+j)x]e^{-i\pi(m+j)t'}
\nonumber\\ && +
\langle \hat{a}_m^{\dag}\hat{a}_j\rangle
\cos[\pi(m-j)x]e^{i\pi(m-j)t'}
\Big] .
\label{W}
\eea
The same expression (\ref{W}) can be obtained if one calculates the mean
value of the {\em normally ordered\/} (with respect to the operators
$\hat{a}^{\dag}_n$ and $\hat{a}_n$) counterpart of the operator (\ref{hatW})
(cf. \cite{LawPRL}).
Then the total energy (without the vacuum part) assumes the usual form
\be
{\cal E}=\int_0^{L_0}W(x,t)dx
=\sum_{n=1}^{\infty}\omega_n\left\langle\hat
a^{\dag}_n\hat a_n\right\rangle
\label{Etot}
\ee
which justifies the choice of the normalization in (\ref{Ast}) and
(\ref{vecpotfin}).

Since the initial quantum state was defined with respect to the
`in'-operators $\hat{b}_n^{\dag}$ and $\hat{b}_n$, we must express
the `out' operators $\hat{a}_m^{\dag}$ and $\hat{a}_m$ in terms of
$\hat{b}_n^{\dag}$ and $\hat{b}_n$ by means of formula (\ref{Bogol}).
Thus we arrive at the expression containing
a combination of the mean values
$\langle\hat{b}_n\hat{b}_k\rangle$,
$\langle\hat{b}_n^{\dag}\hat{b}_k^{\dag}\rangle$,
$\langle\hat{b}_n^{\dag}\hat{b}_k\rangle$
and $\langle\hat{b}_n\hat{b}_k^{\dag}\rangle$ calculated in the
initial quantum state.
For the initial vacuum state defined according to the relations
$\hat{b}_n |0\rangle =0$, $n=1,2,\ldots$,
the only nonzero mean values are
$\langle\hat{b}_n\hat{b}_k^{\dag}\rangle =\delta_{nk}$.
Then (\ref{W}) is transformed into the triple sum
\bea
W_0(x,t)&=&\pi\sum_{n,m,j=1}^{\infty}\frac{mj}{n}\mbox{Re}\Big\{
\cos[\pi(m-j)x]e^{i\pi(m-j)t'} \rho_{-m}^{(n)}\rho_{-j}^{(n)*}
\nonumber\\ && -
\cos[\pi(m+j)x]e^{i\pi(m+j)t'} \rho_{-m}^{(n)}\rho_j^{(n)*}
\Big\}    .
\label{W-0}
\eea
Evidently, $W_0(x,t)=0$ for $t\le 0$. For an arbitrary initial state
the energy density can be written as a sum of the `vacuum' and
`nonvacuum' contributions
\be
W=W_0 + W_1, \quad W_1= \pi\sum_{n,k=1}^{\infty}\frac{1}{\sqrt{nk}}\mbox{Re}
\left[ \langle\hat{b}_n\hat{b}_k\rangle B^{(nk)}
+\langle\hat{b}_n^{\dag}\hat{b}_k\rangle \tilde{B}^{(nk)} \right]
\label{W-gen}
\ee
where
\bea
&&B^{(nk)}=
\sum_{m,j=1}^{\infty}{mj}\Bigg\{
\cos[\pi(m+j)x]\left[
e^{-i\pi(m+j)t'} \rho_m^{(n)}\rho_j^{(k)}
\right.\nonumber\\ && \left.
+e^{i\pi(m+j)t'} \rho_{-m}^{(n)}\rho_{-j}^{(k)}\right]
\nonumber\\ && -
\cos[\pi(m-j)x]\left[
e^{-i\pi(m-j)t'} \rho_m^{(n)}\rho_{-j}^{(k)}
+e^{i\pi(m-j)t'} \rho_{-m}^{(n)}\rho_{j}^{(k)}\right]\Bigg\}
\label{def-B}
\eea
\bea
&&\tilde{B}^{(nk)}=
\sum_{m,j=1}^{\infty}{mj}\Bigg\{
\cos[\pi(m-j)x]\left[
e^{-i\pi(m-j)t'} \rho_{-m}^{(n)*}\rho_{-j}^{(k)}
\right.\nonumber\\ && \left.
+e^{i\pi(m-j)t'} \rho_{m}^{(n)*}\rho_{j}^{(k)}\right]
 \nonumber\\ &&
-\cos[\pi(m+j)x]\left[
e^{-i\pi(m+j)t'} \rho_{-m}^{(n)*}\rho_j^{(k)}
+e^{i\pi(m+j)t'} \rho_{m}^{(n)*}\rho_{-j}^{(k)}\right] \Bigg\}
\label{def-Btil}
\eea
Making the change of the summation index $j\to -j$ in the first term of
(\ref{W-0}) we can write
\[
W_0(x,t)= -\pi\mbox{Re}\sum_{n=1}^{\infty}\sum_{m=1}^{\infty}
\sum_{j=-\infty}^{\infty}\frac{mj}{n}
\cos[\pi(m+j)x]e^{i\pi(m+j)t'} \rho_{-m}^{(n)}\rho_j^{(n)*}.
\]
Similarly, changing the indices $m\to -m$ or $j\to -j$
in (\ref{def-B}) and (\ref{def-Btil})
we can reduce 4 sums with apparently different summands and
the indices running from $1$ to $\infty$ to the unified sums whose
two indices run from $-\infty$ to $\infty$:
\[
B^{(nk)}=
\sum_{m,j=-\infty}^{\infty}{mj}
\cos[\pi(m+j)x]
e^{-i\pi(m+j)t'} \rho_m^{(n)}\rho_j^{(k)}
\]
\[
\tilde{B}^{(nk)}=
\sum_{m,j=-\infty}^{\infty}{mj}
\cos[\pi(m-j)x]
e^{i\pi(m-j)t'} \rho_{m}^{(n)*}\rho_{j}^{(k)}
\]
Now, replacing the cosine function by the sum of two imaginary
exponentials we see that $W(x,t)$ is actually the sum of two
identical functions of the light-cone variables:
\be
W(x,t)=\frac{\pi}{2}\left[F(u;\tau) +F(v;\tau)\right],
\quad   u=t'+x, \quad v=t'-x,
\label{W-F}
\ee
where
\be
F= F_0 +
\sum_{n,k=1}^{\infty}\frac{1}{\sqrt{nk}}\mbox{Re}
\left[ \langle\hat{b}_n\hat{b}_k\rangle F^{(nk)}
+\langle\hat{b}_n^{\dag}\hat{b}_k\rangle \tilde{F}^{(nk)} \right]
\label{def-FFF}
\ee
\be
F_0(u;\tau)= -\mbox{Re}\sum_{n=1}^{\infty}\sum_{m=1}^{\infty}
\sum_{j=-\infty}^{\infty}\frac{mj}{n}
e^{i\pi(m+j)u} \rho_{-m}^{(n)}(\tau)\rho_j^{(n)*}(\tau)
\label{def-Fu}
\ee
\bea
F^{(nk)}(u;\tau)&=& \sum_{m=-\infty}^{\infty}
\sum_{j=-\infty}^{\infty}\,{mj}
e^{-i\pi(m+j)u} \rho_{m}^{(n)}(\tau)\rho_j^{(k)}(\tau)
\label{def-Funk}           \\
\tilde{F}^{(nk)}(u;\tau)&=& \sum_{m=-\infty}^{\infty}
\sum_{j=-\infty}^{\infty}\,{mj}
e^{i\pi(m-j)u} \rho_{m}^{(n)*}(\tau)\rho_j^{(k)}(\tau)
\label{def-tilFunk}
\eea
The extra argument $\tau$ in the above expressions is introduced
in order to emphasize that the energy density depends not only on
the value of the current time variable $t$
(which must satisfy the condition $t>(1+\delta)T$),
but also on the moment of time $T$ when the wall stopped to move.
It is worth mentioning that the variables $t$ and $\tau$ are independent,
as well as $u$ and $\tau$ or $v$ and $\tau$.

Evidently, the double sums (\ref{def-Funk}) and (\ref{def-tilFunk}) are
factorized to the products of independent sums over $m$ and $j$:
\[
F^{(nk)}(u;\tau)=G^{(n)}(u;\tau)G^{(k)}(u;\tau), \quad
\tilde{F}^{(nk)}(u;\tau)=G^{(n)*}(u;\tau)G^{(k)}(u;\tau),
\]
\be
G^{(n)}(u;\tau)= \sum_{m=-\infty}^{\infty}\; m e^{-i\pi mu}
\rho_{m}^{(n)}(\tau)=\frac{i}{\pi}\frac{\partial}{\partial u}
\sum_{m=-\infty}^{\infty}\; e^{-i\pi mu}\rho_{m}^{(n)}(\tau)
\label{G-u}
\ee
The last sum in (\ref{G-u}) can be easily expressed in terms of the
generating function (\ref{defR}) if one writes $n=j+kp$, $m=j+lp$ and
$z=\exp(-i\pi pu)$. Thus we obtain
\be
G^{(n)}(u;\tau)=\left.\frac{nz
\left[z g(p\tau)+\sigma S(p\tau)\right]^{n/p-1}}
{\left[g^*(p\tau)+ z\sigma S(p\tau)\right]^{n/p+1}}
\,\right|_{\,z=\exp(-i\pi pu)}\,= n{f}^{1/2} \Lambda^{n/p},
\label{G-n}
\ee
where
\be
f(u;\kappa)=\left|g^*(p\tau)+ z\sigma S(p\tau)\right|^{-4}
=\frac{\left(1-\kappa^2\right)^2}
{\left[1+ \kappa^2 +2\sigma\kappa\cos(p\pi u -\varphi)\right]^2} ,
\label{def-f}
\ee
\be
\Lambda=
\frac{z g(p\tau)+\sigma S(p\tau)}{g^*(p\tau)+ z\sigma S(p\tau)}
=e^{i(2\varphi -\pi pu)}
\frac{1 +\sigma \kappa\exp[i(\pi pu -\varphi)]}
{1 + \sigma \kappa\exp[i(\varphi -\pi pu)]},
\label{def-Lam}
\ee
\[
\kappa=\frac{S(p\tau)}{\sqrt{1+S^2(p\tau)}}, \quad
\exp(i\varphi) =\sqrt{1-\gamma^2\kappa^2} +i\gamma\kappa\,.
\]

In the `vacuum' contribution (\ref{def-Fu}) we have some asymmetry
between the indices $m$ and $j$, since $m$ runs from $1$ to $\infty$,
whereas $j$ runs from $-\infty$ to $\infty$.
This asymmetry can be eliminated if one differentiates both sides of
equation (\ref{def-Fu}) with respect to the independent variable $\tau$
at a fixed value of $u$ and
performs the summation over the superscript $n$ with the aid of the
recurrence relations (\ref{recrho}) and (\ref{recrho1}).
It is easy to verify that all the summands
with $n\ge p$ are canceled, so the infinite series over $n$ can be
reduced to the finite sum from $1$ to $(p-1)$:
\bea
\frac{\partial F_0(u;\tau)}{\partial\tau}&=& -\sigma\mbox{Re}
\sum_{n=1}^{p-1}\sum_{m=1}^{\infty}
\sum_{j=-\infty}^{\infty}\;{mj} e^{i\pi(m+j)u}
\nonumber\\ &&\times
\left[ \rho_{-m}^{(n)}(\tau)\rho_{-j}^{(p-n)}(\tau)
+\rho_{m}^{(p-n)*}(\tau)\rho_{j}^{(n)*}(\tau)\right]
\label{res1}
\eea
Making the change of summation indices $m\to -m$, $j\to -j$, $n\to p-n$
in the first product inside the square brackets one can reduce two
sums in the right-hand side of (\ref{res1}) to the single series where both
the indices $m$ and $j$ run from $-\infty$ to $\infty$. Moreover, the
sums over $m$ and $j$ become completely independent,
giving rise to the equation
\be
\frac{\partial F_0(u;\tau)}{\partial\tau}= -\sigma\mbox{Re}
\sum_{n=1}^{p-1} G^{(n)}(u;\tau)G^{(p-n)}(u;\tau),
\label{F-G}
\ee
where
$G^{(n)}(u;\tau)$ is given by (\ref{G-u}). Due to (\ref{G-n})
the sum in the right-hand side of (\ref{F-G}) is reduced to the sum
$\sum_{n=1}^{p-1} \,n(p-n)=\frac16 p(p^2-1)$.
Introducing the variable $\eta=\exp(2ap\tau)$ we obtain
the explicit expression
\be
\frac{\partial F_0(u;\eta)}{\partial\eta}=-\,\frac
{(p^2-1)a^4\eta\left[\eta^2(1+\alpha +\beta) +\alpha -\beta -1\right]}
{12\left[\eta^2(1+\alpha +\beta) -2\eta(\gamma^2 +\beta) +1 +\beta -\alpha
\right]^3}\,,
\label{der}
\ee
where $\alpha =\sigma a\cos(p\pi u)$ and $\beta=\sigma\gamma\sin(p\pi u)$.
Integrating (\ref{der}) with the initial condition $F_0=0$ at $\tau=0$
(or $\eta=1$) we arrive after some algebra at the simple expression
\be
F_0(u;\kappa)={\cal B}\left[ f(u;\kappa) -1 \right],  \quad
{\cal B}\equiv \left(p^2-1\right)/24\,,
\label{F0-fin}
\ee
where function $f(u;\kappa)$ is given by (\ref{def-f}).
Finally we obtain the following expression for the function $F(u;\tau)$
defined by equation (\ref{W-F}):
\be
F= -{\cal B} + f(u;\tau) \left\{ {\cal B} +
  \sum_{n,k=1}^{\infty} \sqrt{nk}
\mbox{Re}\left[ \langle\hat{b}_n\hat{b}_k\rangle \Lambda^{\frac{n+k}{p}}
+\langle\hat{b}_n^{\dag}\hat{b}_k\rangle \Lambda^{\frac{k-n}{p}} \right]
\right\}.
\label{W1-fin}
\ee
In the special case of the initial states whose density matrix is
diagonal in the Fock basis, so that
$\langle\hat{b}_n^{\dag}\hat{b}_k\rangle=\nu_n\delta_{nk}$ and
$\langle\hat{b}_n\hat{b}_k\rangle=0$ (for example, the Fock or thermal
states; $\nu_n$ is the mean number of quanta in the $n$th mode),
the sum in (\ref{W1-fin}) is proportional
to the initial total energy ${\cal E}_0$
in all the modes (above the Casimir level):
\be
F^{(diag)}(u;\tau)=
-{\cal B} + f(u;\tau) \left[ {\cal B}+{\cal N}_0\right],
\label{Fdiag}
\ee
\[
{\cal N}_0=\sum_{n=1}^{\infty}\, n\nu_n ={\cal E}_0/\pi
\equiv {\cal E}_0/(\hbar\omega_1)\,.
\]

\subsection{Packet formation}

Now let us analyse the expressions for the energy density obtained above.
 For the initial vacuum state we see
immediately from equations  (\ref{def-f}) and (\ref{F0-fin})
that in the generic case the function $W_0(x,t)$ with the fixed value of
the `fast time' $t$ has $p$ peaks in the interval $0\le x\le 1$,
whose positions are determined by
the equations $\sigma\cos(p\pi u -\varphi)=-1$ and
$\sigma\cos(p\pi v -\varphi)=-1$.
Obviously, for $t>T$ the energy density is a periodic function
of the time variable $t$, with the period $\Delta t=1$ if $p$ is an even
number and $\Delta t=2$ if $p$ is odd.

For the even values of the resonance multiplicity $p$ we have $p/2$ peaks
moving (with the light speed) in the positive direction
and $p/2$ peaks moving in the negative direction. If $p$ is odd, then
the numbers of peaks of each kind differ by $1$.
All the peaks have the same height
\be
W_{max}^{(vac)}= 2\pi{\cal B}\kappa/(\kappa-1)^2=
\frac{\pi}{2}{\cal B}\left( e^{4p\tau} -1\right)
\label{max}
\ee
(in this section the expressions containing $\tau$
are related to the special case of the strict resonance $\gamma=0$),
excepting some distinguished instants of
time when two peaks moving in the opposite directions merge, forming
a peak with double the height.

If $\kappa\to 1$ (i.e. $\tau>1$ and $\gamma<1$), then
the energy density can be approximated in the vicinity of each peak  by
the Lorentz-like distribution
\be
W^{(vac)}(\delta x)=\frac{W_{max}^{(vac)}}
{\left[1 + \left(2\delta x/\Delta_{1/4}\right)^2 \right]^{2}},
\label{min14}
\ee
where the width
\[
\Delta_{1/4}=\frac{2}{p\pi}\frac{1-\kappa}{\sqrt{\kappa}}
\approx \frac{4}{p\pi} e^{-2p\tau}
\]
 of each peak is defined as the double
distance between the position of the maximum and the point where
the energy density decreases 4 times.
One can introduce also the `energy width' of each peak
by means of the relation $W_{max}\Delta_E={\cal E}(\tau)/p$.
For $\kappa\to 1$ we obtain
\[
\Delta_E\approx (1-\kappa)/(2\pi p)
\approx (\pi p)^{-1}e^{-2p\tau}.
\]
Except for narrow regions of the length
$\Delta_+ \approx (\pi p)^{-1}\sqrt{1-\kappa}
\approx \sqrt2(\pi p)^{-1}e^{-p\tau}$
nearby the peaks  the `dynamical' energy density is less than
its initial
vacuum value, in agreement with the results of \cite{Dal99,LawPRL,Cole,Wu}
obtained in the framework of different approaches.
The minimum values of $W_0$ far off the peaks are given by
(taking into account the contributions of
both the functions $F_0(u)$ and $F_0(v)$)
\be
W_{min}= -4\pi{\cal B}\frac{\kappa}{(\kappa+1)^2}=
\pi{\cal B}\left( e^{-4p\tau} -1\right).
\label{min}
\ee
If $\kappa\to 1$,
$ W_{min}\to -\pi\left(p^2-1\right)/24 $. Adding to this expression the
initial Casimir energy (\ref{WCas}) we obtain the total asymptotical
minimum value (cf. \cite{LawPRL})
\be
\tilde{W}_{min}^{(as)}= -\pi p^2/24 .
\label{lim}
\ee

For an arbitrary initial state the energy density has, besides the `vacuum'
part, the additional terms given in equation (\ref{W1-fin}).
Since these terms are proportional to the same functions $f(u;\kappa)$
or $f(v;\kappa)$ which determine the structure of the `vacuum' part,
the positions of the peaks are not changed (remember that $|\Lambda|=1$).
For the initial states with diagonal density matrices in the Fock basis
(in particular, for the thermal states) all the peaks
still have equal heights, increased by the quantity
$\Delta W_{max}=\frac12{\cal E}_0 (1+\kappa)^2/(1-\kappa)^2$, compared with
the vacuum case.
However, the asymptotical {\em minimal\/}
value of the energy density at $\kappa\to 1$ does not depend on the initial
state, being given by formula (\ref{lim}) in all the cases.

If the initial density matrix in the Fock basis has nonzero
off-diagonal elements (as happens, in particular, for any pure state
different from the Fock one, e.g., for the coherent states),
different terms in the sum (\ref{W1-fin}) can interfere.
As a consequence, the peaks acquire some kind of `fine structure'.
For example, if only the first mode
was excited initially in the coherent state $|\alpha\rangle$,
$\alpha=|\alpha|\exp(i\psi)$, then for $p=2$ and $\gamma=0$ (the strict
resonance) we have
\[
\Delta W\equiv W-W^{(vac)}= \pi|\alpha|^2 \frac{\left(1-\kappa^2\right)^2
\left[\kappa\sin(z +\psi) +\sin(z -\psi)\right]^2}
{\left[(1-\kappa)^2 +4\kappa\sin^2 z\right]^3}
\]
where $z\equiv \pi \left(u -u_*\right)$ and $u_*$ is the position of
the `vacuum' peak determined above.
If $\psi=\pi/2$, then we have the high maximum
$\Delta W_{\pi/2}^{max}=\pi|\alpha|^2(1+\kappa)^2/(1-\kappa)^2$
at $z=0$. However, if $\psi=0$, then instead of a maximum we have the
minimum $\Delta W=0$ at the same point $z=0$, and the peak is split
in two symmetric humps with equal maximal heights
$\Delta W_0^{max}= \left([1+\kappa]^2/27\kappa\right)
\Delta W_{\pi/2}^{max}$ located at the points
$\sin z = \pm (1-\kappa)/\sqrt{8\kappa}$.
In the intermediate case $0<\psi<\pi/2$ asymmetric forms of the peaks
are observed.
If $p>2$ or several modes were excited initially, the interference
between different terms in (\ref{W1-fin}) can result in different
heights of the peaks and more complicated `fine structures'
(provided $\langle \hat{b}_n\hat{b}_k\rangle \neq 0$ for some $n$ and $k$).

If the detuning $\gamma$ is different from zero, then some deformations
of the form of peaks are observed, although the maximal heights are still
of the same order of magnitude as in the case $\gamma=0$, as far as
$\gamma\le 1$.
But if the detuning exceeds the critical value
$\gamma=1$, the energy becomes an oscillating function of
the `slow time' $\tau$, the amplitude of oscillations being
proportional approximately to $\left(\gamma^2-1\right)^{-1}$ \cite{Djpa,DA}.
The peaks become rather wide and low,
since the parameter $\kappa$ is limited by the inequality
$\kappa\le \gamma^{-1}$ if $\gamma>1$.
The illustrations can be found in \cite{AD00}.

Since the components of the energy-momentum tensor $T_{00}$ and $T_{11}$
are given by similar expressions in the case of single space dimension,
the force acting on each wall has the same time dependence
as the energy density at the points $x=0$ and $x=1$.
For the most part of time during the period of field oscillations
$2L_0/c$ (where $L_0$ is the distance between the walls at rest)
this force is negative, being less than the static
Casimir force, with the maximal amplification coefficient $p^2$.
However, the average value of the force over the period is positive due
to the creation of real photons inside the cavity.

If the walls possess some small transmission coefficient, then a small
part of radiation accumulated inside the cavity can leave it. In this case
one could observe sharp pulses of radiation outside the cavity
\cite{Lamb-puls},
whose amplitudes must be proportional to the heights of the peaks inside the
cavity multiplied by the small transmission coefficient. The intensity of
these pulses can be significantly increased, if the initial state is
different from vacuum and possesses sufficient energy, like thermal
states \cite{Lamb-puls,Plun00} or coherent states.
However, to describe the form of the pulses exactly it is necessary to
develop a more general theory which would take into account the
boundary conditions corresponding to the partially transmitting walls
(because the nonzero transmission coefficient can change significantly
the pulse shape, just as the nonzero detuning deformed
the form of packets in the examples considered above).

\subsection{Total energy }

The total energy (\ref{Etot}) of the field inside the cavity
(above the initial Casimir level)  can be obtained by
integrating the density $W(x)$ (\ref{W-F}) over $x$.
The contribution of the vacuum (function $F_0$ in (\ref{def-FFF}))
and `diagonal' terms (given by the partial sum in (\ref{def-FFF}) over $n=k$)
can be calculated with the aid of the formula
\[
\int_0^{\pi}\frac{dx}{(a+b\cos x)^2} =
\frac{\pi a}{\left(a^2-b^2\right)^{1/2}}.
\]
To find the contribution of `nondiagonal' terms ($n\neq k$) it is convenient
to replace the integration over $x$ by the integration in the complex
$z$-plane ($z=\exp[-i\pi pu]$ or $z=\exp[-i\pi pv]$)
over the circle $|z|=1$. One can
check that this circle is passed $p$ times when
$x$ goes from $0$ to $1$ (if one takes into account both the `$u$'-
and `$v$'-contributions).
It turns out that the integrals of the `non-diagonal' terms
are different from zero if and only if
the corresponding integrands in the $z$-plane have simple poles inside
the circle $|z|=1$. This happens only when $k+n=p$ in the first term
inside the square brackets in (\ref{W1-fin}) and $k-n=p$ in the second term
inside the same brackets.

Finally we obtain a simple expression
\be
{\cal E}(\tau)= {\cal E}_0 +2S^2(p\tau)
 \left[{\cal E}_0 +\pi{\cal B} +\frac{\gamma\sigma }{2}
 {\rm Im}({\cal G})\right]
-\frac{\sigma}{2}S(2p\tau){\rm Re}({\cal G}),
\label{ansEtot}
\ee
where
\begin{equation}
{\cal G} = 2\pi\sum_{n=1}^{\infty}
\sqrt{n(n+p)}\langle\hat {b}_n^{\dag}\hat {b}_{n+p}\rangle
+\pi\sum_{n=1}^{p-1}\sqrt{n(p-n)}\langle\hat {b}_n\hat {b}_{p-n}\rangle
\label{defcalG}
\end{equation}
(if $p=1$, the last sum in (\ref{defcalG}) should be replaced by zero).
Formula (\ref{ansEtot}) was found for the first time in a different way
in \cite{Djpa}.
In particular, for the initial vacuum state of field we have
\begin{equation}
{\cal E}^{(vac)}(\tau)= \frac{p^2-1}{12a^2}\sinh^2(pa\tau)\,.
\label{Etotvac}
\end{equation}
The total energy increases exponentially at $\tau\to\infty$,
provided $\gamma< 1$.
In the special case $\gamma=0$ such asymptotical behaviour of the total
energy was obtained also in the frameworks of other approaches in
\cite{LawPRL,Cole,Mep,Law-new}. Here we have found the explicit
dependence of the total energy on time in the whole interval
$0\le \tau<\infty$, as well as a nontrivial dependence on the initial state
of field, which is contained in the constant parameter ${\cal G}$.
This parameter is equal to zero for initial Fock or
thermal states of the field. However, in a generic case ${\cal G}\neq 0$,
 and it can affect significantly the total energy,
if ${\cal E}(0)\gg 1$. Consider, for example, the case $p=2$.
If initially the first mode ($n=1$) was
in the coherent state $|\alpha\rangle$ with $\alpha=|\alpha|e^{i\phi}$,
$|\alpha|\gg 1$, and all other modes were not excited,
then ${\cal E}(0)=|\alpha|^2$, ${\cal G}=\alpha^2$, so
for $\tau\gg 1$ and $\gamma=0$ (exact resonance) we have
${\cal E}(\tau\gg 1)\approx \frac14 |\alpha|^2 e^{4\tau}
\left[2-\cos(2\phi)\right]$. The maximal value of the energy in this case
is three times bigger than the minimal one, depending on the phase $\phi$.

According to (\ref{ansEtot}), the initial stage of the evolution does not
depend on the detuning parameter $\gamma$ for all states which yield
Im$({\cal G})=0$, since at $\tau\to 0$ one has
\begin{equation}
{\cal E}(\tau) \approx {\cal E}(0)
-\sigma{\rm Re}({\cal G}) p\tau
+2\left[{\cal E}(0) +\frac{p^2-1}{24} +\frac{\gamma\sigma }{2}
 {\rm Im}({\cal G})\right] (p\tau)^2
\label{Etot-0}
\end{equation}
 Formula (\ref{Etot-0}) is {\it exact\/} in the case of $\gamma=1$.

If $\gamma>1$, then one should replace each function $\sinh(ax)/a$ in
(\ref{ansEtot}) by its trigonometrical counterpart
$\sin(\tilde{a}x)/\tilde{a}$: see Eq. (\ref{def-tilgS}).
In this case the total energy {\it oscillates\/} in time with
the period $\pi/(p\tilde{a})$, returning to the initial value at the
end of each period. For a large detuning $\gamma\gg 1$
the amplitude of oscillations decreases as $\gamma^{-1}$ if
Re${\cal G}\neq 0$ and as $\gamma^{-2}$ otherwise.

Note that the total ``vacuum'' and ``nonvacuum'' energies increase
exponentially with time, if $\gamma<1$ and $\tau>1$,
whereas the total number of photons
increases only as $\tau^2$ and $\tau$, respectively, under the same
conditions.
The origin of such a great difference in the behaviours of the total
energy and the total number
of photons becomes clear, if one looks at the asymptotical formulae
(\ref{asrate})-(\ref{asratebigqS}). They show that the rate of photon
generation in the $m$th completely excited
mode decreases approximately as $\dot{\cal N}_m \sim 1/m$
(excepting the modes whose numbers
are multiples of $p$), so the stationary rate of the {\it energy\/}
generation $\dot{\cal E}_m = m\dot{\cal N}_m $ asymptotically
almost does not depend on $m$.
In turn, the number of the effectively excited modes increases
in time exponentially. These two factors lead to
the exponential growth of the total energy
(see also \cite{Klim} in the special case $\gamma=0$).

\section{Three-dimensional nondegenerate cavity}

\subsection{Empty cavity}

Now let us proceed to the three-dimensional case.
For definiteness we choose a rectangular cavity with dimensions
$L_x$, $L_y$, $L_z$ (briefly designated by symbol $\{L\}$).
If these dimensions do not depend on time, each field mode is determined
by three integers $m,n,l$, responsible for the eigenfrequency
\begin{equation}
\omega_{mnl}=\pi \left[\left(m/L_x\right)^2+\left(n/L_y\right)^2
+\left(l/L_z\right)^2\right]^{1/2},
\label{freq}\end{equation}
and by two orthogonal directions of polarization.
In order to simplify the exposition and to get rid of extra
unessential indices, let us consider the case when $L_z\ll L_x\sim L_y$.
Then the frequencies with $l\neq 0$ are much greater than those with $l=0$.
It is clear that the interaction between low- and high-frequency modes in the
nonstationary case is weak. Consequently, studying
the excitation of the {\em lowest modes\/} we may confine ourselves to
the case of $l=0$. Then the only possible polarization of the vector
potential is along $z$-axis, so the low-frequency part of
the {\em Heisenberg field operator\/} at $t<0$ reads
\be
\hat {A}_z(x,y,t<0)=\sum_{\bf n}\left(2\pi/\omega_{\bf n}\right)^{1/2}
\psi_{\bf n}\left(x,y|\{L\}\right)
\left[\hat b_{\bf n}e^{-i\omega_{\bf n}t}
+\hat b^{\dag}_{\bf n}e^{i\omega_{\bf n}t}\right].
\label{Ast3}
\ee
The difference from the similar expression (\ref{Ast}) is that now the
suffix $n$ is replaced by its ``vector'' counterpart ${\bf n}=(m,n)$,
and the function $\psi_{\bf n}\left(x,y|\{L\}\right)$ depends on two
space coordinates:
\[\psi_{\bf n}\left(x,y|\{L\}\right)=2\left(L_xL_yL_z\right)^{
-1/2}\sin\frac {m\pi x}{L_x}\sin\frac {n\pi y}{L_y}.\]
The coefficients in Eq. (\ref{Ast3}) are chosen again in correspondence
with the standard form of the field Hamiltonian (\ref{Ham}).

Now let the dimension $L_x$ to depend on time according to the given law
$L(t)$. To satisfy the boundary conditions
\[\left.A_z\right|_{x=0}=\left.A_z\right|_{x=L(t)}=\left.A_z\right
|_{y=0}=\left.A_z\right|_{y=L_y}=0\]
we write the field operator at $t>0$ in the same functional form
(\ref{Ast3}), but with the time-dependent parameter $L(t)$:
\begin{equation}\hat {A}_z(x,y,t)=2\sqrt {\pi}\sum_{\bf n}\psi_{\bf n}\left
(x,y|L(t),L_y\right)\hat {Q}_{\bf n}(t).\label{Anst}\end{equation}
In the stationary case the operators $\hat {Q}_{\bf n}(t)$ coincide with
the (coordinate) quadrature components of the field mode operators.
Putting (\ref{Anst}) into the wave equation
\[\partial^2A_z/\partial t^2\,-\Delta A_z=0\]
we arrive at the equation looking just as Eq. (\ref{Qeq}):
\bea
\ddot {Q}_{\bf k}^{({\bf n})}+\omega_{\bf k}^2(t)Q_{\bf k}^{({\bf n})}
&=& 2\lambda (t)\sum_{\bf j}g_{{\bf k}{\bf j}}\dot {Q}_{\bf j}^{({\bf n})}
+\dot{\lambda }(t)\sum_{\bf j}
g_{{\bf k}{\bf j}}Q_{\bf j}^{({\bf n})}
\nonumber\\&&
+\lambda^2(t)\sum_{{\bf j}{\bf l}}g_{{\bf j}{\bf k}}g_{{\bf j}{\bf l}}
Q_{\bf {\bf l}}^{({\bf n})}.
\label{Qeq-3}
\eea
Now $\lambda (t)={\dot {L}(t)}/{L(t)}$,
all the indices are ``two-vectors'', the
frequencies are given by Eq. (\ref{freq}) with $l=0$ and $L(t)$ instead of
$L_x$, and the {\it constant\/} numerical coefficients $g_{{\bf kj}}$
are given by the integrals
\[g_{{\bf kj}}=L\int_0^L\mbox{d}x\int_0^{L_y}\mbox{d}y\int_0^{L_z}\mbox{d}
z\,\psi_{\bf j}({\bf r}|L)\frac {\partial\psi_{\bf k}({\bf r}|L)}
{\partial L}.\]
The explicit form of ``two-dimensional'' coefficients $g_{{\bf kj}}$
is more complicated  than a simple formula
(\ref{gkj}). However, these coefficients remain antisymmetrical:
$g_{{\bf kj}}=-g_{{\bf jk}}$,  due to the normalization of functions
$\psi_{\bf k}$,
\[\int_0^{\{L\}}\mbox{d}{\bf r}\,
\psi_{\bf m}\psi_{\bf n}=\delta_{{\bf mn}},\]
and due to zero boundary conditions at $x=L$.
(Moreover, they do not depend on the cavity dimensions.)

Although we use the same notation as in the 1D case, the operators
$\hat {Q}_{\bf n}(t)$ in Eq. (\ref{Anst}) differ from their analogs in
a similar decomposition (\ref{psit}).
Now $\hat {Q}_{{\bf n}}(t)$ means the
Hermitian operator coinciding with the (``coordinate'') {\it quadrature
component\/} of the field mode operator.

Suppose for simplicity that the wall oscillates at twice
the eigenfrequency of some unperturbed mode,
\[L(t)=L_0\left[1-\epsilon_{L}\cos(2\omega_{\bf m}t)\right],
\quad |\epsilon_{L}|\ll 1\]
and let us look for the solution to Eq. (\ref{Qeq-3}) in the form
\begin{equation}Q_{\bf k}(t)=\xi_{\bf k}(\epsilon_{L} t)\exp\left(
-i\omega_{\bf k}t\right)+\eta_{\bf k}(\epsilon_{L} t)\exp\left(
i\omega_{\bf k}t\right)\label{slow}\end{equation}
(we have omitted ``hats'' over operators).
Contrary to the one-dimensional case,
 now {\it all the terms\/} on the right-hand side of Eq. (\ref{Qeq-3})
disappear after averaging over fast oscillations,
since the spectrum $\omega_{\bf j}$ {\em is not equidistant}.  Indeed, the
first and the second sums on the right-hand side do not contain
functions $Q_{\bf k}$ due to the antisymmetricity of coefficients
$g_{{\bf kj}}$, whereas the last sum is proportional to
$\lambda^2\sim\epsilon_{L}^2$.
Consequently, after the multiplication by the proper exponential functions, 
the right-hand side will consist of the terms containing the factors
such as $\exp\left(i\left[\pm\omega_{\bf j}\pm\omega_{\bf k}\pm
2\omega_{\bf m}\right]t\right)$ with ${\bf j}\neq {\bf k}$. After
averaging all these terms turn into zero. (Strictly speaking,
the frequency spectrum (\ref{freq}) contains the ``equidistant subset'',
corresponding to the indices $m,n,l$ multiplicated by the same
integral factors. However, this fact does not change the conclusion,
because the ``coupling constants'' $g_{\bf kj}$
between such modes are equal to zero.)

Consequently, in the {\em resonance case\/} the field problem is
reduced \cite{D95,DK96}
to that of a one-dimensional parametric oscillator with the
time dependence of the eigenfrequency in the form 
\begin{equation}\omega (t)=\omega_0\left[1+2\tilde\varepsilon\cos(2\omega_0t)\right],
\label{omeg-t}\end{equation}
$\omega_0\equiv\omega_{mn}$ being the unperturbed eigenfrequency of the
resonance mode. Here the frequency modulation depth $\tilde\varepsilon$
is related to the
cavity length modulation depth $\epsilon_{L}$ as follows:
\[\tilde\varepsilon =\frac12\epsilon_{L}\left[1+\left(nL_0/mL_y\right)^2\right]
^{-1/2}.\]
We use the notation $\tilde\varepsilon$ in order not to confuse the
dimensionless modulation parameters in the one-dimensional and
three-dimensional cases.

At this point we may abandon the Heisenberg picture
and proceed to the Schr\"odinger representation.  Of course,
both representations are equivalent, as soon as the field
problem has been reduced to studying a finite-dimensional quantum 
system.  However, the most of numerous investigations of the 
time-dependent quantum oscillator, since Husimi's paper
\cite{Husimi}, were performed in the Schr\"odinger picture.  So it is 
natural to use the known results.
According to \cite{183,Husimi,DM79}), all the
characteristics of the {\em quantum\/} oscillator are determined completely 
by the {\em complex\/} solution of the {\em classical\/} oscillator
equation of motion
\begin{equation}\ddot{u }+\omega^2(t)u =0,
\label{eqeps}\end{equation}
satisfying the normalization condition
\begin{equation}\dot{u}u^{*}-\dot{u}^{*}u =2i.\label{Wrons}\end{equation}
Let us assume that function $\omega (t)$ takes the constant value 
$\omega_0$ at $t\le 0$ and at $t>t_f>0$. Moreover, it is convenient to choose
the initial conditions for $u$-function as follows:
\begin{equation}u (0)= 1/\sqrt{\omega_0},\qquad\dot{u}(0)=i\sqrt {\omega_0}.
\label{initeps}\end{equation}
Then the quantum mechanical average number of photons created 
from the ground state due to the time dependence of the frequency 
in the interval of time $0<t<t_f$ is given by the formula
\begin{equation}\langle n\rangle =\frac 1{4\omega_0}\left(|\dot
u |^2+\omega_0^2|u |^2\right)-\frac 12.
\label{number3D}\end{equation}
Looking for the solution of Eq. (\ref{eqeps}) in the parametric resonance
case (\ref{omeg-t}) in the form
\begin{equation}
u (t)=\frac 1{\sqrt {\omega_0}}\left[\xi(t)
e^{i\omega_0t}+\eta(t)e^{-i\omega_0t}\right]
\label{aseps}\end{equation}
(the opposite signs in the arguments of the exponential functions in
Eqs. (\ref{slow}) and (\ref{aseps}) are due to the different representations:
the former equation is written in the Heisenberg picture, while the
latter -- in the Schr\"odinger one)
and using the method of averaging over fast oscillations,
one can easily obtain the first order differential
equations for the amplitudes (provided that $|\tilde\varepsilon |\ll 1$),
\begin{equation}\dot {\xi}=i\omega_0\tilde\varepsilon \eta,\qquad
\dot {\eta}=-i\omega_0\tilde\varepsilon \xi.\label{eqampl}\end{equation}
Their solutions satisfying initial conditions (\ref{initeps}) (up to the
terms of the order of $\tilde\varepsilon$) read
\cite{Yar66,paramp,Takah,Raif1,DMRud}
\begin{equation}\xi(t)=\cosh(\omega_0\tilde\varepsilon t),\qquad
\eta(t)=-i\sinh(\omega_0\tilde\varepsilon t).\label{solampl}\end{equation}
Due to Eqs.  (\ref{number3D}), (\ref{aseps}) and (\ref{solampl}), the
average number of photons (and the total energy in the cavity)
grows in time exponentially:
\begin{equation}
\langle n\rangle =|\eta|^2=\sinh^2\left(\omega_0\tilde\varepsilon
t\right).
\label{num-1}
\end{equation}
It is well known that the initial vacuum state of the oscillator is 
transformed into the {\em squeezed vacuum state}, if the frequency 
depends on time (see, e.g.,  reviews \cite{Kiel,183} and numerous
references therein). Moreover, looking at Eq. (\ref{num-1})
one can immediately recognize the combination
$\omega_0\tilde\varepsilon t$ as the so called {\em squeezing parameter\/}.
Therefore the probability to registrate $n$ photons exhibits typical 
oscillations:
\begin{equation}{\cal P}_{2m}=\frac {\left[\tanh\left(\omega_0\tilde\varepsilon
t\right)\right]^{2m}}{\cosh\left(\omega_0\tilde\varepsilon t\right)}\frac {(
2m)!}{\left(2^mm!\right)^2},\qquad {\cal P}_{2m+1}=0.\label{prob-1}
\end{equation}
This distribution possesses the photon number variance 
$\sigma_n=\frac 12\sinh^2(2\omega_0\tilde\varepsilon t)$.
Similar formulas for the amount of photons
created in a cavity filled with a medium with a time-dependent
dielectric permeability (and stationary boundaries) were found in
\cite{DKNPR}.
The quadrature variances change in time as
 (now $\tau=\tilde\varepsilon\omega_0 t$)
\be
U=\frac12 e^{-2\tau}, \quad V=\frac12 e^{2\tau}.
\label{idsqz}
\ee
An unlimited squeezing can be achieved in this case due to the absence of
interaction with other modes.

\subsection{Interaction with a probe oscillator inside the cavity}

The situation changes drastically, if the field mode can interact with
some detector placed inside the cavity. Following \cite{D95,DK96}
 we demonstrate the effect in the framework of a simplified model,
when a {\em harmonic oscillator\/} tuned to the frequency of the resonant 
mode is placed at the point of maximum of the amplitude mode 
function $\psi_{mn}\left(x,y|\{L\}\right)$ in the 3D rectangular cavity.

Assuming the interaction between the
oscillator and the field to be described by means of the standard minimal
coupling term $-(e/mc){\bf p}{\bf A}$, we arrive at
the following two-dimensional Hamiltonian governing the evolution of the
coupled system ``field oscillator + detector'':
\begin{equation}
H=\frac 12\left[P^2+\omega^2(t)Q^2+p^2+
\omega_0^2q^2-4\omega_0\kappa pQ\right].
\label{ham}\end{equation}
Here $P,Q$ are the quadrature components of the field oscillator, and
$p,q$ are those of the probe oscillator.
We neglect the interaction with
nonresonant modes, since it is reasonable to suppose that under the
resonance conditions their contribution is not essential at
$\varepsilon\ll 1$.

In general, the dimensionless coupling coefficient $\kappa$ must
depend on time, due to the decomposition (\ref{Anst}).  However, 
since this coefficient is small, its variations of the order of
$\varepsilon\kappa$ can be neglected in comparison with the relative
variation of the eigenfrequency $\delta\omega/\omega\sim\varepsilon$.
So $\kappa$ is assumed to be constant.

Suppose that the lowest cavity mode is resonant.  Then one can 
evaluate the dimensionless coupling constant as
$\kappa\sim\left(e^2/2\pi mc^2L\right)^{1/2}$ (here we return to the
dimensional variables).  The maximum value of
parameter $\varepsilon$ is (see the discussion in section \ref{Discuss})
$\varepsilon_{max}\sim\delta_{max}v_s/2\pi c$, where
$\delta_{max}\sim 0.01$ is the maximal
possible relative deformation in the material of the wall,
and $v_s\sim 5\cdot 10^3$~m/s is the sound velocity inside the wall.
Then the ratio $\tilde\varepsilon /\kappa$ cannot exceed the value
$\delta_{max}\left(mv_s^2L/8\pi e^2\right)^{1/2}\sim 0.05$ for
$L\sim 1$~cm and $m\sim$ the mass of electron (for these parameters
$\kappa\sim 2\cdot10^{-7}$).
Consequently, one may believe that in the real conditions
$\tilde\varepsilon /\kappa\ll 1$.

In the time-independent case, $\omega (t)=const=\omega_0$, we have two
eigenfrequencies
\begin{equation}\omega_{\pm}=\omega_0(1\pm\kappa )\label{om-pm}\end{equation}
(provided that $|\kappa |\ll 1$).  Let us assume that the wall vibrates exactly at
twice the lower frequency $\omega_{-}$:
\begin{equation}\omega (t)=\omega_0\left[1+2\tilde\varepsilon\cos(2\omega_{-}
t)\right].\label{-omeg}\end{equation}
Then the lower and upper modes practically do not interact in the
limit of $\tilde\varepsilon\ll\kappa$.

The Schr\"odinger equation with
Hamiltonian (\ref{ham}) can be solved in the framework of the
general theory of multidimensional quantum systems with 
arbitrary quadratic Hamiltonians, first proposed in \cite{Int-75} and
exposed in detail, e.g., in \cite{183}.
In particular, if both for the field and the probe oscillators were
initially in their ground states,
\begin{equation}
\psi (Q,q,0)=\sqrt {\frac 1{\pi}}\exp\left[-\frac
12\left(Q^2+q^2\right)\right],
\label{psi-in}\end{equation}
then the wave function of the coupled ``field $+$ probe oscillator''
system at $t>0$ can be written as \cite{DK96}
\be
\psi (Q,q,t)=\sqrt {\frac 1{\pi\cosh\mu}}\exp\left(
-it-\frac 12\left[a(t)Q^2
+b(t)q^2-2c(t)qQ\right]\right),
\label{psi-fin}\ee
with the following coefficients:
\[
a(t)=1+i\tanh\mu e^{-2i\varphi_{-}}
-i\kappa e^{i\Phi}\left[\tanh\mu e^{-i\Phi}\left(1+\tanh\mu\sin\Phi
e^{i\varphi}\right)-\sin\varphi\right],
\]
\[
b(t)=1-i\tanh\mu e^{-2i\varphi_{-}}
-i\kappa e^{i\Phi}\left[\tanh\mu e^{-i\Phi}\left(1-\tanh\mu\sin\Phi
e^{i\varphi}\right)+\sin\varphi\right],
\]
\[
c(t)=\tanh\mu e^{-2i\varphi_{-}}
+i\kappa\left[1-\cos\varphi e^{-i\Phi}+i\tanh^2\mu\sin\Phi
e^{i(\varphi -2\Phi )}\right].
\]
Here
\[
\Phi=\left(\omega_{+}+\omega_{-}\right)t=2\omega_0t,\quad
\varphi=\left(\omega_{+}-\omega_{-}\right)t=2\omega_0\kappa t=\kappa
\Phi ,
\]
\[
\mu=\tilde\varepsilon\omega_0t,\quad
\Delta=1-\cosh\mu\cos\Phi .
\]
In all the formulas above, the terms of the order of $\kappa^2$ were
neglected, as well as the terms proportional to $\tilde\varepsilon$ (excepting,
of course, the arguments of the hyperbolic functions).
Evidently, $\Phi\gg\varphi\gg\mu$.
Hereafter we confine ourselves to the most interesting
{\em long-time limit\/} case, when $\mu\gg 1$.
Then all the terms proportional to $\kappa$ can be neglected, so
one can write
\[a(t)=1+i\chi ,\quad b(t)=1-i\chi ,\quad c(t)=\chi ,\quad
\chi =\tanh\mu e^{-2i\varphi_{-}}.\]

Eq.  (\ref{psi-fin}) shows that  the coupled system turns out in a
{\em two-mode squeezed state\/} at $t>0$.
The properties of this state, as well as of any {\em Gaussian state\/}
are determined completely by its {\em covariance matrix \/}
\[
{\bf M}=\left\Vert {\cal M}_{\alpha\beta}\right\Vert =
\left\Vert\begin{array}{cc}
{\bf M}_{\pi\pi}&{\bf M}_{\pi x}\\
{\bf M}_{x\pi}&{\bf M}_{xx}\end{array}
\right\Vert ,\qquad
{\cal M}_{\alpha\beta}=\frac 12\langle\hat {z}_{\alpha}\hat {z}_{
\beta}+\hat {z}_{\beta}\hat {z}_{\alpha}\rangle ,
\]
where the $4$-dimensional (in the present case) vector ${\bf z}$ is defined 
as follows:  ${\bf z}=(\pi ,{\bf x})=(P,p,Q,q)$ (evidently, $\langle 
{\bf z}\rangle =0$ in the case 
under study).  Using the general formulas for multidimensional 
Gaussian states given in \cite{183}, we have obtained \cite{DK96}
the following explicit expressions for the two-dimensional blocks of 
matrix ${\bf M}$ in the long-time limit $\mu\gg 1$:
\[
{\bf M}_{\pi\pi}=\frac 12\cosh^2\mu\left\Vert\begin{array}{cc}
1+\tanh\mu\sin\phi&-\tanh\mu\cos\phi\\
-\tanh\mu\cos\phi&1-\tanh\mu\sin\phi\end{array}
\right\Vert ,
\]
\[
{\bf M}_{xx}=\frac 12\cosh^2\mu\left\Vert\begin{array}{cc}
1-\tanh\mu\sin\phi&\tanh\mu\cos\phi\\
\tanh\mu\cos\phi&1+\tanh\mu\sin\phi\end{array}
\right\Vert ,
\]
\[
{\bf M}_{\pi x} = \widetilde{\bf M}_{ x \pi}
=\frac 14\sinh(2\mu )\left\Vert\begin{array}{cc}
-\cos\phi&-\tanh\mu -\sin\phi\\
\tanh\mu -\sin\phi&\cos\phi\end{array}
\right\Vert ,
\]
where $\phi =2\varphi_{-}$, and $\widetilde{\bf M}$ means the transposed
matrix.
Consequently, there exists a strong correlation
between the field and probe oscillators in the long-time limit. For 
instance, the correlation coefficient between the quadrature 
components reads
\[r_{qQ}\equiv\frac {\langle qQ\rangle}{\sqrt {\langle q^2\rangle
\langle Q^2\rangle}}=\frac {\sinh\mu\cos\phi}{\sqrt {1+(\sinh\mu\cos
\phi )^2}}.\]
(If $\phi\approx\pi /2$, this coefficient, as well as other analogous 
elements of the covariance matrix, does not turn exactly into zero; 
in such a special case $r_{qQ}\sim\kappa$, due to neglected terms of 
the order of $\kappa$.)

It is clear that the {\em density matrix\/} of the probe oscillator (which 
is obtained from the density matrix of the total system 
$\rho\left(Q,q;Q',q'\right)=\psi\left(Q,q\right)\psi^{*}\left(Q',
q'\right)$ by putting $Q=Q'$ and integrating 
over $Q$) has also the Gaussian form. Its properties are determined 
completely by the reduced covariance matrix
(accidentally, it coincides with ${\bf M}_{xx}$ when $\omega_0=1$)
\begin{equation}{\bf M}_{pr}=\frac 12\cosh^2\mu\left\Vert\begin{array}{cc}
1-\tanh\mu\sin\phi&\tanh\mu\cos\phi\\
\tanh\mu\cos\phi&1+\tanh\mu\sin\phi\end{array}
\right\Vert .\label{var-pr}\end{equation}
A similar matrix for the field oscillator can be obtained from Eq.
(\ref{var-pr}) by means of changing the sign of parameter $\mu$. As
was shown in \cite{1mod}, the photon statistics in Gaussian
one-mode states is determined completely by two invariants of the 
covariance matrix,
\be
d=\det {\bf M},\qquad T=\mbox{Tr}\,{\bf M}.
\label{def-dT}
\ee
Evidently, parameter $T$ is twice the energy of quantum 
fluctuations.  The parameter $d$ characterizes the {\em degree of
purity\/} of the quantum state, due to the relation \cite{167} 
\be
\mbox{Tr}\hat{\rho}^2=\frac 1{2\sqrt d},
\label{d-pur}
\ee
where $\hat{\rho}$ is the statistical operator of the system.
The {\em degree of squeezing}, defined as
the minimal possible value of the variance of some
quadrature component, normalized by its vacuum value
(which is equal to $\hbar/(2m\omega)$
for an oscillator with mass $m$ and frequency $\omega$), is
determined jointly by both parameters, $T$ and $d$, according to the 
relation \cite{Pol} [cf. Eq. (\ref{def-uv})]
\be
s\equiv 2\langle q^2\rangle =T-\sqrt{T^2-4d}.
\label{s-min}
\ee
Both subsystems have identical invariants:
\[ T=4d=\cosh^2\mu,\]
so for $\mu\gg 1$ they appear in highly mixed quantum states,
with rather moderate degree of squeezing, which tends asymptotically to
$50\%$ (cf. the one-dimensional case described in subsection \ref{p=2}):
\[
s=e^{-\mu}\cosh\mu =\frac12\left(1+e^{-2\mu}\right).
\]
The average number of quanta in each subsystem equals
\[ \langle n\rangle=\frac12(T-1)=\frac12\sinh^2\mu,\]
i.e., twice less than in the case of an empty cavity.
The variance of the number
of quanta (photons) equals
\[
\sigma_n \equiv \langle n^2\rangle - \langle n\rangle^2
=\frac14\left(2T^2-4d-1\right)
=\frac14\sinh^2\mu\cosh(2\mu).
\]
Mandel's parameter (defined by Eq. (\ref{Mand}))
turns out much greater than unity for $\mu\gg 1$, indicating that
the photon statistics is highly super-Poissonian:
\[ {\cal Q} \approx \sinh^2(\mu).\]

The photon distribution
function can be expressed in terms of the Legendre polynomials,
according to the general formula (\ref{fotdis}):
\begin{equation}
{\cal P}_n=\frac{2(iz)^n}{\sqrt{1+3\cosh^2\mu}}P_n(-iz),
\label{prob-2}\end{equation}
where
\[z=\frac{\sinh\mu}{\sqrt{1+3\cosh^2\mu}}.\]
Actually the right-hand side of Eq. (\ref{prob-2}) is a polynomial of 
degree $n$ with respect to the variable $z^2$, due to the recurrence relation
\[ n{\cal P}_n=z^2\left[(2n-1){\cal P}_{n-1}+(n-1){\cal P}_{n-2}\right].\]
If $\mu\gg 1$, then $z^2\approx 1/3$.
The behavior of the distribution function (\ref{prob-2}) was shown in
\cite{DK96}. Since the argument of the Legendre polynomial is pure
imaginary, ${\cal P}_n$
 has no oscillations, in contradistinction to the vacuum squeezed state.

\subsection{Interaction with a two-level detector}

Another model of the detector, which has only two energy levels,
was considered in \cite{D95}. The most
significant features can be described, in the rotating wave approximation,
in the framework of the following generalization of the Jaynes-Cummings
Hamiltonian:
\begin{equation}
H=a^{\dagger}a + \frac12\Omega\sigma_z + \kappa\left(a\sigma_{+} +
a^{\dagger}\sigma_{-}\right) + \frac{\tilde\varepsilon}{2}\sin(\omega_w t)
\left[a^2 + \left(a^{\dagger}\right)^2\right]. \label{JC}\end{equation}
The eigenfrequency of the unperturbed mode is assumed $\omega_0=1$, $\Omega$
is the energy level difference of the detector, $\kappa$ and $\tilde\varepsilon\ll\kappa$
have the same meaning as above; $a,a^{\dagger}$ and $\sigma_{+},
\sigma_{-}, \sigma_z$ are the standard photon and spin operators.
The wave function of the system ``field + detector'' can be written as
\[
\psi(t)=\sum_{n=0}^{\infty}
\left( c_n^{(-)}(t)|n,-\rangle +c_n^{(+)}(t)|n,+\rangle\right),
\]
with a clear meaning of the symbols. If $\tilde\varepsilon=0$, then the known
solution of the JC-model reads \cite{Louisell}
\begin{eqnarray}
c_0^{(-)}(t)&=&c,\label{c0}\\
c_{n+1}^{(-)}(t)&=&a_n\cos\vartheta_n\exp\left[-itE_n^{+}\right]-
b_n\sin\vartheta_n\exp\left[-itE_n^{-}\right],
\label{c-}\\
c_n^{(+)}(t)&=&a_n\sin\vartheta_n\exp\left[-itE_n^{+}\right]+
b_n\cos\vartheta_n\exp\left[-itE_n^{-}\right],\label{c+}
\end{eqnarray}
where $n=0,1,2,\ldots$,
\[E_n^{\pm}=n+\frac12 \pm\lambda_n, \quad
\lambda_n=\left[\frac14(1-\Omega)^2+\kappa^2(n+1)\right]^{1/2},
\]
\[
\tan\vartheta_n=\left(\frac{2\lambda_n-1+\Omega}{2\lambda_n+1-\Omega}
\right)^{1/2}.\]
We suppose that initially the system was in the ground state with the
only nonzero coefficient $c_0^{(-)}(0)=1$, and that the frequency of wall's
vibrations is close to twice the frequency of the unperturbed mode:
$\omega_w=2-\nu$. Looking for the solution at $\tilde\varepsilon\neq 0$
in the same form (\ref{c0})-(\ref{c+}), but with time-dependent coefficients,
and neglecting the rapidly oscillating terms containing $\exp(2it)$, we get
the following equation for the coefficient $c(t)$:
\[ \dot c =\frac{\sqrt{2}}{4}\tilde\varepsilon\left\{b_1\sin\vartheta_1\exp\left[
it\left(\lambda_1-\nu\right)\right] - a_1\cos\vartheta_1\exp\left[
-it\left(\lambda_1+\nu\right)\right]\right\}.\]
Assuming $\nu=\lambda_1=\left[\frac14(1-\Omega)^2+2\kappa^2\right]^{1/2}$
and neglecting the terms oscillating with the frequencies of the order of
$\kappa$, one can check that the infinite system of equations for $a_n$ and
$b_n$ is reduced to the following {\em two} equations:
\[ \dot c =\frac{\sqrt{2}}{4}\tilde\varepsilon\sin\vartheta_1\, b_1,\qquad
 \dot b_1 =-\frac{\sqrt{2}}{4}\tilde\varepsilon\sin\vartheta_1\, c.\]
Consequently, in the resonance case we have only three nonzero amplitudes:
\[ c_0^{(-)}=\cos(\alpha t),\quad
c_2^{(-)}=\sin\vartheta_1\sin(\alpha t)\exp\left[-itE_1^{(-)}\right],
\]
\[
c_1^{(+)}=-\cos\vartheta_1\sin(\alpha t)\exp\left[-itE_1^{(-)}\right],
\]
where $\alpha=\sin\vartheta_1\,\tilde\varepsilon\sqrt2/4$.
No more than two photons can be created, and the probability
of finding the detector in an excited state ${\cal P}^{(+)}$ is always less
than $1/2$. All the probabilities, in contradistinction to the first example,
are periodically oscillating functions of time:
\[{\cal P}_1={\cal P}^{(+)}=\cos^2\vartheta_1\sin^2(\alpha t),\qquad
{\cal P}_2=\sin^2\vartheta_1\sin^2(\alpha t).\]
It is interesting that the upper level of the detector never can be populated
with 100\% probability, since $\vartheta_n>0$ for all values of parameters.
For $\Omega=1$, $\vartheta_n=\pi/4$, and
${\cal P}_1={\cal P}_2={\cal P}^{(+)}=\frac12\sin^2(\tilde\varepsilon t/4)$. Large
detuning, $1-\Omega\gg\kappa$, results in increasing ${\cal P}^{(+)}_
{\mbox{max}}$, since $\vartheta_1\to 0$. However, in such a case
$\alpha\to 0$, as well, and the applicability of JC-model to the description
of the interaction between the detector and field becomes questionable.
Recently, a more general model was considered in \cite{Fedot00}.

\section{Influence of damping}

The complete theory of
the field quantization in media with moving nonideal boundaries
is not available at present. The
field quantization in spatially inhomogeneous but {\it nonabsorbing\/}
dielectrics was studied in \cite{Lewen,Tip}, and the same problem for
nonabsorbing media with time dependent parameters was considered in
\cite{BB87,Lob91a,Lob91b,DKNPR}. The case of absorbing media was analysed in
\cite{Mat95,Mat96,Grun1,Grun2}.
The theory of the field quantization in leaky cavities was developed in
\cite{Uji,Gea,Knoll}.
However, in all those studies the boundaries were fixed.
Due to the complexity of the problem, only a few simplest models
have been considered up to now in the case of {\em moving walls}.

For example, one can try, as the first step,
to neglect coupling between different field modes inside the cavity.
Such an approximation can be justified, e.g.
for an adiabatic motion of the cavity walls, when the characteristic
mechanical frequency, $\omega_m$, is many orders of magnitude less than the
electromagnetic field eigenfrequency, $\omega_e$.
However, no new photons can be created under
the condition $\omega_m\ll\omega_e$ (and the photon number distribution
cannot be changed, as well), since the photon number operator is the
adiabatic invariant in this case.

Fortunately, as was shown in the preceding section,
the interaction between different field modes can be neglected
also in the case of a three-dimensional cavity with a
nonequidistant spectrum of the field eigenfrequencies,
under the parametric resonance condition $\omega_m\approx 2\omega_e$.
In such a case, one can infere some quantitative information
on the behavior of the field in the cavity, studying the problem
of the parametrically excited oscillator with damping.

The influence of an environment on the parametric amplification
was considered in detail, e.g., in \cite{Raif}, where an explicit coupling
with a heat bath consisting of harmonic oscillators was introduced.
More general models of the environment were studied, e.g.,
in the framework of the influence functional approach
\cite{Hot}, mainly in connection with cosmological problems.
It was shown in \cite{Gea,Knoll,Coll,Carm87} that the influence of the
``modes of the
universe'' outside the cavity with {\it fixed mirrors\/} can be described
effectively in the framework of the Heisenberg-Langevin equation of
motion for the photon annihilation and creation operators
$\hat{a}, \hat{a}^{\dag}$.
An equivalent description in the Schr\"odinger picture is achieved
in the framework of the ``standard master equation''
\cite{master1,master2,master3,Dek,Gard1,Gard2}, whose simplest form reads
\begin{equation}
\dot{\hat\rho}=\frac{i}{\hbar}[\hat\rho,\hat H] +\frac{\gamma}{2}\left[
2\hat{a}\hat\rho \hat{a}^{\dag}
- \hat{a}^{\dag}\hat{a}\hat\rho - \hat\rho \hat{a}^{\dag}\hat{a} \right],
\label{stan}
\end{equation}
where $\hat\rho$ is the statistical operator of the distinguished field mode,
$\hat H$ is the Hamiltonian, and
the damping coefficient $\gamma$ absorbs all the details
of the loss mechanism: the transmissivity of the mirror, the coupling
between the field and the atoms inside the wall, etc., so that it is
proportional to the reciprocal of the dissipation time scale.
It was assumed in \cite{Manci,Bose} that Eq. (\ref{stan}) can be used
as well in the case of the {\it moving mirrors\/}, with the same value of
the damping coefficient as in the case of the fixed boundaries.
Following the same line as in \cite{Manci,Bose}, we also assume that the time
evolution of the mixed quantum
state of the resonance field mode is governed by a linear master equation
(although one cannot exclude a possibility that such an approach is
oversimplified: see, e.g., \cite{Leung94,Ching98}).
However, instead of using the operator equation like (\ref{stan}),
we consider the most general linear equation of the Fokker--Planck type for
the {\it Wigner function\/} $W(q,p,t)$ \cite{167,Dek,Conn}
($q,p$ are the quadrature components of the field mode),
\begin{eqnarray}
\frac{\partial W}{\partial t}&=&
\frac{\partial }{\partial q}\left([\gamma_q q -p]W\right)
+\frac{\partial }{\partial p}\left([\gamma_p p +\omega^2(t)q]W\right)
\nonumber\\ &+&
D_{qq}\frac{\partial^2 W}{\partial q^2}
+D_{pp}\frac{\partial^2 W}{\partial p^2}
+2D_{qp}\frac{\partial^2 W}{\partial q\partial p}.
\label{FP}
\end{eqnarray}
The coefficients $\gamma_i$ and $D_{ij}=D_{ji}$
depend on the concrete
form of the microscopic interaction between the system involved and
an environment \cite{master1,master2,master3,Dek,JRLR}.
For example, the simplest models of the damped optical oscillator
with a constant frequency $\omega_0$ yield the following set of coefficients
 \cite{Dek,Glaub}:
\begin{equation}
\gamma_p=\gamma_q= \pi s(\omega_0)|g(\omega_0)|^2,
\label{gamma}\end{equation}
\begin{equation}
D_{pp}=\omega_0^2 D_{qq}=
 \gamma_q {\cal E}_{eq}(\omega_0), \quad D_{pq}=0,
\label{D}\end{equation}
where $s(\omega)$ is the density of states
of the reservoir, $g(\omega_0)$ is the function describing the intensity of
coupling between
the distinguished oscillator and the reservoir degrees of freedom,
and  ${\cal E}_{eq}(\omega_0)$ is the equilibrium energy of the
oscillator with frequency $\omega_0$ at temperature $T$,
\begin{equation}
{\cal E}_{eq}\left(\omega_0\right)
=\frac12 \hbar\omega_0
\coth\left(\frac{\hbar\omega_0}{2 k_B T}\right) .
\label{Eeq}
\end{equation}
Choosing different couplings between the oscillator under study and the
reservoir, one can obtain various other sets of the drift and diffusion
coefficients \cite{Dek,JRLR}, but all of them
must obey the constraint
\cite{167,Zven,Barch,Vals,Sand}
\begin{equation}
D_{pp}D_{qq}-D_{qp}^2 \ge \hbar^2\left(\gamma_p +\gamma_q\right)^2/16,
\label{diffcond}
\end{equation}
which guarantees an absence of nonphysical solutions violating the uncertainty
relations and corresponding to nonpositively definite density matrices.
Besides, the coefficients $D_{pp}$ and $D_{qq}$ must be positive.
However, it will be shown in this section,
that in the case of a weak damping,
the evolution depends on two combinations of the damping and diffusion
coefficients only,
\begin{equation}
\gamma=\frac12\left(\gamma_p+\gamma_q\right), \quad
{\cal E}_*= \frac1{2\gamma}\left(D_{pp} +\omega_0^2D_{qq} \right).
\label{def-D*}
\end{equation}
Due to the fluctuation-dissipation theorem,
the diffusion coefficients are proportional to the damping coefficients,
therefore ${\cal E}_*$ does not depend on $\gamma$, at least up to small
corrections of an order of $\gamma^2$.
The physical meaning of the parameters $\gamma$ and ${\cal E}_*$
is elucidated below: $2\gamma$ is the
reciprocal energy relaxation time of the cavity due to all possible
mechanisms (a real dissipation in the walls and the leakage through the
boundaries), i.e.
$2\gamma=\omega_0/Q$, $Q$ being the cavity quality factor, whereas
${\cal E}_* ={\cal E}_{eq}\left(\omega_0\right)$.

Solutions to Eq. (\ref{FP}) with a {\it constant\/} frequency and
different sets of constant diffusion coefficients were obtained by many
authors; they were analysed in detail in \cite{167,Dek}, where other
references can be found.
Since Eq. (\ref{FP}) looks like a two-dimensional Schr\"odinger equation
with a quadratic (although nonhermitian) Hamiltonian,
its propagator can be calculated
in the framework of the method of quantum integrals of motion \cite{Int-75}.
The explicit form of this propagator in the generic case of time dependent
coefficients can be found in \cite{167,Phys-78}.
Here we confine ourselves to calculating
the second order statistical moments and the energy (the number of photons)
of the field oscillator.

\subsection{Evolution of the energy and the second order moments}

The time dependence of the energy and the second order statistical
moments (variances) of the field mode quadrature components,
$ \sigma_{ab}\equiv \frac12\langle ab+ba\rangle -
\langle a\rangle \langle b\rangle$,
is governed by the equations following from Eq. (\ref{FP}),
\begin{eqnarray}
\dot\sigma_{qq}&=&2\sigma_{pq} -2\gamma_q \sigma_{qq} +2D_{qq},
\label{eq-sig-qq}\\
\dot\sigma_{pq}&=& \sigma_{pp} -\omega^2(t)\sigma_{qq}
-2\gamma\sigma_{pq} +2D_{pq},
\label{eq-sig-pq}\\
\dot\sigma_{pp}&=&  -2\omega^2(t)\sigma_{pq}
-2\gamma_p \sigma_{pp} +2D_{pp}.
\label{eq-sig-pp}
\end{eqnarray}
We assume that the oscillator eigenfrequency depends on time as
\begin{equation}
\omega(t)=\omega_0\left[1+2\tilde\varepsilon\sin(\Omega t)\right],
\quad
\Omega=2(\omega_0 +\delta), \quad
|\delta|\ll \omega_0, \quad |\tilde\varepsilon|\ll 1,
\label{omeg-t3}
\end{equation}
where $\omega_0$ is the unperturbed field eigenfrequency, and $\Omega$ is
the frequency of the wall vibrations.
Then one could suppose that the damping and
diffusion coefficients must depend on time, as well.
We argue, however, that in the case under study
the coefficients $\gamma_a$ and $D_{ab}$ can be considered as time
independent.
For example, let us look at the expressions (\ref{gamma}) and (\ref{D}).
In the case of a vibrating cavity,
the time variable could enter the coefficients $\gamma_a$ and $D_{ab}$
through the coupling function $g(\omega_0)$, which can depend on
the variable length of the cavity
$L(t)=L_0\left[1+\xi_L\tilde\varepsilon\sin(\Omega t)\right]$,
$\xi_L $ being a numerical coefficient.
Then one could expect a similar
time dependence of the coefficients of the Fokker--Planck equations,
$\gamma(t)=\gamma_0\left[1+\xi_{\gamma}\tilde\varepsilon\sin(\Omega t)\right]$,
$D(t)=D_0\left[1+\xi_D\tilde\varepsilon\sin(\Omega t)\right]$,
$\xi_{\gamma}$ and $\xi_{D}$ being some other numerical coefficients.
One should remember, however, that the modulation parameter $\tilde\varepsilon$
is very small under the realistic conditions: its absolute value cannot
exceed $10^{-8}$ \cite{D95,DK96}.
The set of equations (\ref{eq-sig-qq})-(\ref{eq-sig-pp}) contains three small
dimensionless parameters, $\tilde\varepsilon$, $\delta/\omega_0$,
and $\gamma_0/\omega_0$, and we are
interested in the weak damping case, when these parameters are of the same
order of magnitude. Under this condition, the time dependent parts of the
coefficients $\gamma_a$ and $D_{ab}$ are proportional to the products
$\tilde\varepsilon\gamma_0/\omega_0\sim{\cal O}(\tilde\varepsilon^2)\sim 10^{-16}$,
so it seems reasonable to neglect these extremely small terms.

A significant time dependence of the damping and diffusion coefficients
could arise in the specific case of an {\it unstable\/} reservoir,
provided that the
reservoir oscillators having the frequencies close to $\omega_0$ could be
also excited due to some resonance processes between the vibrating surface
of the wall and the
reservoir. In such a case, the energy of the resonant oscillators
would increase in time, resulting in increasing values of the damping and
diffusion coefficients (see Eq. (\ref{D})). As a consequence, we would obtain
an additional amplification of the energy of the field mode due to the
interaction with the reservoir.
However, such a model seems unrealistic, since it implies that some
distinguished degrees of freedom of the reservoir are practically
isolated from the rest of the reservoir. This conjecture contradicts the
usual concept of the
reservoir consisting of a great number of strongly interacting particles,
so that the state of the reservoir is not sensitive to small external
perturbations.
Therefore, we assume that the only time dependent coefficient in Eqs.
(\ref{eq-sig-qq})-(\ref{eq-sig-pp}) is $\omega(t)$.

The set of equations (\ref{eq-sig-qq})-(\ref{eq-sig-pp})
 is equivalent to the third order differential equation for the
variance $\sigma_{qq}\equiv\sigma$
\begin{equation}
\frac{d^3\sigma}{dt^3} +6\gamma\frac{d^2\sigma}{dt^2}
+4\omega^2(t)\frac{d\sigma}{dt}
+4\omega_0(\dot\omega +2\gamma\omega_0)\sigma = 8\gamma{\cal E}_*,
\label{eq-sig3}
\end{equation}
where the coefficients $\gamma$ and ${\cal E}_*$ are given by
Eq. (\ref{def-D*})
(we neglect the terms of the second order with respect to $\gamma$,
$D_{ab}$, and $\tilde\varepsilon$).
To find an approximate explicit solution to Eq. (\ref{eq-sig3})
with function $\omega(t)$ given by Eq. (\ref{omeg-t3}),
we use the well known method of slowly varying amplitudes, which was
exposed, e.g.
in textbooks \cite{Louis,Land,Bogol} and applied to the quantum parametric
oscillator in \cite{Raif1,Raif}.
Following this method, we write
\begin{equation}
 \sigma_{qq}(t)=A(t) +B(t)\cos(\Omega t) +C(t)\sin(\Omega t).
\label{sig-abc}
\end{equation}
so that the function (\ref{sig-abc}) with $A,B,C=const$ is an exact solution
to Eq. (\ref{eq-sig3}) for $\tilde\varepsilon=\gamma=\delta=0$.
Supposing that the dimensionless small parameters $\gamma/\omega_0$ and
$\delta/\omega_0$ have the same orders of magnitude as the small parameter
$\tilde\varepsilon$, i.e. $\gamma=\tilde{\gamma}\tilde\varepsilon\omega_0$,
$\delta=\tilde{\delta}\tilde\varepsilon\omega_0$,
$\tilde{\gamma},\tilde{\delta}\sim {\cal O}(1)$,
we assume that the amplitude coefficients $A,B,C$ are functions of the
``slow time'' $\tau=\tilde\varepsilon t$, so that the time derivatives
$d^k A/dt^k, d^k B/dt^k, d^k c/dt^k$ are proportional to $\tilde\varepsilon^k$
($k=1,2,3$).
Then we put the function (\ref{sig-abc}) into Eq. (\ref{eq-sig3})
and neglect the terms proportional to $\tilde\varepsilon^2$ and $\tilde\varepsilon^3$.
Besides, we perform averaging over fast oscillations with the frequency
$\Omega$, in order to eliminate higher harmonics with frequences $m\Omega$,
$m=2,3,\ldots$, whose amplitudes are proportional to $\tilde\varepsilon^{m-1}$
\cite{Land,Bogol}).
Finally, we arrive at the following set of equations for the slowly varying
amplitudes (here the overdot means the derivative with respect to
the real time $t$, and the new parameter $\kappa\equiv \tilde\varepsilon\omega_0$
has the dimensionality of frequency, like $\gamma$ and $\delta$):
\begin{eqnarray}
\dot{A}&=& -2\gamma A +2\kappa B +2\gamma{\cal E}_*,
\label{eq-a}\\
\dot{B}&=& 2\kappa A-2\gamma B -2\delta C,
\label{eq-b}\\
\dot{C}&=& 2\delta B-2\gamma C .
\label{eq-c}
\end{eqnarray}
This system can be easily solved for an arbitrary time dependent function
$\gamma(t)$, since the substitution
$A(t)=\tilde A(t)\exp\left[-2\int\gamma(\tau)d\tau\right]$
removes the function $\gamma(t)$ from the homogeneous parts of Eqs.
(\ref{eq-a})-(\ref{eq-c}). However, we consider the case of constant
coefficients only, due to the physical reasons discussed above.

To simplify the formulas, we assume hereafter $\hbar=\omega_0=1$; thence
$\Omega=2$ in the amplitude coefficients.
Neglecting small terms
of the order of $\gamma/\Omega$, $\kappa/\Omega$,
and $\delta/\Omega$ in the amplitude coefficients, one can express the
variances $\sigma_{pp}$ and $\sigma_{pq}$ as follows
\begin{eqnarray}
 \sigma_{pp}(t)&=&A(t) -B(t)\cos(\Omega t) -C(t)\sin(\Omega t),
\label{sigpp-abc} \\
 \sigma_{pq}(t)&=&C(t)\cos(\Omega t) -B(t)\sin(\Omega t).
\label{sigpq-abc}
\end{eqnarray}
Then the initial conditions for the set (\ref{eq-a})-(\ref{eq-c}) read
\[
A(0)=\left[\sigma_{qq}(0)+\sigma_{pp}(0)\right]/2, \quad
B(0)=\left[\sigma_{qq}(0)-\sigma_{pp}(0)\right]/2, \quad
C(0)=\sigma_{pq}(0).
\]
Evidently, $A(t)$ coincides with the mean energy to within small corrections
of the order of $\tilde\varepsilon$, $\gamma$, and $\delta$:
\[
{\cal E}(t) \equiv \left(\sigma_{pp}+\sigma_{qq}\right)/2=A(t).
\]

In order to elucidate the meaning of the coefficients $B$ and $C$, consider
the determinant of the invariance matrix (cf. Eq. (\ref{def-dT}))
\begin{equation}
d\equiv \sigma_{pp}\sigma_{qq} -\sigma_{pq}^2 \ge \hbar^2/4.
\label{SR}
\end{equation}
Here the last inequality holds due to
 the Schr\"odinger--Robertson uncertainty relation \cite{Kur,183,S,R}.
The meaning of the parameter $d$ as the {\em universal quantum invariant\/}
is discussed in \cite{D-inv00}.
The {\em minimal invariant variance\/} (\ref{def-uv})
can be expressed in terms of ${\cal E}$ and $d$ as
\begin{equation}
u ={\cal E} -\sqrt{{\cal E}^2-d}=\frac{d}{{\cal E} +\sqrt{{\cal E}^2-d}}\,,
\label{sEd}
\end{equation}
and one can easily verify the relations
\be
d=A^2-B^2-C^2, \quad
u =A -\sqrt{B^2+C^2}.
\label{d-ABC}
\ee

The solutions to Eqs. (\ref{eq-a})-(\ref{eq-c}) read
\begin{eqnarray}
A(t) &=& A_* + e^{-2\gamma t}
 \left[a_0\delta + a_{+}\kappa e^{2\nu t} + a_{-}\kappa e^{-2\nu t} \right],
\label{A(t)}\\
B(t) &=& B_* + e^{-2\gamma t}
 \left[a_{+}\nu e^{2\nu t} - a_{-}\nu e^{-2\nu t} \right],
\label{B(t)}\\
C(t) &=& C_* + e^{-2\gamma t}
 \left[a_0\kappa + a_{+}\delta e^{2\nu t} + a_{-}\delta e^{-2\nu t} \right],
\label{C(t)}
\end{eqnarray}
where $\nu=\sqrt{\kappa^2-\delta^2}$, and
\begin{eqnarray}
a_0 &=&\frac1{\nu^2} \left[\kappa C(0) -\delta A(0) +{\cal E}_*\delta
\right],
\label{a0}\\
a_{\pm}&=& \frac1{2\nu^2}\left[\kappa A(0) -\delta C(0) \pm \nu B(0)
-  \frac{\kappa \gamma{\cal E}_*}{\gamma\mp\nu} \right],
\label{a+-}
\end{eqnarray}
\[
A_*= \frac{\gamma^2+\delta^2}{\gamma^2-\nu^2}{\cal E}_*, \quad
B_*= \frac{\kappa \gamma{\cal E}_*}{\gamma^2-\nu^2}, \quad
C_*= \frac{\kappa\delta {\cal E}_*}{\gamma^2-\nu^2}.
\]
The meanings of the parameters $\gamma$ and ${\cal E}_*$ become clear,
if one considers the
special case of the oscillator with time independent coefficients,
$\kappa=\delta=\nu=0$. Then we see that $2\gamma$ is the energy relaxation
coefficient, so that it can be expressed in terms of the cavity $Q$-factor
by means of the relation $2\gamma=\omega_0/Q$. The energy of the oscillator
in this special case tends to ${\cal E}_*$ as $t\to\infty$, and this value
can be identified with the thermodynamic equilibrium oscillator energy
${\cal E}_{eq}$ given by Eq. (\ref{Eeq}) (up to corrections of the order of
$(\gamma/\omega_0)^2$ \cite{167,Dek}.

The sign of the difference $\nu-\gamma$ determines the regions of stable
and unstable solutions of Eq. (\ref{eq-sig3}) in the space of parameters
$\kappa,\delta,\gamma$ (for small values of these parameters). The stable
(limited in time) solutions exist for large values of the damping or
detuning coefficients, $\nu <\gamma$, i.e.
\begin{equation}
\gamma^2 +\delta^2 > \kappa^2.
\label{thresh}
\end{equation}
In this case, the final state of the oscillator does not depend on the initial
conditions. The asymptotical values of the energy, the $d$-factor,
and the minimal invariant variance read
\be
{\cal E}(\infty)= \frac{\gamma^2+\delta^2}{\gamma^2-\nu^2}{\cal E}_{eq},
   \quad
d(\infty)=
 \frac{\gamma^2+\delta^2}{\gamma^2-\nu^2}{\cal E}_{eq}^2,
\quad
u(\infty) =\frac{\sqrt{\gamma^2 +\delta^2}}
{\kappa +\sqrt{\gamma^2 +\delta^2}}{\cal E}_{eq}.
\label{sinf}
\ee
At zero temperature, when $\kappa$ tends to the threshold, the minimal
variance $u$ goes to the value $1/4$, which is twice less than in the
coherent state.

Above the threshold, i.e. in the instability region $\nu >\gamma$,
the energy (or the number of photons) increases
exponentially for $(\nu-\gamma)t\gg 1$,
\begin{equation}
{\cal E}(t)= a_{+}\kappa e^{2(\nu-\gamma) t},
\label{Eas}
\end{equation}
and it depends on the initial conditions through the coefficient $a_+$.
Using Eq. (\ref{a+-}) and taking into account the uncertainty relation
$d\ge 1/4$ (for $\hbar=1$), one can verify that
the coefficient $a_+$ is bounded from below by a positive value,
\[
a_+ > \frac{\kappa\gamma{\cal E}_{eq}}{2\nu^2(\nu -\gamma)}.
\]
Consider a special case of initial thermal equilibrium state.
Then
\begin{equation}
a_{\pm}^{(eq)}= \frac{\kappa{\cal E}_{eq}}{2\nu(\nu \mp\gamma)},
\quad a_0^{(eq)}=0,  \quad d(0)={\cal E}_{eq}^2,
\label{a+eq}
\end{equation}
thus $d$-factor (\ref{d-ABC}) depends on time as
\be
d(t)=  \frac{{\cal E}_{eq}^2 }{\nu(\nu^2-\gamma^2)}\left[
2\kappa^2\gamma e^{-2\gamma t}\sinh(2\nu t) +\kappa^2\nu e^{-4\gamma t}
- \nu(\gamma^2 +\delta^2)\right].
\label{d(t)}
\ee
If $\gamma>0$ and $(\nu-\gamma)t\gg 1$, then due to Eq. (\ref{sEd}),
the minimal variance tends asymptotically to a constant value
\begin{equation}
u_{\infty} =\frac{\gamma}{\gamma +\nu}{\cal E}_{eq}.
\label{stinf}
\end{equation}
Consequently, a large squeezing can be achieved even for a high
temperature initial state, if $\nu\gg\gamma$.
If $\gamma=0$, then $d$ does not depend on time, $d\equiv{\cal E}_{eq}^2$,
and for $\nu t\gg 1$ the minimal variance goes asymptotically to zero as
$u\approx{\cal E}_{eq}(\nu/\kappa)^2\exp(-2\nu t)$.
One should remember, nonetheless, that the
solutions indicating the exponential growth of the energy are justified until
$t\ll t_2\sim (\omega_0\tilde\varepsilon^2)^{-1}$, since
for larger times the neglected second order terms in Eqs. (\ref{eq-sig3}),
(\ref{eq-a})-(\ref{eq-c}) could become important.
However, the time $t_2$ is very large under the realistic conditions.

\section{Discussion}\label{Discuss}

We have demonstrated a significant progress in our understanding and
{\it quantitative\/} description of quantum processes in cavities with
moving boundaries, achieved 30 years after the pioneer paper by
Moore \cite{Moore} and almost 80 years after the first papers on
{\it classical electrodynamics\/} in such cavities by Nicolai and
Havelock \cite{Nic1,Nic2,Hav}).
We have shown that quanta of electromagnetic field can be created from
vacuum in a cavity with vibrating walls {\it under the resonance condition\/}
(and cannot be generated for nonresonance nonrelativistic laws of boundary
motion, in particular, in the case of a large detuning from the resonance),
and the quantum state of field exhibits the ``nonclassical'' properties.

The possibility of observing the effect
depends crucially on the achievable values of the wall
displacement amplitude. For the cavity dimensions of the order of
$1\div100$ centimeters, the resonance frequency $\omega_0/\pi$ varies from
30 GHz to 300 MHz. It is difficult to imagine that the wall could be forced
to oscillate as a whole at such a high frequency. Rather, one could think on
the oscillations of the {\em surface} of the cavity wall. In such a case one
has to find a way of exciting a sufficiently strong standing acoustic wave at
frequency $\omega_w=2\omega_0$  inside the wall.
The amplitude $a$ of this wave (coinciding with the amplitude of
oscillations of the free surface) is connected with the relative deformation
amplitude $\delta$ inside the wall as $\delta =\omega_w a/v_s$,
where $v_s$ is the sound velocity. Since the usual materials cannot bear the
deformations exceeding the value $\delta_{max}\sim 10^{-2}$
\cite{Table}, the maximal possible velocity of the boundary appears
$v_{max}\sim \delta_{max}v_s \sim 50$~m/s (independent on the
frequency). Thus the maximal dimensionless displacement $\varepsilon=a/L_0$
is $\varepsilon_{max}\sim(v_s/2\pi c)\delta_{max}\sim 3\cdot 10^{-8}$
for the lowest mode with the frequency $\omega_0\sim c\pi/L_0$. It also
does not depend on the frequency. Consequently,
the maximal rate of photon generation in the principal mode of a 1D cavity
can be estimated as
\begin{equation}
\left(\frac{\mbox{d}{\cal P}_1}{\mbox{d}t}\right)_{max}=
\frac{4}{\pi^2} \frac{v_s}{c}\delta_{max}\frac{\omega_1}{2\pi}
\sim 6\cdot 10^{-8}\omega_1/2\pi.\label{max-phot}
\end{equation}
It is proportional to the frequency. For $\omega_1/2\pi = 10$~GHz
(corresponding to a distance between the plates of the order of several
centimeters) we get 600 photons/sec.

This number can be significantly increased in a 3D cavity, due to the
exponential law (\ref{num-1}). For the same frequency
$\omega_0/2\pi = 10$~GHz, the maximal value of parameter
$\mu=\gamma\omega_0 t$ equals $\mu_{max}\sim 600 t$, time $t$
being expressed in seconds. Even if the amplitude of the vibrations
were 100 times less than the maximal possible value, in $t=1$~s one could get
about $\sinh^2(6)\approx 4\cdot10^{4}$ photons in an empty cavity.
Obviously, the concrete shape of a 3D cavity
is not important. The significant requirements are:
{\it i}) the nondegenerate
character of the eigenfrequency spectrum, and {\it ii}) the condition
of the parametric resonance between the oscillating wall and
some electromagnetic mode.
The {\it total} energy of
photons created in the 1D cavity is approximately the same as in the
three-dimensional case. The difference is that
in the 1D case this energy is spread over many
interacting modes, resulting in moderate numbers of quanta in each mode.
The rate of
photon generation in the $m$-th (odd) mode of the 1D cavity is approximately
$m$ times less than in the principal mode with $m=1$
(in the asymptotical regime).

To create the above-mentioned 600 or $4\cdot 10^4$ photons, one should
vibrate the wall for not less than $1$~second. The necessary Q-factor of
the cavity must be $Q\sim 3\cdot 10^{10}$. This value was achieved in
experiments already several years ago \cite{Wal90}.
An unsolved problem is how to excite the high-frequency surface vibrations
with a sufficiently large amplitude. One could think, for instance, on using
some kind of piezoeffect. This method was successfully applied in early
experiments devoted to solving the mode-locking and pulse production problems
in lasers with the aid of vibrating mirrors. The displacements of the mirror
from 0.1~$\mu$m to 0.7~$\mu$m at the frequency 500~KHz were achieved in
\cite{500KHza,500KHzb}.
In \cite{Henneb} the resonance vibrations of the mirror
in a laser with the length 250~cm (i.e., at a frequency about 100~MHz)
were excited with the aid of a quarz transducer. However, for our purposes
the frequency 100~MHz is too small, since the parameter $\mu$ becomes 100
times less, comparing with the estimations given above (remember that
$\varepsilon_{max}$ does not depend on the frequency).

Fortunately, the results of recent studies \cite{DA,Lamb-puls,Plun00,Hui00}
show that the influence of temperature is not so significant as it
could appear at first glance. Moreover, in certain situations the
initial temperature fluctuations could be used to amplify the
effect \cite{Plun00}. However, the resonance requirements are rather
hard: for the values of the frequency $\omega\sim 10^{10}$~Hz
and the maximal possible modulation parameter $\varepsilon \sim 10^{-8}$
the admissible detuning should not exceed $100$~Hz during the whole
time interval $\sim 1\,$s, necessary to accumulate the resonance effect.

One of the reasons of the studies on the dynamical Casimir effect
for the last few
years was Schwinger's hypothesis \cite{Sch1,Sch2,Sch3,Sch4,Sch5}
that this effect could explain the {\em sonoluminescence\/} phenomenon,
i.e., the emission of bright short pulses of the visible light from
the gas bubbles in the water, when the bubbles
pulsate due to the pressure oscillations
in a strong standing acoustic wave.
(For the reviews and numerous references related to this effect see, e.g.,
\cite{Lamor99,sonrev,Barb97,son-2}.)
There are several publications, e.g., \cite{Sas,Eb-son1,Eb-son2},
whose authors considered the models giving tremendous numbers of photons
which could be produced even in the visible range due to the fast
motion of the boundaries. However, the analysis of these models shows
that they are based on such laws of motion of the boundaries which
imply the superluminal velocities, so they are not realistic.

Although the results of this chapter, being obtained in the framework of
simplified one-dimensional and three-dimensional models,
cannot be applied directly to the analysis of the sonoluminescence
problem, they are not in favor
of Schwinger's hypothesis. The main difficulty is
connected with quite different time scales of the phenomena.
The accumulation of the `dynamical Casimir
energy' is a very slow process, which needs a great number of wall
oscillations, whereas the sonoluminescence pulses (containing up to $10^7$
photons) have the duration of the order of picoseconds.
Moreover, the wall oscillations must be in extremely fine tuned resonance
with the field eigenfrequencies, since the detuning $\delta>\vep$
completely destroys the energy growth \cite{Djpa}. In particular,
if the frequency of the wall oscillations $\omega_{wall}$ is much less
than the minimal field eigenfrequency $\omega_{1}$, then
the field variation is adiabatic, and the mean number
of created photons is proportional to \cite{DK96}
$\vep^2\left(\omega_{wall}/\omega_1\right)^4 \ll 1$.
These features survive in the three-dimensional model considered in this
chapter, too. Therefore,
it is difficult to believe that very specific conditions of the parametric
resonance described
above could arise naturally in the sonoluminescence case.
For other discussions of the problem see, e.g., \cite{Esq} and the
contributions in Ref. \cite{Cas50}.

Actually, the main obstacle to produce the `Casimir light'
is the very low ratio of the wall velocity to the speed of
light in possible laboratory experiments.
If the velocity of the boundary were of the order of $c$, then a
sufficient number of photons could be created from vacuum practically
for any law of motion. For the nonrelativistic velocities the only
possibility is to accumulate the effect gradually under the resonance
conditions.
Nonetheless, perhaps, the experimental situation could
be improved in the case of using some kinds of `effective mirrors',
such as, e.g., the layers made of the electron-hole plasma \cite{Loz},
or some others. Therefore, we cannot exclude a possibility that
in a not very remote future one could assist a show of
``quantum magics,'' when some ``quantum magician'' takes an empty box,
then shakes it well, opens --- and an astonished audience would see
a great number of photons which have appeared ``from nothing''
due to the Nonstationary Casimir Effect.

\subsection*{Acknowledgments}
I would like to thank my coauthors V. I. Man'ko, A. B. Klimov and
M. A. Andreata for the long-term fruitful collaboration. Also,
I would like to express my sincere gratitude to Prof. S. S. Mizrahi and
the Physics Department of the Federal University of S\~ao Carlos, since
this review could not be written without their great support and help.

\end{document}